\documentclass[reqno,10pt, centertags,draft]{amsart}
\usepackage{amsmath,amsthm,amscd,amssymb,latexsym,upref}

\newcommand{\e}{\hbox{\rm e}}

\newcommand{\bbN}{{\mathbb{N}}}
\newcommand{\bbR}{{\mathbb{R}}}

\newcommand{\bbZ}{{\mathbb{Z}}}
\newcommand{\bbC}{{\mathbb{C}}}

\newcommand{\cA}{{\mathcal A}}

\newcommand{\cK}{{\mathcal K}}
\newcommand{\cL}{{\mathcal L}}
\newcommand{\cM}{{\mathcal M}}
\newcommand{\cN}{{\mathcal N}}

\newcommand{\cP}{{\mathcal P}}

\newcommand{\cS}{{\mathcal S}}

\newcommand{\cV}{{\mathcal V}}

\newcommand{\Oh}{O}

\newcommand{\bi}{\bibitem}
\newcommand{\no}{\nonumber}
\newcommand{\lb}{\label}
\newcommand{\f}{\frac}
\newcommand{\ul}{\underline}
\newcommand{\ol}{\overline}
\newcommand{\hatt}{\widehat}  
\newcommand{\ti}{\tilde}
\newcommand{\wti}{\widetilde  }

\renewcommand{\Re}{\text{\rm Re}}
\renewcommand{\Im}{\text{\rm Im}}
\renewcommand{\ln}{\text{\rm ln}}

\newcommand{\floor}[1]{\lfloor#1\rfloor}

\DeclareMathOperator{\KdV}{KdV}
\DeclareMathOperator{\sKdV}{s-KdV}
\DeclareMathOperator{\shKdV}{s-\widehat{KdV}}

\allowdisplaybreaks
\numberwithin{equation}{section}

\newtheorem{theorem}{Theorem}[section]
\newtheorem{lemma}[theorem]{Lemma}
\newtheorem{corollary}[theorem]{Corollary}

\theoremstyle{definition}
\newtheorem{definition}[theorem]{Definition}
\newtheorem{remark}[theorem]{Remark}
\newtheorem{example}[theorem]{Example}

\begin{document}

\title[Calogero--Moser Systems]{An Explicit Characterization \\ of
Calogero--Moser Systems}
\author[F.\ Gesztesy, K.\ Unterkofler, and R.\ Weikard]{Fritz Gesztesy,
Karl Unterkofler, and Rudi Weikard}
\address{Department of Mathematics,
University of Missouri, Columbia, MO 65211, USA}
\email{fritz@math.missouri.edu}
\urladdr{http://www.math.missouri.edu/people/fgesztesy.html}
\address{Department of Computer Science,
Applied Mathematics Group,
FH-Vorarlberg,
A--6850 Dornbirn,
Austria}
\email{karl.unterkofler@fh-vorarlberg.ac.at}
\address{Department of Mathematics,
University of Alabama at Birmingham, \\
Birmingham, AL  35294--1170, USA}
\email{rudi@math.uab.edu}
\urladdr{http://www.math.uab.edu/rudi}
\date{May 28, 2003}
\thanks{Based upon work supported by the
US National Science Foundation under Grant No.~DMS-9970299.}

\begin{abstract}
Combining theorems of Halphen, Floquet, and Picard and a
Frobenius type analysis, we characterize  rational, meromorphic simply
periodic, and elliptic KdV potentials. In particular, we explicitly
describe the proper extension of the Calogero--Moser locus associated
with these three classes of  algebro-geometric solutions of the KdV
hierarchy with special  emphasis on the case of multiple collisions
between the poles of solutions.
\end{abstract}

\maketitle

\section{Introduction} \lb{s1}

The principal purpose of this paper is to analyze rational,
meromorphic simply periodic, and elliptic (algebro-geometric) solutions
of the Korteweg-de Vries (KdV) hierarchy of nonlinear evolution
equations and the associated Calogero--Moser-type models. In
particular, we derive an explicit characterization of the (properly
extended) Calogero--Moser locus for stationary rational, periodic
and elliptic solutions of the KdV hierarchy. Our techniques rely
on a combination of a Frobenius-type analysis with results of
Halphen, Floquet, and Picard in the rational, simply periodic,
and elliptic case, respectively.

Next we describe this topic in more detail. We freely use
the notation introduced in Appendix \ref{sA} in connection with the
KdV hierarchy. In particular, we will often call a solution $q$ of some
equation of the stationary KdV hierarchy (and hence of infinitely many
such equations) a {\it KdV potential}.

We first consider the case of rational solutions $q$ of the stationary
KdV hierarchy. All such (nonconstant) solutions $q$ are known to be
necessarily of the form
\begin{equation}
q(z)=q_0-\sum_{\ell=1}^M s_\ell(s_\ell+1)(z-\zeta_\ell)^{-2}, \lb{1.1}
\end{equation}
where
\begin{align}
\begin{split}
& q_0\in\bbC, \quad \{\zeta_\ell\}_{1\leq\ell\leq
M}\subset\bbC, \; \zeta_\ell^\prime\neq\zeta_\ell \text{ for }
\ell^\prime\neq\ell, \, 1\leq \ell, \ell' \leq M, \\
& s_\ell\in\bbN, \; 1\leq\ell\leq M,
\quad \sum_{\ell=1}^M s_\ell(s_\ell+1)=g(g+1)
\text{ for some } g\in\bbN. \lb{1.2}
\end{split}
\end{align}
The underlying spectral curve is then of the simple rational type
\begin{equation}
y^2=(E-q_0)^{2g+1}. \lb{1.3}
\end{equation}
\noindent (To avoid annoying case distinctions we will in almost all
circumstances exclude the trivial case $N=g=0$ in this paper.)

On the other hand, not every $q$ of the type  \eqref{1.1}, \eqref{1.2} is
an algebro-geometric solution of the KdV hierarchy. In general, the
points $\zeta_\ell$ must satisfy a set of intricate constraints. In
fact, necessary and sufficient  conditions on $\zeta_\ell$ for $q$ in
\eqref{1.1} to be a rational KdV solution are given by
\begin{equation}
\sum_{\substack{\ell^\prime=1\\ \ell^\prime\neq \ell}}^M
\frac{s_{\ell^\prime}(s_{\ell^\prime}+1)}{(\zeta_{\ell}
-\zeta_{\ell^\prime})^{2k+1}}=0
    \text{ for $1\leq k\leq s_{\ell}$ and $1\leq \ell\leq M$.} \lb{1.4}
\end{equation}
This result was first derived by Duistermaat and Gr\"unbaum
\cite[p.\ 199]{DG86} in 1986, as a by-product of their
investigations of bispectral pairs of differential operators. An
elementary alternative derivation of this result on the basis of
Halphen's theorem, describing the structure of fundamental systems of
solutions of differential equations with rational coefficients (and a
growth restriction at infinity), and an explicit Frobenius-type analysis
were recently provided in our paper \cite{GUW00}. For the convenience
of the reader we will summarize these results in Section~\ref{s2}.

For a fixed $g\in\bbN$, \eqref{1.2} and
\eqref{1.4} yield a complete parametrization of all rational KdV
potentials belonging to the  spectral curve \eqref{1.3}. In other
words, they provide a complete  characterization of the
isospectral class of KdV potentials corresponding to \eqref{1.3}.
The constraints \eqref{1.4} represent the proper  generalization
of the locus of poles introduced by Airault, McKean, and Moser
\cite{AMM77} in the sense that they explicitly describe the
situation where poles are permitted to collide (i.e., where some
of  the $s_\ell >1$). In this context it seems appropriate to
recall the collisionless case associated with the traditional
rational  Calogero--Moser locus. In that case $q$ is of the form,
\begin{equation}
q(z)=q_0-2\sum_{j=1}^N (z-z_j)^{-2}, \lb{1.4a}
\end{equation}
where
\begin{align}
\begin{split}
& q_0\in\bbC, \quad N=g(g+1)/2 \text{ for some $g\in\bbN$,} \lb{1.4b}\\
&\{z_j\}_{1\leq j\leq N}\subset\bbC,
\quad z_j\neq z_{j'} \text{ for } j\neq j', \; 1\leq j,j'\leq N
\end{split}
\end{align}
and the corresponding Calogero--Moser locus is then given by
\begin{equation}
\sum_{j'=1, j'\neq j}^N (z_j-z_{j'})^{-3}=0, \quad 1\leq j\leq N.
\lb{1.4c}
\end{equation}
Equations \eqref{1.1}, \eqref{1.2}, and \eqref{1.4} are then the proper
extensions of the traditional equations \eqref{1.4a}, \eqref{1.4b}, and
\eqref{1.4c} in the presence of collisions, where some of the $z_j$
are permitted to cluster in groups of $s_\ell(s_\ell+1)/2$ mutually
distinct points $\zeta_\ell$ with $\sum_{\ell=1}^M
s_\ell(s_\ell+1)=2N$, $s_\ell\in\bbN$, $1\leq\ell\leq M$.

In the case of elliptic solutions of the stationary KdV hierarchy it
is known that all such (nonconstant) solutions $q$ are necessarily of
the form
\begin{equation}
q(z)=q_0-\sum_{\ell=1}^M s_\ell(s_\ell+1)\wp(z-\zeta_\ell), \lb{1.5}
\end{equation}
where
\begin{align}
\begin{split}
& q_0\in\bbC, \quad \{\zeta_\ell\}_{1\leq\ell\leq
M}\subset\bbC, \; \zeta_\ell^\prime\neq\zeta_\ell \text{ for }
\ell^\prime\neq\ell,  \, 1\leq \ell, \ell' \leq M, \\
& s_\ell\in\bbN, \; 1\leq\ell\leq M. \lb{1.6}
\end{split}
\end{align}
(Here $\wp$ denotes the Weierstrass elliptic function, cf.\
\cite[Ch.\ 18]{AS72} and Appendix \ref{sB}.)

On the other hand, as in the rational context, not every $q$ of the type
\eqref{1.5}, \eqref{1.6} is an algebro-geometric solution of the KdV
hierarchy. Again, the points $\zeta_\ell$ must satisfy an analogous set
of intricate constraints. In fact, combining a Frobenius-type analysis
and a theorem of Picard, describing the structure of solutions of
differential equations with elliptic coefficients, we derive necessary
and sufficient  conditions on $\zeta_\ell$ for $q$ in \eqref{1.5} to be
an elliptic solution. More precisely, our principal result, to be
proven in Section \ref{s2}, reads as follows.

\begin{theorem} \lb{t1.1}
Let $q$ be an elliptic function. Then $q$ is a Picard
potential, that is, the differential equation $y''+qy=Ey$ has a
meromorphic fundamental system of solutions $($w.r.t.~$z$$)$ for
each value of the complex  spectral parameter $E\in\bbC$, if and only if
there are $M\in\bbN$, $s_\ell\in\bbN$,
$1\leq\ell\leq M$, $q_0\in\bbC$, and pairwise distinct
$\zeta_\ell\in\bbC$, $1\leq \ell\leq M$, such that
\begin{equation}
q(z)=q_0-\sum_{\ell=1}^M s_\ell(s_\ell+1)\wp(z-\zeta_\ell) \lb{1.7}
\end{equation}
and
\begin{equation}
\sum_{\substack{\ell^\prime=1\\ \ell^\prime\neq \ell}}^M
    s_{\ell^\prime}(s_{\ell^\prime}+1) \wp^{(2k-1)}(\zeta_{\ell}
-\zeta_{\ell^\prime})=0
\text{ for $1\leq k \leq s_{\ell}$ and $1\leq \ell\leq M$.} \lb{1.8}
\end{equation}
Moreover, $q$ is an elliptic KdV potential if and only if $q$ is of
the type \eqref{1.7} and the constraints \eqref{1.8} hold.
\end{theorem}

Since $\wp(z)$ converges to $1/z^2$ in the limit as both of its periods
tend to infinity, condition \eqref{1.4} is the rational analog of
\eqref{1.8}.

To the best of our knowledge, the characterization \eqref{1.8} of the
extended elliptic Calogero--Moser locus appears to be new in spite of
the extensive attention this topic has received over the past two
decades. A discussion of the pertinent literature will be provided in
Section
\ref{s2}.

Again we briefly comment on the collisionless case associated with
the traditional elliptic Calogero--Moser locus. In that case $q$ is of
the form,
\begin{equation}
q(z)=q_0-2\sum_{j=1}^N \wp(z-z_j), \lb{1.8a}
\end{equation}
where
\begin{equation}
   q_0\in\bbC, \quad \{z_j\}_{1\leq j\leq N}\subset\bbC,
\quad z_j\neq z_{j'} \text{ for } j\neq j', \; 1\leq j,j'\leq N
\lb{1.8b}
\end{equation}
and the corresponding Calogero--Moser locus is then given by
\begin{equation}
\sum_{j'=1, j'\neq j}^N \wp'(z_j-z_{j'})=0, \quad 1\leq j\leq N.
\lb{1.8c}
\end{equation}
Equations \eqref{1.7} and \eqref{1.8} are then the proper extensions of
the traditional equations \eqref{1.8a}, \eqref{1.8b}, and
\eqref{1.8c} in the presence of collisions, where some of the $z_j$
are permitted to cluster in groups of $s_\ell(s_\ell+1)/2$ mutually
distinct points $\zeta_\ell$ with $\sum_{\ell=1}^M
s_\ell(s_\ell+1)=2N$, $s_\ell\in\bbN$, $1\leq\ell\leq M$.

We also prove the analog of Theorem \ref{t1.1} for the case of simply
periodic meromorphic KdV potentials bounded near the ends of the period
strip, by combining the same kind of Frobenius-type analysis with a
variant of Floquet's theorem, describing the structure of solutions of
differential equations with simply periodic meromorphic coefficients.

In Section \ref{s2} we provide the necessary background for rational,
simply periodic, and elliptic KdV potentials and present our principal
result on the extended Calogero--Moser locus in Theorem \ref{t3.5}.
Section \ref{s3} provides additional results on the extended
Calogero--Moser locus in the rational and simply periodic cases. In
particular, in these cases we prove that the extended Calogero--Moser
locus is the closure of the traditional Calogero--Moser locus in an
appropriate (in fact, canonical) topology. We also provide a
detailed discussion of the isospectral manifold of simply
periodic meromorphic KdV potentials in Section \ref{s3}. Our
final Section \ref{s4} then provides some applications to the
time-dependent KdV hierarchy and the dynamics of poles of
rational, simply periodic, and elliptic KdV solutions with
particular emphasis on collisions of poles. Appendix \ref{sA}
reviews basic facts on the KdV hierarchy, Appendix \ref{sB}
summarizes essentials of elliptic functions, Appendix
\ref{sC} recalls some results on symmetric products of Riemann
surfaces, and Appendix \ref{sD} provides the proof of Theorem \ref{t2.13}.

Although this paper is not directly concerned with the
Kadomtsev-Petviashvili (KP) hierarchy and its connection with
Calogero--Moser-type systems, it is clear that this connection is
responsible for much of the fascination surrounding this circle of ideas.
In this context we refer to \cite{AKV02}, \cite{Ch79}, \cite{Ka95},
\cite{Kr78}--\cite{KBBT95}, \cite{Or98}, \cite{Pe94},
\cite{Sh94}, \cite{TV93}, \cite{Va99}, \cite{Va99a}, \cite{Wi98}.

\section{Rational, simply periodic, and elliptic solutions \\ of
the stationary KdV hierarchy} \lb{s2}

In this section we recall an application of Halphen's theorem to
rational solutions of the KdV hierarchy recently presented in
\cite{GUW00} and then extend these arguments to simply periodic and
elliptic KdV potentials using corresponding theorems by Floquet and
Picard. More precisely, we revisit stationary rational KdV potentials
bounded near infinity (cf.\ \cite{AS78}, \cite{AM78}, \cite{Ai78},
\cite{AMM77}, \cite{CC77}, \cite{EK82}, \cite{Gr82}, \cite{Kr74},
\cite{Mc79}, \cite{Mo77}, \cite{Oh88}, \cite{So78}, \cite{VP00},
\cite{We99} and the literature cited therein), stationary simply periodic
KdV potentials bounded near the ends of the period strip (cf.\
\cite{AMM77}, \cite{Th76}, \cite{We99}), and stationary elliptic KdV
solutions (cf.\ \cite{AV01}, \cite{AMM77}, \cite{Bi87}, \cite{Ca78},
\cite{CC77}, \cite{DS00}--\cite{DN75}, \cite{En83}--\cite{EK94},
\cite{GP99}, \cite{IE86}, \cite{Kr94}, \cite{Sm89}, \cite{Sm94},
\cite{Tr89}--\cite{VP00} and the literature cited therein).
In particular, we completely characterize the so called {\it locus} of
Calogero--Moser-type systems employing an elementary Frobenius analysis.
The time-dependent case (including a discussion of collisions) will be
presented in the next section.

The principal results on the stationary KdV hierarchy, as needed in
this section, are summarized in Appendix \ref{sA}, and we
freely use these results and the notation established there in what
follows.

We start by describing Halphen's original result. Consider the following
$n$th-order differential equation
\begin{align}
    q_n(z)y^{(n)}(z)+q_{n-1}(z) y^{(n-1)}(z)+\dots+
q_0(z) y(z) =0, \label{2.1A}
    \end{align}
where $q_j$, $0\leq j\leq n$, are polynomials, and the order of
$q_n$ is at least the order of $q_j$ for  all  $0\leq j\leq (n-1)$,
that is,
\begin{subequations} \lb{2.2A}
\begin{align}
& q_m \text{ are polynomials, $0\leq m\leq n$,} \lb{2.3A} \\
& q_m/q_n \text{ are bounded near $\infty$ for all
$0\leq m \leq n-1$}. \lb{2.4A}
\end{align}
\end{subequations}
Then the following theorem due to Halphen holds.

\begin{theorem} $($Halphen \cite{Ha85}, Ince
\cite[p.~372--375]{In56}$)$ \lb{t2.1}
Assume \eqref{2.2A} and suppose the differential equation \eqref{2.1A} has
a meromorphic fundamental system of solutions. Then the general solution
of \eqref{2.1A} is of the form
\begin{align}
y (z ) =  \sum_{m=1}^{n} c_m r_m(z) e^{\lambda_m z}, \lb{2.5A}
    \end{align}
where   $r_m$ are rational functions, $\lambda_m\in\bbC$, $1\leq m\leq n$,
and $c_m$, $1\leq m\leq n$ are arbitrary complex constants. \\
Conversely, suppose $r_m$ are rational functions and
$\lambda_m, c_m\in\bbC$, $1\leq m\leq n$. If $r_1(z)e^{\lambda_1 z},
\dots,r_n(z)e^{\lambda_n z}$ are linearly independent, then
\begin{align}
y (z ) =  \sum_{m=1}^{n} c_m r_m(z) e^{\lambda_m z} \lb{2.6A}
\end{align}
is the general solution of an $n$th-order equation of the type
\eqref{2.1A}, whose  coefficients satisfy \eqref{2.2A}.
\end{theorem}

For an extension of Theorem \ref{t2.1} to first-order $n\times n$ systems
and the explicit structure of the corresponding fundamental system of
solutions we refer to our recent paper \cite{GUW00}.

Next, we treat the case of a simply periodic meromorphic
potential. Halphen's theorem is then replaced by a variant of Floquet's
theorem which we state below as Theorem \ref{t4.1}.

First, we recall a few basic facts from the theory of meromorphic, simply
periodic functions (for more information see, e.g., Markushevich
\cite[Ch.\ III.4]{Ma85}): If $f$ is a meromorphic periodic function
with period $2\pi$, then $f^*(t)=f(-i\,\ln(t))$ is meromorphic on
$\bbC\backslash\{0\}$. If $f$ is entire then $f^*$ is analytic on
$\bbC\backslash\{0\}$. We call a function simply periodic if it is
periodic but not doubly periodic. A meromorphic simply periodic function
$q$ with period $\omega\in\bbC\backslash\{0\}$ which is bounded as
$|\Im(z/\omega)|$ tends to infinity, is of the form
\begin{equation}
q(z)=\frac{a_0+a_1\e^{2\pi iz/\omega}+\cdots+a_m \e^{2\pi imz/\omega}}
   {b_0+b_1\e^{2\pi iz/\omega}+\cdots+b_m \e^{2\pi imz/\omega}}.
\end{equation}
In particular, such functions have only finitely many poles in the period
strip
\begin{equation}
\cS_\omega=\{z\in\bbC\,|\, 0\leq\Re(z/\omega)<1\}.
\end{equation}
We will call such functions {\it bounded near the ends of the period
strip $\cS_\omega$}. Note that
\begin{equation}
\lim_{\Im(z/\omega)\to\infty} q(z)=\frac{a_0}{b_0}=q^*(0)
\end{equation}
and
\begin{equation}
\lim_{\Im(z/\omega)\to-\infty} q(z)=\frac{a_m}{b_m}=q^*(\infty).
\end{equation}

Next, consider the $n$th-order differential equation
\begin{equation}
y^{(n)}(z)+q_{n-1}(z) y^{(n-1)}(z)+\dots+ q_0(z) y(z) =0, \label{2.8A}
\end{equation}
where $q_j$, $0\leq j\leq n-1$, are meromorphic, simply periodic
functions with a common period $\omega\in\bbC\backslash\{0\}$, bounded
near the ends of the period strip $\cS_\omega$. Then the following variant
of Floquet's theorem holds.

\begin{theorem} $($Weikard \cite{We00}$)$ \label{t4.1}
Suppose the differential equation \eqref{2.8A} has a meromorphic
fundamental system of solutions. Then there exists a solution $y_1$ of
the differential equation \eqref{2.8A} of the form
\begin{equation}
y_1(z)=R(\e^{2\pi iz/\omega})\exp(i\lambda z),
\end{equation}
where $R$ is a rational function and $\lambda$ satisfies
\begin{equation}
(i\lambda)^n+q_{n-1}^*(0)(i\lambda)^{n-1}+\cdots+q_0^*(0)=0.
\end{equation}
\end{theorem}

\begin{remark} \lb{r4.2}
$(i)$ This version of Floquet's theorem differs from the standard one by
imposing considerably stronger hypotheses on the coefficients $q_j$ and
the nature of all solutions of \eqref{2.8A}. In return it provides a
considerably stronger conclusion with regard to the explicit form of the
solution $y_1$. An extension of Theorem \ref{t4.1} to the case of
first-order systems is discussed in \cite{We00b}. \\
$(ii)$ The conditions in Theorem \ref{t4.1} apply to stationary soliton
solutions of the Gelfand--Dickey hierarchy which are periodic with a
purely imaginary period.
\end{remark}

Finally, we turn to elliptic KdV solutions and start by describing
Picard's original result. Consider the following
$n$th-order differential equation
\begin{align}
    y^{(n)}(z)+q_{n-1}(z) y^{(n-1)}(z)+\dots+
q_0(z) y(z) =0, \label{3.1}
    \end{align}
where $q_j$, $0\leq j\leq n-1$, are elliptic functions associated with
the same period lattice generated by the fundamental half-periods
$\omega_1$, $\omega_3$ with $\Im(\omega_3/\omega_1)>0$.

Assuming the fundamental system of solutions of \eqref{2.1A} to be
meromorphic, the following theorem due to Picard holds.

\begin{theorem} $($Picard \cite{Pi79}--\cite{Pi81},
Ince \cite[p.~372--375]{In56}$)$ \lb{t3.1}
Suppose the differential equation \eqref{3.1} has a meromorphic
fundamental system of solutions. Then there exists a solution
$y_1$ of \eqref{3.1} which is elliptic of the second kind, that is, $y_1$
is meromorphic and there exist constants $\rho_j\in\bbC$, $j=1,2$, such
that
\begin{equation}
y_1(z+2\omega_j)=\rho_j y_1(z), \quad j=1,3, \; z\in\bbC. \lb{3.2}
\end{equation}
If in addition, the characteristic equation corresponding to the
translation $z\to z+2\omega_1$ or $z\to z+2\omega_3$ $($see \cite[p.
358, 376]{In56}$)$ has distinct roots, then there exists a fundamental
system of solutions of \eqref{3.1} which are elliptic functions of the
second kind.
\end{theorem}

The characteristic equation associated with the substitution
$z\mapsto z +2\omega_j$, $j=1,2$, alluded to in Theorem \ref{t3.1}, is
given by
\begin{equation} \label{3.3}
\det [A-\rho I]=0,
\end{equation}
where
\begin{equation} \label{3.4}
\phi_{\ell}(z+2\omega_j) = \sum^n_{k=1} a_{\ell,k} \phi_k(z), \quad
A=(a_{\ell,k})_{1\le \ell, k \le n}
\end{equation}
and $\phi_1,\dots,\phi_n$ is any fundamental system of solutions of
\eqref{3.1}.

What we call Picard's theorem following the usual convention in
\cite[p.\ 182--185]{Ak90}, \cite[p.\ 338--343]{Bu06},
\cite[p.\ 536--539]{Ha88}, \cite[ p.\ 181--189]{Kr97}, appears, however,
to have a much longer history.  In fact, Picard's investigations
\cite{Pi79}--\cite{Pi81} were inspired by earlier work of
Hermite in the special case of Lam\'e's equation \cite[p.\ 118--122,
266--418, 475--478]{He12} (see also \cite[Sect.\ 3.6.4]{BBEIM94},  and
\cite[p.\ 570--576]{WW86}).  Further contributions were made by
Mittag-Leffler \cite{Mi80}, and Floquet \cite{Fl84}--\cite{Fl84b}.
Detailed accounts of Picard's differential equation can be found in
\cite[p.\ 532--574]{Ha88}, \cite[p.\ 198--288]{Kr97}.

For an extension of Theorem \ref{t3.1} to first-order $n\times n$ systems
and the explicit structure of the corresponding fundamental system of
solutions we refer to \cite{GS98}.

We continue by quoting a number of known results on stationary
rational, simply periodic, and elliptic  KdV potentials. To simplify
notations in what follows we introduce the unifying notation $\cP$ to
denote
\begin{equation}
\cP(z)=\begin{cases} z^{-2} & \text{in the rational case,} \\
\f{\pi^2}{\omega^{2}}
   \Big([\sin(\pi z/\omega)]^{-2}-\f{1}{3}\Big) & \text{in the simply
periodic case,} \\
\wp(z) & \text{in the elliptic case.} \end{cases} \lb{P}
\end{equation}
We note for later purposes that the three cases depicted in
\eqref{P} can be viewed as specializations of the elliptic case in the
following sense: we recall the invariants $g_2$ and $g_3$ associated
with
$\wp(\cdot)=\wp(\cdot\,|g_2,g_3)$ as introduced in \eqref{B.2}. Then
(cf.\ \cite[p.\ 652]{AS72}),
\begin{equation}
\cP(z)=\begin{cases} \wp(z|0,0) & \text{in the rational case,}
\\
\wp\big(z|[2\pi^2/\omega^2]^2/3,[2\pi^2/\omega^2]^3/27\big) & \text{in
the simply periodic case,} \\
\wp(z|g_2,g_3) & \text{in the elliptic case.} \end{cases} \lb{WP}
\end{equation}
Here and in the following, the rational case always refers to
rational potentials bounded near infinity, and similarly the simply
periodic case always refers to meromorphic simply periodic potentials
bounded near the ends of the period strip.

\begin{theorem} \lb{t3.2}
Let $N\in\bbN$, $\{z_j\}_{1\leq j\leq N}\subset\bbC$ and define $\cP$
as in \eqref{P}. \\
\noindent $(i)$ $($Airault, McKean, and Moser
\cite{AMM77}, Gesztesy and Weikard \cite{GW95b}$)$
Any rational, simply periodic $($bounded near the ends of the period
strip$)$, or elliptic solution $q$ of some equation
$($and hence infinitely many equations$)$ of the stationary KdV
hierarchy, or equivalently, any rational, simply-periodic $($bounded near
the ends of the period strip$)$, or elliptic algebro-geometric KdV
potential $q$, is necessarily of the form
\begin{equation}
q(z)=q_0 -2\sum_{j=1}^N \cP(z-z_j),  \lb{3.6}
\end{equation}
for some $q_0\in\bbC$ and $N\in\bbN$. In the rational case, $N$ is of the
special type\footnote{$N\in\bbN$ is called triangular if there  is a
$g\in\bbN$ such that $N=g(g+1)/2$.} $N=g(g+1)/2$ for some $g\in\bbN$. \\
\noindent $(ii)$ $($Airault, McKean, and Moser
\cite{AMM77}, Gesztesy and Weikard \cite{GW95b}, Weikard \cite{We99}$)$ If
one allows for ``collisions'' between the
$z_j$, that is, if the set $\{z_j\}_{1\leq j\leq N}$ clusters into
groups of points, then the corresponding algebro-geometric potential $q$
is necessarily of the form
\begin{align}
q(z)&=q_0 -\sum_{\ell=1}^M s_\ell(s_\ell +1)\cP(z-\zeta_\ell),
\lb{3.7} \\
&=q_0-2\sum_{j=1}^N \cP(z-z_j), \lb{3.7A}
\end{align}
where
\begin{align}
\begin{split}
&\{z_j\}_{1\leq j\leq
N}=\{\zeta_\ell\}_{1\leq\ell\leq M}\subset\bbC \text{ with $\zeta_\ell$
pairwise distinct,}  \\
&  s_\ell\in\bbN,\,\,\, 1\leq\ell \leq M, \quad
\sum_{\ell=1}^M s_\ell(s_\ell +1)=2N. \lb{3.8 }
\end{split}
\end{align}
\noindent $(iii)$ The extreme case of all
$z_j$ colliding into one point,  say $\zeta_1$, that is,
$\{z_j\}_{1\leq j\leq N}=\{\zeta_1\}\subset\bbC$
yields an algebro-geometric KdV potential $($also called Lam\'e potential
in the elliptic case, cf.\ \cite{GW95}, \cite{GW98a} and the extensive
literature therein$)$ of the form
\begin{equation}
q(z)=q_0 -g(g+1)\cP(z-\zeta_1), \quad g\in\bbN \lb{3.9}
\end{equation}
and no additional constraints on $\zeta_1\in\bbC$. \\
\noindent $(iv)$ If $q$ is a KdV potential, the underlying
hyperelliptic curve $\cK_g$ is of the form
\begin{equation}
\cK_g \colon y^2=\prod_{m=0}^{2g}(E-E_m)
\text{ for some $\{E_m\}_{0\leq m\leq 2g}\subset\bbC$.} \lb{3.10}
\end{equation}
If $q$ is a simply periodic meromorphic KdV potential of period
$\omega$, bounded near the ends of the period strip $\cS_\omega$, one
infers $($Weikard \cite{We00}$)$
\begin{align}
& E_0=q^*(0)=e_0, \quad E_{2p-1}=E_{2p}=e_p, \; 1\leq p\leq g
\text{ for some $\{e_m\}_{0\leq m\leq g}\subset\bbC$,} \no \\
& e_m\neq e_{m'} \text{ for $m\neq m'$, $0\leq m,m'\leq g$,} \\
&q^*(e^{2\pi iz/\omega})=q(z), \lb{2.8C}
\end{align}
and the corresponding simply periodic $($singular$)$ hyperelliptic
curve $\cK_g$ in \eqref{3.10} reduces to the special form
\begin{equation}
\cK_g \colon y^2=(E-e_0)\prod_{p=1}^g (E-e_p)^2. \lb{2.8B}
\end{equation}
In the special case where $q$ is a rational KdV potential, one
obtains
\begin{equation}
E_0=\cdots=E_{2g}=q_0
\end{equation}
and hence \eqref{3.10} reduces to the especially simple form of a
rational curve
\begin{equation}
\cK_g \colon y^2=(E-q_0)^{2g+1}. \lb{2.8}
\end{equation}
In particular, the KdV potentials \eqref{3.6}, \eqref{3.7}, and
\eqref{3.9} are all isospectral. \\
\noindent $(v)$ $($Gesztesy and Weikard \cite{GW96}, Weikard
\cite{We99}$)$ Suppose $q$ is a rational function, or a meromorphic
simply periodic
function bounded near the ends of the period strip, or an elliptic
function. Then $q$ is a KdV potential if and only if $\psi''+(q-
E)\psi=0$ has a meromorphic fundamental system of solutions
$($w.r.t.~$z$$)$ for all values of the spectral
parameter $E\in\bbC$. \\
\noindent $(vi)$ $($Gesztesy, Unterkofler, and Weikard
\cite{GUW00}$)$ If $q$ is a rational KdV potential of the form
\eqref{3.7}, then $y''+qy=Ey$ has linearly independent solutions of
the Baker--Akhiezer-type
\begin{align}
&\psi_\pm(E,z)=\big(\pm E^{1/2}\big)^{-g}\Bigg(\prod_{p=1}^g
\big[\pm E^{1/2}-\nu_p(z)\big]\Bigg) e^{\pm E^{1/2} z}, \lb{2.8a} \\
&\hspace*{5.65cm} E\in\bbC\backslash\{q_0\}, \,\, z\in\bbC \no
\end{align}
with $\mu_p(z)=\nu_p(z)^2$, $1\leq p\leq g$, the zeros of $F_g(z,x)$
as defined in \eqref{A.9} and the elementary symmetric functions
of $\nu_p$, $1\leq p\leq g$, are rational functions. \\
\noindent $(vii)$ $($Weikard \cite{We00}$)$ If $q$ is a simply
periodic meromorphic KdV potential of period $\omega$, bounded near the
ends of the period strip $\cS_\omega$, of the form \eqref{3.7}, then
$y''+qy=Ey$ has linearly independent solutions of the
Baker--Akhiezer-type
\begin{align}
&\psi_\pm(E,z)=\bigg(\sum_{m=0}^g r_{m}\big(e^{2\pi
iz/\omega}\big) (\pm\lambda)^m\bigg) e^{\pm \lambda z}, \quad
r_g \big(e^{2\pi iz/\omega}\big)=1, \lb{2.8D} \\
&\hspace*{.9cm} E\in\bbC\backslash\{e_m\}_{0\leq
m\leq g}, \,\, e_0=q^*(0), \,\, E=\lambda^2 + q^*(0), \,\, z\in\bbC,
\no
\end{align}
where $r_m$, $0\leq m\leq g-1$, are rational functions.
\end{theorem}

\begin{remark} \lb{r2.3}
$(i)$ In connection with Theorem \ref{t3.2}\,(v) we note that the
``only if'' part follows {}from the explicit theta function representation
of the Baker--Akhiezer function due to Its and Matveev \cite{IM75} in the
special case where $\cK_g$ is nonsingular and {}from the loop group and
$\tau$-function approach of Segal and Wilson \cite{SW85} in the general
case of possibly singular hyperelliptic curves. \\
$(ii)$ Strictly speaking, the version of Theorem \ref{t3.2}\,(v) in the
rational case  proven in \cite{We99} assumes in addition to $q$ being
rational, that $q$ is bounded near infinity. However, a simple inductive
argument using \eqref{A.1} proves that a rational function $q$ unbounded
near infinity cannot satisfy any of the stationary KdV equations $($cf.\
\cite{GUW00}$)$. \\
$(iii)$ While \eqref{2.8B} is not explicitly recorded in
\cite{We00}, it immediately follows from \eqref{2.8D} by noting
that the curve is of the form
\begin{equation}
\cK_g \colon y^2
=W(\psi_+(\lambda,\cdot),\psi_-(\lambda,\cdot))^2=(E-q^*(0))
\prod_{p=1}^g (E-e_p)^2 \lb{2.8E}
\end{equation}
for some $e_p\in\bbC$, $1\leq p\leq g$.
\end{remark}

\begin{remark} \lb{r2.3a}
Combining the explicit form of the rational and simply periodic
hyperelliptic curves \eqref{2.8} and \eqref{2.8B} with
\cite[Theorem\ 2.3]{GH00} shows that all rational and meromorphic
simply periodic KdV potentials (bounded near the ends of the period
strip) satisfying $\sKdV_g (q)=0$ (cf.\ \eqref{A.10b}), can be
generated from the genus zero case $q(x)=q_0$, respectively,
$q(x)=q^*(0)$, by precisely $g$ Darboux transformations. This is in sharp
contrast to the elliptic case and will play an important role in
Section
\ref{s3}.
\end{remark}

\begin{remark} \lb{r2.3b}
For future purposes we note the following $\tau$ function
representation of the function $q$ in \eqref{3.6}. In accordance with
the three cases discussed in \eqref{P} we now define
\begin{equation}
\nu(z)=\begin{cases} \sigma(z|0,0)=z & \text{in the rational case,} \\
\sigma\big(z|[2\pi^2/\omega^2]^2/3,[2\pi^2/\omega^2]^3/27\big) \\
=(\omega/\pi) \sin(\pi z/\omega)\exp[\pi^2z^2/(6\omega)^2]
& \text{in the simply periodic case,} \\
\sigma(z|g_2,g_3) & \text{in the elliptic case} \end{cases} \lb{nu}
\end{equation}
with $\sigma(\cdot)=\sigma(\cdot\,|g_2,g_3)$ the Weierstrass
$\sigma$-function in the elliptic case associated with the invariants
$g_2$ and $g_3$ (cf.\ \cite[Sect.\ 18.1]{AS72}), and
\begin{equation}
\tau(z;z_1,\dots,z_N)=\prod_{j=1}^N \nu(z-z_j). \lb{tau}
\end{equation}
Then obviously,
\begin{align}
q(z)&=q_0 -2\sum_{j=1}^N \cP(z-z_j),  \no \\
&= q_0+2[\ln(\tau(z;z_1,\dots,z_N))]''.  \lb{qt}
\end{align}
\end{remark}

Theorems \ref{t2.1}--\ref{t3.1} motivate the following definition.

\begin{definition} \lb{d2.4}
$(i)$ Let $q$ be a rational function. Then $q$ is called a \textit{Halphen
potential} if it is bounded near infinity and if $y''+qy=Ey$ has a
meromorphic fundamental system of solutions {\rm(}w.r.t.~$z${\rm)} for
each value of the complex  spectral parameter $E\in\bbC$. \\
$(ii)$ Let $q$ be a simply periodic meromorphic function. Then $q$ is
called a \textit{Floquet potential} if it is bounded near the ends of
the period strip and if $y''+qy=Ey$ has a meromorphic fundamental system
of solutions $($w.r.t.~$z$$)$ for each value of the complex spectral
parameter $E\in\bbC$. \\
$(iii)$ Let $q$ be an elliptic function. Then $q$ is called a
\textit{Picard potential} if $y''+qy=Ey$ has a meromorphic fundamental
system of solutions $($w.r.t.~$z$$)$ for each value of the complex
spectral parameter $E\in\bbC$.
\end{definition}

\noindent By Theorem \ref{t3.2}\,(v), $q$ is a Halphen (respectively,
Floquet or Picard) potential if and only if $q$ is a rational
(respectively, simply periodic meromorphic (bounded near the ends of the
period strip) or elliptic) KdV potential, or equivalently, if and only if
it satisfies one and hence infinitely many of the equations of the
stationary KdV hierarchy (cf.\ Definition \ref{dA.1}).

Next, we turn to the principal aim of this paper, the precise restrictions
on the set of poles $\{z_j\}_{1\leq j\leq N}=\{\zeta_\ell\}_{1\leq\ell\leq
M}$ of $q$ in \eqref{3.6} to be a KdV potential. We start with the
following known fact.

\begin{lemma} \label{l3.3}
Suppose $q$ is meromorphic in a neighborhood of $z_0\in\bbC$ with a
Laurent expansion about the point $z_0$ of the type
\begin{equation}
q(z)=\sum_{j=0}^\infty q_j (z-z_0)^{j-2}, \lb{3.30}
\end{equation}
where $q_0=-s(s+1)$ and, without loss of generality, $\Re(2s+1)\geq0$.
Define for $\sigma\in\bbC$,
\begin{align}
f_0(\sigma)&=-\sigma(\sigma-1)-q_0=(s+\sigma)(s+1-\sigma), \lb{3.31}\\
c_0(\sigma)&=\begin{cases} 1 & \text{ if $2s+1\notin\bbN$}, \\
\prod_{j=1}^{2s+1} f_0(\sigma+j) & \text{ if $2s+1\in\bbN$},
\end{cases} \lb{3.32} \\
c_j(\sigma)&=\frac{\sum_{m=0}^{j-1}
      q_{j-m} c_m(\sigma)}{f_0(\sigma+j)}, \;\, j\in\bbN, \label{crec} \\
w(\sigma,z)&=\sum_{j=0}^\infty c_j(\sigma) (z-z_0)^{\sigma+j},
\lb{3.33} \\
    v(\sigma,z)&=\frac{\partial w}{\partial\sigma}(\sigma,z)
    =\sum_{j=0}^\infty \big(c_j^\prime(\sigma)
     +c_j(\sigma) \ln(z-z_0)\big) (z-z_0)^{\sigma+j}
\text{ if $(2s+1)\in\bbN_0$}. \lb{3.34}
\end{align}
If $(2s+1)\notin\bbN_0$, then $y''+qy=0$
has the linearly independent solutions $y_1=w(s+1,\cdot)$ and
$y_2=w(-s,\cdot)$. If $(2s+1)\in\bbN_0$, then $y''+qy=0$ has the
linearly independent solutions $y_1=w(s+1,\cdot)$ and $y_2=v(-s,\cdot)$.

Moreover, $y''+qy=0$ has a meromorphic fundamental system of solutions
near $z_0$ if and only if $s\in\bbN_0$ and $c_{2s+1}(-s)=0$.
\end{lemma}
This is a classical result in ordinary differential equations (cf., e.g.,
\cite[Chs.~XV, XVI]{In56}). A recent proof adapted to the present context
can be found in Section 3 of \cite{We99}. We note that $q$ is not assumed
to be rational, simply periodic, or elliptic in Lemma \ref{l3.3}.

Our principal new result on simply periodic and elliptic
solutions of the stationary KdV hierarchy then reads as follows (we recall
our notational convention \eqref{P} to unify the rational, simply
periodic, and elliptic cases by the symbol $\cP$).

\begin{theorem} \lb{t3.5}
Let $q$ be a rational function bounded near infinity, or a simply
periodic function bounded near the ends of the period strip, or an
elliptic function. Then $q$ is a Halphen, or a Floquet, or a Picard
potential if and only if there are $M\in\bbN$, $s_\ell\in\bbN$,
$1\leq\ell\leq M$, $q_0\in\bbC$, and pairwise distinct
$\zeta_\ell\in\bbC$, $1\leq \ell\leq M$, such that
\begin{equation}
q(z)=q_0-\sum_{\ell=1}^M s_\ell(s_\ell+1)\cP(z-\zeta_\ell), \lb{3.35}
\end{equation}
and
\begin{equation}
\sum_{\substack{\ell^\prime=1\\ \ell^\prime\neq \ell}}^M
    s_{\ell^\prime}(s_{\ell^\prime}+1) \cP^{(2k-1)}(\zeta_{\ell}
-\zeta_{\ell^\prime})=0
    \text{ for $1\leq k\leq s_{\ell}$ and $1\leq \ell\leq M$.} \lb{3.36}
\end{equation}
Moreover, $q$ is a rational, simply periodic $($bounded near the ends of
the period strip$)$, or elliptic KdV potential if and only if $q$ is of
the type
\eqref{3.35} and the constraints
\eqref{3.36} hold. \\
In the particular rational case, for fixed $g\in\bbN$,
the constraints \eqref{3.36} characterize the isospectral class of all
rational KdV potentials associated with the curve $y^2=(E-q_0)^{2g+1}$,
where $g(g+1)=\sum_{\ell=1}^M s_\ell(s_\ell+1)$.
\end{theorem}
\begin{proof}
The proof of the current theorem is analogous to the one presented in
the rational case in \cite{GUW00}. However, we use this opportunity to
improve the presentation of the proof and to remove some inaccuracies.
As pointed out at the end of the proof, it is sufficient to focus on the
elliptic case.

By Theorem~\ref{t3.2}\,(v), it suffices to prove the characterization
of Picard potentials. Suppose that $q$ is a nonconstant Picard
potential. Then a pole $z_0$ of $q$ is a regular singular point of
$y''+qy=Ey$ and hence
\begin{equation}
q(z)-E=\sum_{j=0}^\infty Q_j (z-z_0)^{j-2}
\end{equation}
in a sufficiently small neighborhood of $z_0$, where $Q_2$ is a
polynomial of first degree in $E$, while $Q_j$ for $j\neq2$ are
independent of $E$. The indices associated with $z_0$, defined as
the
roots of $\sigma(\sigma-1)+Q_0=0$ (hence they are $E$-
independent),
must be distinct integers whose sum equals one. We denote them
by $-s$
and $s+1$, where $s\in\bbN$, and note that $Q_0=-s(s+1)$. We
intend to
prove that $Q_{2j+1}=0$ whenever $j\in\{0,\dots,s\}$ by applying
Lemma
\ref{l3.3}. Proceeding by way of contradiction, we thus  assume that
for some nonnegative integer $k\in\{0,\dots,s\}$, $Q_{2k+1}\neq 0$
and
$k$ is the smallest such integer.

We note that $f_0(\cdot+j)$ are positive in $(-s-1,-s+1)$ for
$j=1,\dots,2s$, whereas $f_0(\cdot+2s+1)$ has a simple zero at $-s$
and
its derivative is negative at $-s$. Next one defines
\begin{equation}
\gamma_0(\sigma)=\prod_{j=1}^{2s+1} f_0(\sigma+j). \lb{3.38}
\end{equation}
Note that $\gamma_0$ has a simple zero at $-s$ and that
$\gamma_0'(-s)$
is negative.

The functions $c_0=\gamma_0$ and
$c_1=Q_1\gamma_0/f_0(\cdot+1)$ are
polynomials with respect to $E$. Actually, $c_0$ has degree zero in
$E$
and $c_1$ has degree at most zero ($c_1$ might be equal to zero).
Hence
the relations \eqref{ce}, \eqref{gammae}, \eqref{co}, and
\eqref{gammao} below are satisfied for $j=1$ if we let
$\gamma_1(\sigma)=\gamma_0(\sigma)/f_0(\sigma+1)$. Next let
$\ell$ be
some integer in $\{1,\dots, s\}$. Assume that there are suitable
coefficients $\gamma_p$, $p=0,..., 2\ell-1$ such that the functions
$c_0$,\dots, $c_{2\ell-1}$ are polynomials in $E$ satisfying the
relations
\begin{align}
     c_{2j-2}(\sigma)&=\gamma_{2j-2}(\sigma) Q_2^{j-1}+ O(E^{j-2}),
      \label{ce}\\
      \gamma_{2j-2}(-s)&=0, \quad \gamma_{2j-2}'(-s)<0,
\label{gammae}\\
     c_{2j-1}(\sigma)&=\gamma_{2j-1}(\sigma) Q_{2k+1} Q_2^{j-k-1}
       + O(E^{j-k-2}), \label{co} \\
      \gamma_{2j-1}(-s)&=0, \quad \gamma_{2j-1}'(-s)\leq0
\label{gammao}
\end{align}
for $1\leq j\leq\ell$ as $E$ tends to infinity. Using the recursion
relation \eqref{crec} we then obtain that $c_{2\ell}(\sigma)$ and
$c_{2\ell+1}(\sigma)$ are polynomials in $E$ and that
\begin{align}
     c_{2\ell}(\sigma)
      &=\frac{\gamma_{2\ell-2}(\sigma)}{f_0(\sigma+2\ell)} Q_2^\ell
      +O(E^{\ell-1}),\\
     c_{2\ell+1}(\sigma)&=
       \frac{\gamma_{2\ell-1}(\sigma)+\gamma_{2(\ell-k)}(\sigma)}
        {f_0(\sigma+2\ell+1)}
         Q_{2k+1} Q_2^{\ell-k}+O(E^{\ell-k-1})
\end{align}
as $E$ tends to infinity. Letting $\gamma_{2\ell}
=\gamma_{2\ell-2}/f_0(\cdot+2\ell)$ and $\gamma_{2\ell+1}
=(\gamma_{2\ell-1}+\gamma_{2(\ell-k)})/f_0(\cdot+2\ell+1)$ we find
that relations \eqref{ce}, \eqref{gammae} and \eqref{co} are satisfied
for $j=\ell+1$. Moreover, relation \eqref{gammao} is also satisfied
unless $\ell=s$. Hence we proved that $c_{2s+1}$ is a polynomial in
$E$
and that
\begin{equation}
c_{2s+1}(\sigma)
      =\frac{\gamma_{2s-1}(\sigma)+\gamma_{2(s-k)}(\sigma)}
      {f_0(\sigma+2s+1)}Q_{2k+1} Q_2^{s-k}+O(E^{s-k-1}).
\end{equation}
But both $\gamma_{2s-1}+\gamma_{2(s-k)}$ and $f_0(\cdot+2s+1)$
have
simple zeros at $-s$. Therefore $\gamma_{2s+1}(-s)$ is different
from
zero. In fact
\begin{equation}
\gamma_{2s+1}(-s)
=-\frac{\gamma_{2s-1}'(-s)+\gamma_{2(s-k)}'(-s)}{2s+1}>0.
\end{equation}
Lemma~\ref{l3.3} then shows that $y''+qy=Ey$ has a solution which
is
not meromorphic whenever $E$ is not a root of the polynomial
$c_{2s+1}(-s)$. This contradiction proves our assumption
$Q_{2k+1}\neq0$ wrong.

Since $Q_1=0$, we proved that if $q$ is a Picard potential with
pairwise distinct poles $\zeta_1,\dots,\zeta_M$, then the principal
part of $q$ about any pole $\zeta_\ell$ is of the form
$-s_\ell(s_\ell+1)/(z-\zeta_\ell)^2$ for an appropriate positive
integer $s_\ell$. Since $q$ is elliptic Theorem \ref{tb2} then proves
\eqref{3.35}. This immediately implies that for $z_0=\zeta_\ell$,
\begin{equation}
Q_{2k+1}=-\sum_{\substack{\ell^\prime=1\\ \ell^\prime\neq \ell}}^M
s_{\ell^\prime}(s_{\ell^\prime}+1) \frac{1}{(2 k-1)!}
\cP^{(2k-1)}(\zeta_\ell -\zeta_{\ell^\prime}). \lb{3.40}
\end{equation}
This proves necessity of the conditions \eqref{3.35} and \eqref{3.36}
for $q$ to be a Picard potential. To prove their sufficiency we now
assume that \eqref{3.35} and \eqref{3.36} hold. Then, if $z_0$
denotes
any of the points $\zeta_\ell$, one infers that the corresponding
$c_{2s_\ell+1}(-s_\ell)=0$.  Lemma~\ref{l3.3} then guarantees that
all
solutions of $y''+qy=Ey$ are meromorphic and hence that $q$ is a
Picard
potential.

The proof for simply periodic or rational potentials is virtually the
same. Indeed, the proof presented in the elliptic case uses only the
fact that $q$ is meromorphic and that elliptic functions allow a
partial fractions expansion, which is true for simply periodic
meromorphic functions, too. In particular, Lemma \ref{l3.3} does not
rely on $q$ being elliptic.
\end{proof}

\begin{remark} \lb{r3.6}
To the best of our knowledge, the explicit characterization \eqref{3.36}
of the simply periodic and elliptic Calogero--Moser locus is
new. This is perhaps a bit surprising since this circle of ideas received
considerable attention in the past two decades. The algebraic curves
associated with various special cases of \eqref{3.35}, \eqref{3.36} have
been extensively studied and we refer, for instance, to
\cite[Sects.\ 7.7, 7.8]{BBEIM94}, \cite{BBME86}--\cite{BE94},
\cite{EE95}--\cite{EK94}, \cite{GP99}, \cite{GW95}--\cite{GW98},
\cite{KE93}, \cite{Sm89}, \cite{Sm94}, \cite{Ta90}, \cite{Ta94},
\cite{Tr89}--\cite{VP00}.
\end{remark}

\begin{remark} \lb{r2.6}
$(i)$ The necessary and sufficient
conditions on $\zeta_\ell$ for $q$ in \eqref{3.35} to be a rational KdV
potential were first obtained by Duistermaat and Gr\"unbaum \cite{DG86}
in their analysis of bispectral pairs of differential operators. Our
approach to proving the locus characterization \eqref{3.36} in
\cite{GUW00} was based on Halphen's theorem and a direct Frobenius-type
analysis exactly along the lines just presented in the elliptic case. \\
$(ii)$ We note that the restrictions
\eqref{3.36} simplify in the absence of collisions, where $s_\ell=1$,
$1\leq\ell\leq N$. In this case \eqref{3.36} reduces to
\begin{equation}
\sum_{\substack{j^\prime=1 \\ j^\prime\neq j}}^N \cP'(z_j
-z_{j^\prime})=0, \quad 1\leq j\leq N, \lb{2.38A}
\end{equation}
which  represents the well-known locus discussed by Airault, McKean, and
Moser \cite{AMM77}. Equation \eqref{3.36} properly extends this locus to
the case of collisions $($i.e., to cases where some of the
$s_\ell>1$$)$. Historically, the locus defined by \eqref{2.38A} is
called the Calogero--Moser $($CM\,$)$ locus. Since the extended
Calogero--Moser locus \eqref{3.36}, that is,
\begin{equation}
\sum_{\substack{\ell^\prime=1\\ \ell^\prime\neq \ell}}^M
    s_{\ell^\prime}(s_{\ell^\prime}+1) \cP^{(2k-1)}(\zeta_{\ell}
-\zeta_{\ell^\prime})=0
    \text{ for $1\leq k\leq s_{\ell}$ and $1\leq \ell\leq M$,} \lb{3.36A}
\end{equation}
was first derived by Duistermaat and Gr\"unbaum in the rational case,
from this point on we will call \eqref{3.36A} the Duistermaat and
Gr\"unbaum $($DG\,$)$ locus. The CM and DG loci will be further
explored in Sections \ref{s3} and \ref{s4} $($cf.\ Theorems \ref{t5.7}
and \ref{t4.4}$)$. \\
$(iii)$ For $k=1$, conditions \eqref{3.36} coincide with the
necessary conditions at collision points found by Airault, McKean, and
Moser \cite{AMM77} in their Remark~1 on p.~113. However, since there
are additional necessary conditions in \eqref{3.36} corresponding to
$k\geq 2$, this disproves the conjecture made at the end of the proof of
their Remark~1. \\
$(iv)$ In the special elliptic case $N=3$, the DG locus
\eqref{3.36A} was explicitly determined by Airault, McKean, and Moser
\cite[p.\ 140]{AMM77} using a different method $($in this case one simply
joins the diagonal $z_1=z_2=z_3$ to the original CM locus, cf.\
\eqref{4.25A}$)$. \\
$(v)$ In the rational case it is known that the CM locus is nonempty if
and only if $N$ is of the type $N=g(g+1)/2$ for some $g\in\bbN$ $($cf.\
\cite{AMM77}, \cite{SW85}$)$. The analogous result in the elliptic case
appears to be open. In the simply periodic case we will derive some
new results in Section \ref{s3}. Various examples in connection with Lam\'e
and Treibich--Verdier potentials and their generalizations, in which the
elliptic CM locus is nonempty, are discussed, for instance, in
\cite{AMM77}, \cite[Sects.\ 7.7, 7.8]{BBEIM94},
\cite{BBME86}--\cite{BE94}, \cite{DS00}, \cite{DN75},
\cite{En83}--\cite{EK94}, \cite{GP99}, \cite{GW95}--\cite{GW96},
\cite{GW98a}, \cite{IE86}, \cite{KE93},
\cite{Kr94}, \cite{Sm89}, \cite{Sm94}, \cite{Ta90}, \cite{Ta94},
\cite{Tr89}--\cite{VP00}.
\end{remark}

Next, we present a result on the KdV recursion coefficients $f_j$ (cf.\
Appendix \ref{sA}), extending Proposition\ 4 in \cite{AMM77}.

\begin{theorem} \lb{t2.12}
Assume that $\{z_j\}_{1\leq j \leq N}\subset\bbC$ are pairwise distinct,
$z_j\neq z_k$ for $j\neq k$, $1\leq j,k\leq N$ and suppose the CM locus
conditions are satisfied, that is,
\begin{equation}
\sum_{\substack{j=1\\ j\neq k}}^N \cP'(z_k-z_j)=0 \lb{2.39}
\text{ for $1\leq k\leq N$.}
\end{equation}
In addition, let $q$ be a rational, simply periodic $($bounded near the
ends of the period strip$)$, or elliptic KdV potential of the form
\begin{equation}
q(z)=q_0-2\sum_{j=1}^N \cP(z-z_j). \lb{2.40}
\end{equation}
Then $q$ satisfies some of the equations of the stationary KdV hierarchy.
Next, define the KdV recursion coefficients $f_j$ as in \eqref{A.1}.
Then, $f_j$ are of the form
\begin{equation}
f_0=1, \quad
f_j(z)=d_j+\sum_{k=1}^N a_{j,k}\cP(z-z_k), \quad j\in\bbN \lb{2.41}
\end{equation}
for some $\{a_{j,k}\}_{1\leq k \leq N}\subset\bbC$ and $d_j\in\bbC$,
$j\in\bbN$. More precisely, $d_j$ is of the
form\footnote{We use the standard abbreviations $(2q-1)!!=1\cdot
3\cdots(2q-1)$,
$q\in\bbN$.}
\begin{equation}
d_j=c_j(\ul E)+\sum_{\ell=1}^j c_{j-\ell}(\ul E)\f{(2\ell-1)!!}{2^\ell
\ell!}q_0^\ell, \quad j\in\bbN \lb{2.41a}
\end{equation}
with $c_\ell(\ul E)$, $\ell\in\bbN_0$ given by \eqref{A.16b}, and
$a_{j,k}$ satisfying the recursion relation
\begin{align}
a_{0,k}&=0, \; 1\leq k\leq N, \quad d_0=1, \lb{2.41b} \\
a_{j+1,k}&=a_{j,k}q_0-d_j-\sum_{\substack{\ell=1\\\ell\neq k}}^N
\big(a_{j,\ell}+2a_{j,k}\big)\cP(z_k-z_\ell), \quad j\in\bbN_0, \;
1\leq k\leq N. \no
\end{align}
\end{theorem}
\begin{proof}
The choice
\begin{equation}
a_{0,k}=0, \; 1\leq k\leq N, \quad d_0=1 \lb{2.42}
\end{equation}
proves \eqref{2.41} for $j=0$. Next, assume that it is valid
for some nonnegative integer $j$ and note that by \eqref{A.1}
\begin{equation}
f_{j+1}'=\frac14 f_j'''+qf_j'+\frac12q'f_j, \quad j\in\bbN_0. \lb{2.43}
\end{equation}
Next, we introduce the asymptotic expansion
\begin{equation}
\cP(z)=z^{-2}+\Oh(z^2) \text{ as $z\to 0$}, \lb{2.43a}
\end{equation}
and define the quantities
\begin{align}
&Q_{j,k,r}=a_{j,k}\sum_{\substack{\ell=1\\ \ell\neq k}}^N
   \cP^{(r)}(z_k-z_\ell), \quad R_{j,k,r}=\sum_{\substack{\ell=1\\ \ell\neq
k}}^N a_{j,\ell} \cP^{(r)}(z_k-z_\ell), \lb{2.44} \\
& \hspace*{6.3cm} j,r\in\bbN_0, \, 1\leq k\leq N. \no
\end{align}
Then $Q_{j,k,1}=0$, $j\in\bbN_0$, $1\leq k\leq N$ by hypothesis
\eqref{2.39} and one computes, as $z$ approaches $z_k$,
\begin{align}
   \frac14 f_j'''(z)&=-6a_{j,k}(z-z_k)^{-5}+O(1), \lb{2.45} \\
   q(z)f_j'(z)&=4a_{j,k}(z-z_k)^{-5}
     +(4Q_{j,k,0}-2a_{j,k}q_0)(z-z_k)^{-3} -2R_{j,k,1}(z-z_k)^{-2} \no \\
     & \quad  -2(R_{j,k,2}-Q_{j,k,2})(z-z_k)^{-1}+O(1), \lb{2.46} \\
   \frac12q'(z)f_j(z)&=2a_{j,k}(z-z_k)^{-5}
     +2(R_{j,k,0}+d_j)(z-z_k)^{-3}+2R_{j,k,1}(z-z_k)^{-2} \no \\
     & \quad +(R_{j,k,2}-Q_{j,k,2})(z-z_k)^{-1}+O(1). \lb{2.47}
\end{align}
Since $f_{j+1}$ is a differential polynomial in $q$, it is a meromorphic
function and hence the residues of its derivative are zero. This
implies that
\begin{equation}
R_{j,k,2}-Q_{j,k,2}=
   \sum_{\substack{\ell=1\\ \ell\neq k}}^N (a_{j,\ell}-a_{j,k})
   \cP''(z_k-z_\ell)=0, \quad j\in\bbN_0, \; 1\leq k\leq N. \lb{2.48}
\end{equation}
Hence, as $z$ approaches $z_k$,
\begin{equation}
f_{j+1}'=(4Q_{j,k,0}+2R_{j,k,0}+2d_j-2a_{j,k}q_0)
   (z-z_k)^{-3}+O(1). \lb{2.49}
\end{equation}
Define
\begin{equation}
a_{j+1,k}=a_{j,k}q_0 -d_j-2Q_{j,k,0}-R_{j,k,0} \lb{2.50}
\end{equation}
and
\begin{equation}
p_{j+1}(z)=\sum_{k=1}^N a_{j+1,k}\cP(z-z_k). \lb{2.51}
\end{equation}
This implies that the function $f_{j+1}'-p_{j+1}'$, as well as its
antiderivative $f_{j+1}-p_{j+1}$, are entire. Since $f_{j+1}$ is
a differential polynomial in $q$, $f_{j+1}-p_{j+1}$ is equal to a
constant, say $d_{j+1}$, in the elliptic case. In the simply periodic
meromorphic, or rational case, $f_{j+1}-p_{j+1}$ is simply periodic
meromorphic, or rational, and one arrives at the same conclusion by
considering the behavior of $f_{j+1}-p_{j+1}$ at infinity. This proves
\begin{equation}
f_{j+1}=d_{j+1}+\sum_{k=1}^N a_{j+1,k} \cP(z-z_k) \lb{2.52}
\end{equation}
and hence \eqref{2.41}. By induction on $j$ one verifies from
\eqref{2.43} that $\hat f_j$, $j\in\bbN$, contains the term $\alpha_j
q^j$, where
\begin{equation}
\alpha_{j+1}=\f{2j+1}{2j+2}\alpha_j, \; j\in\bbN, \quad \alpha_1=1/2,
\lb{2.53}
\end{equation}
implying
\begin{equation}
\alpha_j=\f{(2j-1)!!}{2^j j!}, \quad j\in\bbN. \lb{2.54}
\end{equation}
Since according to \eqref{A.5D},
\begin{equation}
f_j=\sum_{k=0}^j c_{j-k} \hat f_k \lb{2.54a}
\end{equation}
with $c_\ell=c_\ell(\ul E)$ as defined in \eqref{A.17}, one infers that
the constant term in $f_j$ is of the form
\begin{equation}
c_j+\sum_{k=1}^j c_{j-k}\alpha_k q_0^k. \lb{2.54b}
\end{equation}
Together with \eqref{2.54} this proves \eqref{2.41a}, which in turn proves
\eqref{2.41b} because of \eqref{2.50}.
\end{proof}

Finally, we derive the analog of \eqref{2.41} for $f_j$ in the presence
of collisions. To the best of our knowledge, this is a new result.
However, since the corresponding proof based on induction is a bit lengthy
(even though the arguments involved are quite elementary), we defer its
proof to Appendix \ref{sD} .

\begin{theorem} \lb{t2.13}
Assume $M\in\bbN$, $s_\ell\in\bbN$, $1\leq\ell\leq M$, $q_0\in\bbC$,
and suppose $\zeta_\ell\in\bbC$, $\ell=1,\dots,M$, are pairwise
distinct. Consider
\begin{equation}
q(z)=q_0-\sum_{\ell=1}^M s_\ell(s_\ell+1)\cP(z-\zeta_\ell), \lb{2.66}
\end{equation}
and suppose the DG locus conditions
\begin{equation}
\sum_{\substack{\ell^\prime=1\\ \ell^\prime\neq \ell}}^M
    s_{\ell^\prime}(s_{\ell^\prime}+1) \cP^{(2k-1)}(\zeta_{\ell}
-\zeta_{\ell^\prime})=0
    \text{ for $1\leq k\leq s_{\ell}$ and $1\leq \ell\leq M$} \lb{2.67}
\end{equation}
are satisfied. Then
\begin{equation}
f_0=1, \quad
f_j(z)=d_j+\sum_{\ell=1}^M \sum_{k=1}^{\min(j,s_\ell)} a_{j,\ell,k}
\cP(z-\zeta_\ell)^k, \quad j\in\bbN \lb{2.68}
\end{equation}
for some $\{a_{j,\ell,k}\}_{1\leq k \leq \min(j,s_\ell), 1\leq \ell\leq
M} \subset\bbC$ and $d_j\in\bbC$, $j\in\bbN$.
\end{theorem}

\section{Additional results on the CM and DG loci} \lb{s3}

The principal purpose of this section is a closer examination of the
locus of poles with special emphasis on collisions. In particular, we
will prove in the rational and simply periodic cases that the DG locus
is the closure of the CM locus in an appropriate (in fact,
canonical) topology.

Following our strategy of describing the rational, simply periodic, and
elliptic cases simultaneously whenever possible, we first introduce
\begin{equation}
X=\begin{cases} \bbC &\text{in the rational case,} \\
\bbC/\Lambda_\omega &\text{in the simply periodic case,} \\
\bbC/\Lambda_{2\omega_1,2\omega_3} &\text{in the elliptic case,}
\end{cases}
\lb{4.14}
\end{equation}
where $\Lambda_\omega$ denotes the period lattice
\begin{equation}
\Lambda_\omega=\{m\omega\in\bbC\, | \, m\in\bbZ \}, \quad
\omega\in\bbC\backslash\{0\} \lb{4.15}
\end{equation}
in the simply periodic case, and $\Lambda_{2\omega_1,2\omega_3}$ denotes
the period lattice
\begin{equation}
\Lambda_{2\omega_1,2\omega_3}=\{2m\omega_1+2n\omega_3\in\bbC\, | \,
(m,n)\in\bbZ^2\}, \quad
\omega_1, \omega_3\in\bbC\backslash\{0\}, \; \Im(\omega_3/\omega_1)>0
\lb{4.16}
\end{equation}
in the elliptic case. In addition to the cartesian product
$X^N=X\times \cdots \times X$ ($N$ factors), $N\in\bbN$, we also need to
introduce the $N$th symmetric product $X^N/S_N$ of $X$ defined
as in \eqref{C.1}, with $S_N$ denoting the symmetric group on $N$ letters
acting as the group of permutations of the factors in the cartesian
product $X^N$. The elements of $X^N/S_N$ are denoted by
$[z_1,\dots,z_N]$ and $X^N/S_N$ will be endowed with the quotient
topology $\tau_{S_N}$ as discussed in Appendix \ref{sC}.

Next, we fix $N\in\bbN$ and define the corresponding  Calogero--Moser
(CM) locus of poles $\cL_N \subset X^N/S_N$ by
\begin{align}
\cL_N&=\Bigg\{[z_1,\dots,z_N]\in X^N/S_N \, \Bigg | \,
\sum_{j'=1,\, j'\neq j}^N \cP'(z_{j}-z_{j'})=0, \; 1\leq j\leq N \no \\
& \hspace*{4.35cm} \text{ and $z_j\neq z_{j'}$ for $j\neq j'$,
$1\leq j,j'\leq N$}\Bigg\}
\lb{4.24}
\end{align}
in the collisionless case.

In the presence of collisions, $\cL_N$ needs to be extended to what we
called the Duistermaat--Gr\"unbaum (DG) locus in Remark \ref{r2.6}\,(ii),
$\hatt\cL_N$, defined by
\begin{align}
\begin{split}
\hatt \cL_1&=X=\cL_1, \\
\hatt\cL_N&=\cL_N\cup\bigcup_{M=1}^{N-1}
\bigcup_{\substack{s_1,\dots,s_M\in\bbN\\ \sum_{\ell=1}^M
s_\ell(s_\ell +1)=2N}}\cM_{s_1,\dots,s_M},
\quad N\geq 2, \lb{4.25}
\end{split}
\end{align}
where
\begin{align}
\cM_{s_1}&=\bigg\{[z_1,\dots,z_N]=[\underbrace{\zeta_1,\dots,\zeta_1}_{N}]
\in X^N/S_N\bigg\}, \lb{4.25A} \\
& \hspace*{1.45cm} s_1\in\bbN, \; s_1(s_1+1)=2N, \; M=1, \no \\
\cM_{s_1,\dots,s_M}&=\Bigg\{[z_1,\dots,z_N]
=[\underbrace{\zeta_1,\dots,\zeta_1}_{s_1(s_1+1)/2},
\underbrace{\zeta_2,\dots,\zeta_2}_{s_2(s_2+1)/2},\dots,
\underbrace{\zeta_M,\dots,\zeta_M}_{s_M(s_M+1)/2}]\in X^N/S_N \, \Bigg |
\no \\
& \hspace*{-3mm} \sum_{\ell'=1,\,
\ell'\neq \ell}^M s_{\ell'}(s_{\ell'}+1)\cP^{(2k-1)}(\zeta_\ell -
\zeta_{\ell'})=0, \quad 1\leq k\leq s_\ell, \; 1\leq \ell\leq M \no \\
& \hspace*{3.4cm} \text{ and } \zeta_\ell\neq \zeta_{\ell'} \text{ for }
\ell\neq \ell', 1\leq \ell,\ell'\leq M \Bigg\}, \lb{4.25a} \\
& \hspace*{1.9cm} s_\ell\in\bbN, \; 1\leq \ell\leq M, \;\,
\sum_{\ell=1}^M s_\ell(s_\ell +1)=2N, \; M\geq 2. \no
\end{align}
In addition to the CM and DG loci we find it convenient to introduce the
following additional locus
\begin{align}
\cA_N&=\Bigg\{[z_1,\dots,z_N]\in X^N/S_N\,\Bigg|\,
\text{$q(z)=-2\sum_{j=1}^N \cP(z-z_j)$ is an} \no \\
& \hspace*{3.8cm} \text{algebro-geometric KdV potential}\Bigg\}
\lb{5.14}
\end{align}

\begin{remark} \lb{r5.2}
$(i)$ Some of the sets $\cM_{s_1,\dots,s_M}$ in the decomposition
\eqref{4.25} of the DG locus \eqref{4.25} may of course be empty. To
illustrate this fact it suffices to consider the simple $g=2$ ($N=4$)
elliptic example
\begin{equation}
q_2(z)=-2\sum_{j=1}^4 \wp(z-\omega_j) \lb{5.15}
\end{equation}
$($here $\omega_4=0$, cf.\ Appendix \ref{sB} for the notation employed in
connection with elliptic functions$)$, a special case of the family of
Treibich--Verdier examples analyzed in detail in \cite{GW95a}. In this
case it is clear that the isospectral manifold of KdV potentials of $q_2$
contains no element of the form $\tilde q_2(z)=-8\wp(z-\zeta_1)$ for
some $\zeta_1\in\bbC$ since $8\neq s_1(s_1+1)$ for any $s_1\in\bbN$. In
particular, there exists no possibility in the corresponding DG locus
associated with the isospectral class of KdV potentials of the form
$\hat q(z)=-2\sum_{j=1}^4 \wp(z-z_j)$ for all $z_1,\dots,z_4$ to
collide at a point $\zeta_1\in\bbC$ and hence $\cM_{s_1}=\emptyset$
in \eqref{4.25}, \eqref{4.25A} in connection with example \eqref{5.15}.
This simple example also shows that for fixed genus $g$, the
corresponding set of elliptic KdV potentials correspond to several DG
loci $\hatt\cL_N$ for different values of $N$, in stark contrast to the
rational case. \\
$(ii)$ Actually, it is easily seen that the situation is even more
complicated in the elliptic case. An analysis of the following KdV
potentials (cf.\ \cite{GW95a})
\begin{align}
q_4 (z) &= -20 \wp (z-\omega_j) -12 \wp (z-\omega_k),\lb{5.16}\\
\hat q_4 (z) &= -20 \wp (z-\omega_j) -6 \wp (z-\omega_k) -6\wp
(z-\omega_\ell),\lb{5.17}\\
q_5 (z) &=-30 \wp (z-\omega_j) -2\wp (z-\omega_k),\lb{5.18}\\
\hat q_5 (z) & =-12 \wp (z-\omega_j) -12\wp (z-\omega_k) -6\wp
(z-\omega_\ell)-2\wp (z-\omega_m), \lb{5.19}
\end{align}
where $j$, $k$, $\ell$, $m\in\{1,2,3,4\}$ are mutually distinct, then
shows the following: The potentials $q_4$ and $\hat q_4$ correspond to
genus $g=4$ while $q_5$ and $\hat q_5$ correspond to $g=5$. However,
we note that all four potentials correspond to $N=16$ in \eqref{2.40}.
In addition, it can be shown that $q_5$ and $\hat q_5$ are isospectral
while $q_4$ and $\hat q_4$ are not. In particular, since $q_4$ and
$\hat q_4$ are not isospectral, there is no KdV flow that deforms
$q_4$ into $\hat q_4$, and one infers that in the elliptic case the DG
locus for fixed $N$ in general consists of several disconnected
components. The latter fact is again in sharp contrast to the rational
case where for fixed $N$ all potentials $($with asymptotic value $q_0$
as $|z|\to\infty$$)$ flow out of $q_g(z)=q_0-g(g+1)z^{-2}$,
$N=g(g+1)/2$. \\
$(iii)$ The simply periodic case is somewhat intermediate between the
rational and elliptic cases. While it is clearly more complex than the
rational case $($e.g., not all simply periodic KdV potentials flow out
of a single potential such as $q_0-g(g+1)z^{-2}$ in the simpler
rational case$)$, it is still possible to describe explicitly the
connected components of the DG locus for fixed genus $g$ $($cf.\
Theorem \ref{t3.15}$)$. This is related to the facts described in
Remark \ref{r2.3a}. \\
$(iv)$ We note that by Theorem \ref{t3.5},
\begin{equation}
\cL_N\subset\hatt\cL_N =\cA_N. \lb{5.20}
\end{equation}
\end{remark}

\medskip

Next we closely investigate the case of rational KdV potentials. In
this case $X=\bbC$ and $\cP(z)=z^{-2}$. We start with the following
known result relating the coefficients and the roots of a polynomial.

\begin{lemma} \label{l5.2}
Fix $N\in\bbN$, assume $r_0=1$, $(r_1,\dots,r_N)\in\bbC^N$ and let
\begin{equation}
\tau_N (z)=\sum_{k=0}^N r_{N-k}z^k=\prod_{j=1}^N (z-z_j),
\quad z\in\bbC \lb{5.22}
\end{equation}
be a monic polynomial of degree $N$ with divisor of zeros
$[z_1,\dots,z_N]\in\bbC^N/S_N$. Introduce the map
\begin{equation}
\Phi_{\tau_N}\colon\begin{cases} \bbC^N\to\bbC^N/S_N \\
(r_1,\dots,r_N)\mapsto [z_1(r_1,\dots,r_N),\dots,z_N(r_1,\dots,r_N)].
\end{cases} \lb{5.22a}
\end{equation}
Then $\Phi_{\tau_N}$ is a homeomorphism.

Let $\Omega\subseteq\bbC^{N}$ and
$\Phi_{\tau_N,\Omega}\colon\Omega\to \Phi_{\tau_N}(\Omega)$ be the
restriction of $\Phi_{\tau_N}$ to $\Omega$. If $\Omega$ and
$\Phi_{\tau_N}(\Omega)$ are both equipped with their relative
topologies, then $\Phi_{\tau_N,\Omega}$ is a homeomorphism.
\end{lemma}
\begin{proof}
Although the first part of this lemma is well-known, we briefly sketch a
proof for completeness: As zeros of a polynomial vary continuously with
the coefficients of a polynomial, $\Phi$ is continuous. The map $\Phi$ is
clearly a bijection. The continuity of the inverse of $\Phi$ is obvious
since the coefficients $r_\ell$ are polynomials (in fact, elementary
symmetric functions) of the roots $[z_1,\dots,z_N]$.

It is clear that $\Phi_{\tau_N,\Omega}$ is a bijection. Let $V$ be an
open set in $\Phi_{\tau_N}(\Omega)$. Then there is an open set
$U\subset\bbC^N/S_N$ such that $V=U\cap \Phi_{\tau_N}(\Omega)$. The
preimage  of $V$ under $\Phi_{\tau_N,\Omega}$ equals
$\Phi^{-1}_{\tau_N}(U)\cap\Omega$. Since $\Phi^{-1}_{\tau_N}(U)$
is open in $\bbC^N$, the preimage $\Phi^{-1}_{\tau_N,\Omega}(V)$ is open
in $\Omega$. Thus $\Phi_{\tau_N,\Omega}$ is continuous. The continuity of
$\Phi^{-1}_{\tau_N,\Omega}$ is shown analogously.
\end{proof}

Next, let $R=\bbC[t_0,\dots,t_{g-1}]$ denote the ring of polynomials in
$t_0,\dots,t_{g-1}$ with coefficients in $\bbC$ and $\tau_N$ a monic
polynomial in $R[z]$ (the ring of polynomials in $z$ with coefficients
in $R$) of degree $N=g(g+1)/2$ for some $g\in\bbN$. The polynomial
$\tau_N$ induces a map
\begin{align}
& \Psi_{\tau_N}\colon\begin{cases} \bbC^g \to X^N/S_N \\
(t_0,\dots,t_{g-1}) \mapsto
[z_1,\dots,z_N], \end{cases} \lb{5.23} \\
& \hspace*{1.3cm} z_j=z_j(r_1(t_0,\dots,t_{g-1}), \dots,
r_N(t_0,\dots,t_{g-1})), \; 1\leq j\leq N, \no
\end{align}
where
\begin{align}
\tau_N (t_0,\dots,t_{g-1},z)&=\sum_{k=0}^N
r_{N-k}(t_0,\dots,t_{g-1}) z^k \no \\
&=\prod_{j=1}^N (z-z_j(r_1(t_0,\dots,t_{g-1}), \dots,
r_N(t_0,\dots,t_{g-1}))), \lb{5.24} \\
& \hspace*{5.5cm} \quad r_0=1, \; z\in\bbC. \no
\end{align}
We note that
\begin{equation}
\Psi_{\tau_N}=\Phi_{\tau_N,\Theta_{\tau_N}(\bbC^g)}\circ\Theta_{\tau_N},
\lb{5.25}
\end{equation}
where
\begin{equation}
\Theta_{\tau_N}\colon\begin{cases}\bbC^g\to\bbC^N \\
   (t_0,\dots,t_{g-1}) \mapsto
(r_1(t_0,\dots,t_{g-1}),\dots,r_N(t_0,\dots,t_{g-1})) \end{cases}
\lb{5.26}
\end{equation}
and
\begin{equation}
\Phi_{\tau_N,\Theta_{\tau_N}(\bbC^g)}
=\Phi_{\tau_N}|_{\Theta_{\tau_N}(\bbC^g)}. \lb{5.28}
\end{equation}

Next, we list the following known results (we recall that $N\in\bbN$ is
called  triangular if there is a $g\in\bbN$ such that $N=g(g+1)/2$).

\begin{theorem} $($Airault, McKean, and Moser \cite[Prop.\
2.2, Cor.\ 3.2]{AMM77}$)$ \lb{t5.3}
If $N$ is triangular then $\cL_N$ $($and hence $\cA_N$$)$ is not
empty. If $N$ is not triangular then $\cA_N$ $($and hence $\cL_N$$)$
is empty.
\end{theorem}

\begin{theorem} $($Airault, McKean, and Moser \cite[Thms.\
3.2]{AMM77}, Adler and Moser \cite[Sect.\ 4]{AM78}$)$ \lb{t5.4}
Suppose $g\in\bbN$ and $N=g(g+1)/2$. Then there exists a
unique monic polynomial $\tau_N\in\bbC[t_0,t_1,\dots,t_{g-1}][z]$ of
degree
$N$ such that the map $\Psi_{\tau_N}\colon\bbC^g\to\cA_N$, defined in
\eqref{5.23}, \eqref{5.24}, is a surjection. The algebro-geometric KdV
potential $q_{\tau_N}$ associated with the divisor
of zeros $[z_1,\dots,z_N]\in\cA_N$ of $\tau_N$ is of the type
\begin{equation}
q_{\tau_N}(t_0,\dots,t_{g-1},z)=q_0
+2[\ln(\tau_N(t_0,\dots,t_{g-1},z))]'' \lb{5.28a}
\end{equation}
with $q_0=\lim_{|z|\to\infty}
q_{\tau_N}(t_0,\dots,t_{g-1},z)$.
\end{theorem}

\begin{theorem} $($Adler and Moser \cite[Lemmas 2.2 and 2.3]{AM78}$)$
\lb{t5.5}  The unique monic polynomial $\tau_N$ in \eqref{5.28a},
\begin{equation}
\tau_N(t_0,\dots,t_{g-1},z)=\sum_{k=0}^N
r_{N-k}(t_0,\dots,t_{g-1}) z^k, \quad r_0=1, \; z\in\bbC,  \lb{5.29}
\end{equation}
has the following properties:\\
$(i)$ Giving $t_m$ weight $2m+1$, $0\leq m\leq g-1$, then $r_j$ is
isobaric of weight $j$, $1\leq j\leq N$.\\
$(ii)$ The coefficient of $t_m$ in $r_{2m+1}$ is not equal to zero.
\end{theorem}

The first part of Theorem \ref{t5.4} can be strengthened as follows.

\begin{theorem} $($Airault, McKean, and Moser \cite[Thm.\
3.2]{AMM77}, Adler and Moser \cite[Sect.\ 4]{AM78}$)$
\label{t5.6}
Let $g$ be a positive integer and $N=g(g+1)/2$. Then
$\Psi_{\tau_N}\colon\bbC^g\to \cA_N$, defined in \eqref{5.23},
\eqref{5.24}, is a homeomorphism.
\end{theorem}
\begin{proof}
For completeness we sketch a proof. It was proven in Lemma \ref{l5.2}
that $\Phi_{\tau_N,\Theta(\bbC^g)}$ is a homeomorphism from
$\Theta_{\tau_N}(\bbC^g)$ to $\Phi_{\tau_N}(\Theta_{\tau_N}(\bbC^g))$.
By Theorem \ref{t5.4},
$\Phi_{\tau_N}(\Theta_{\tau_N}(\bbC^g))=\Psi_{\tau_N}(\bbC^g)=\cA_N$.
Thus, we only have to show that $\Theta_{\tau_N}$ is a homeomorphism from
$\bbC^g$ to $\Theta_{\tau_N}(\bbC^g)$.

Since the $r_j$ are polynomials in $t_0,\dots,t_{g-1}$, continuity
of $\Theta_{\tau_N}$ is obvious. Next we prove by induction that
$t_p\in\bbC[r_1,\dots,r_{2p+1}]$. By Theorem \ref{t5.5} one infers that
$r_1=\alpha_0 t_0$ with $\alpha_0\neq0$. Hence $t_0=r_1/\alpha_0$. So
the claim holds for $p=0$. Next we assume it holds for $p=0,\dots,m-1$.
Again by Theorem \ref{t5.5} one infers that $t_m=(r_{2m+1}-\tilde
r_{2m+1})/\alpha_m$, where $\tilde r_{2m+1}$ is a suitable
polynomial in $\bbC[t_0,\dots,t_{m-1}]$. By the induction
hypothesis $t_0,\dots,t_{m-1}$ are in turn polynomials in
$r_1,\dots,r_{2m-1}$. This completes the induction
step. Thus $\Theta_{\tau_N}$ is injective and
$\Theta^{-1}_{\tau_N}$ is continuous.
\end{proof}

The next theorem contains our principal result in the case of rational
KdV potentials; it details discussions in the literature concerning
the closure of the Calogero--Moser locus (cf., e,g.,
\cite{AMM77}, \cite{Mc79}). To the best of our knowledge, this is the
first explicit characterization of the closure of the rational CM
locus.

\begin{theorem} \label{t5.7}
The DG locus $\hatt\cL_N$ is the closure of the CM locus $\cL_N$ in the
quotient topology $\tau_{S_N}$ of $\bbC^N/S_N$,
\begin{equation}
\cA_N=\hatt\cL_N=\overline{\cL_N}.  \lb{5.30}
\end{equation}
\end{theorem}
\begin{proof}
The statement is trivial if $N$ is not triangular (since all sets are
empty in this case). Hence we suppose for the rest of this proof that
$N=g(g+1)/2$ for some $g\in\bbN$. The first equality in \eqref{5.30} is
then the content of Theorem \ref{t3.5}.

Let $\tau_N$ be the polynomial whose unique existence is guaranteed
by Theorem \ref{t5.4}. First we will prove that
$\overline{\cL_N}\subseteq\hatt\cL_N$. Since, obviously,
$\cL_N\subseteq\hatt\cL_N$, this follows provided that
$\hatt\cL_N$ is closed. But by Theorem \ref{t5.6}
$\hatt\cL_N=\Psi_{\tau_N}(\bbC^g)$ is the preimage of the closed set
$\bbC^g$ under the continuous map $\Psi^{-1}_{\tau_N}$ and hence closed.

Next we prove that $\hatt\cL_N\subseteq\overline{\cL_N}$. Let
\begin{equation}
\Xi=\Bigg[\underbrace{\zeta_1,\dots,\zeta_1}_{s_1(s_1+1)/2},\dots,
\underbrace{\zeta_M,\dots,\zeta_M}_{s_M(s_M+1)/2}\Bigg], \qquad
\sum_{\ell=1}^M s_\ell(s_\ell+1) = 2N  \lb{5.31}
\end{equation}
be an arbitrary point in $\hatt\cL_N$. By Theorem \ref{t5.4} there is a
point $\wti T=(\ti t_0,\dots,\ti t_{g-1})\in\bbC^g$ such that
$\Xi$ represents the roots of $\tau_N (\wti T,\cdot)$. The discriminant
$\Delta_{\tau_N}$ of the polynomial $\tau_N(t_0,\dots,t_{g-1},\cdot)$
is in turn a polynomial in
$\bbC[t_0,\dots,t_{g-1}]$. By Theorem \ref{t5.3}, $\Delta_{\tau_N}$ is
not identically equal to zero, because otherwise $\cL_N$ would be empty.
Let $m$ denote an index for which $\Delta_{\tau_N}$ actually depends on
$t_m$ and define $\delta\in\bbC[s]$ by
\begin{equation}
\delta(s)=
\Delta_{\tau_N}(\ti t_0,\dots,\ti t_{m-1},s,
\ti t_{m+1},\dots,\ti t_{g-1}).
\end{equation}
Then there
is a neighborhood of $\ti t_m$ which contains only one zero of $\delta$
(namely, $\ti t_m$). Let
$t_{n,m}\in\bbC\backslash\{\ti t_m\}$, $n\in\bbN$, be a sequence
of points in this neighborhood which converges to $\ti t_m$ as
$n\to\infty$. Then
\begin{equation}
\Xi_n=\Psi_{\tau_N}(\ti
t_0,\dots,\ti t_{m-1},t_{n,m},\ti t_{m+1},\dots,\ti t_{g-1})
\lb{5.32}
\end{equation}
is in $\cL_N$ and converges to $\Xi$ as $n\to\infty$ by the continuity
of $\Psi_{\tau_N}$. This proves the second equality in \eqref{5.30}.
\end{proof}

In the rational case the issue of the closure of the CM locus $\cL_N$
can also be approached in an alternative manner. Since the actual
details are rather involved, we describe the special case $N=3$ which
reveals some of the underlying mechanism. For this purpose we briefly
recall some facts on elementary symmetric functions. Given $x_j\in\bbC$,
$1\leq j\leq N$, the elementary symmetric functions
$\sigma_j=\sigma_j(x_1,\dots,x_N)$ of $x_1,\dots,x_N$ are defined by
\begin{equation}
\sigma_0 (x_1,\dots,x_N)=1, \quad
\sigma_j(x_1,\dots,x_N)=\begin{cases}
\sum_{\substack{\ell_1=1,\dots,\ell_j=1 \\ \ell_1<\cdots<\ell_j}}^N \,
\prod_{k=1}^j x_{\ell_k}, & 1\leq j\leq N, \\
0, & j\geq N+1. \end{cases} \lb{4.20}
\end{equation}
Alternatively, one can consider $s_j=s_j(x_1,\dots,x_N)$ defined by
\begin{equation}
s_0 (x_1,\dots,x_N) =N, \quad s_j (x_1,\dots,x_N) = \sum_{k=1}^N x_k^j,
\quad j\in\bbN. \lb{4.21}
\end{equation}
We note the following well-known result.

\begin{lemma} \lb{l4.3}
Let $x_j\in\bbC$, $j=1,\dots,N$, and define the
elementary symmetric functions $\sigma_j=\sigma_j(x_1,\dots,x_N)$ and
$s_j=s_j(x_1,\dots,x_N)$, $j\in\bbN_0$ as in \eqref{4.20} and
\eqref{4.21}. Then
\begin{equation}
\sum_{k=0}^{j-1} (-1)^k\sigma_k s_{j-k} + (-1)^j\sigma_j j=0, \quad
j\in\bbN. \lb{4.21A}
\end{equation}
In particular, for $j\in\{1,\dots,N\}$, $s_j$ are polynomials in
$\sigma_1,\dots,\sigma_j$, and for $j\geq N+1$, $s_j$ are polynomials in
$\sigma_1,\dots,\sigma_N$. Conversely, for $j\in\{1,\dots,N\}$,
$\sigma_j$ are polynomials in $s_1,\dots,s_j$. $($All these polynomials
are without constant term.$)$
\end{lemma}

Given these preliminaries we return to the CM locus conditions in
\eqref{2.38A}: They explicitly read for $N=3$,
\begin{equation}
\sum_{\substack{j'=1\\j'\neq j}}^3 (z_j-z_{j'})^{-3}=0, \quad
1\leq j\leq 3, \, \text{ assuming $z_j\neq z_{j'}$ for $j\neq j'$,
$1\leq j,j'\leq 3$.} \lb{5.33}
\end{equation}
Rewriting them in the form
\begin{equation}
[(z_3-z_2)(z_3-z_1)(z_2-z_1)]^{-3} \gamma_j (z_1,z_2,z_3)=0, \quad
1\leq j\leq 3, \lb{5.34}
\end{equation}
where the numerators $\gamma_j$ are certain polynomials in $z_1,z_2,z_3$.
Using the fact that
\begin{equation}
\gamma_j (z_1,z_2,z_3)=0, \;\, 1\leq j\leq 3 \, \text{ is equivalent to}
\,  s_k(\gamma_1,\gamma_2,\gamma_3)=0, \;\, 1\leq k\leq 3, \lb{5.35}
\end{equation}
one infers that
\begin{align}
&s_1(\gamma_1,\gamma_2,\gamma_3)=0 \, \text{ is automatically satisfied
by symmetry},  \lb{5.36} \\
&s_2(\gamma_1,\gamma_2,\gamma_3)=0 \, \text{ is equivalent to \,
$s_1^2(z_1,z_2,z_3)-3s_2(z_1,z_2,z_3)=0$}, \lb{5.37} \\
&s_3(\gamma_1,\gamma_2,\gamma_3)=0 \, \text{ is equivalent to \,
$[s_1^2(z_1,z_2,z_3)-3s_2(z_1,z_2,z_3)]^3$} \no \\
& \hspace*{.3cm} \times [2s_1^2(z_1,z_2,z_3)^3
-9s_1^2(z_1,z_2,z_3)s_2^2(z_1,z_2,z_3)+9s_3^2(z_1,z_2,z_3)]=0.
\lb{5.38}
\end{align}
Thus, the CM conditions \eqref{5.33} reduce to
\begin{equation}
s_1^2(z_1,z_2,z_3)-3s_2(z_1,z_2,z_3)=0 \, \text{ assuming
$z_j\neq z_{j'}$ for $j\neq j'$, $1\leq j,j'\leq 3$.} \lb{5.39}
\end{equation}
One readily verifies that $s_1^2-3s_2=0$ is satisfied in particular on
the diagonal, where all $z_j$ confluent to some point
$z_0\in\bbC$, that is, $z_1=z_2=z_3=z_0$. Since confluence of only two
points $z_j=z_k=z_0$, $z_\ell\neq z_0$ with $j,k,\ell$ pairwise
distinct, is clearly impossible, this readily leads to the fact that
the closure of the CM condition \eqref{5.39} in $\bbC^3/S_3$ is
simply given by
\begin{equation}
s_1^2(z_1,z_2,z_3)-3s_2(z_1,z_2,z_3)=0.  \lb{5.40}
\end{equation}
In other words, the closure of the CM locus $\cL_3$ is obtained by
joining the diagonal $z_1=z_2=z_3$ to $\cL_3$, in agreement with
\eqref{5.30} and the description of $\hatt\cL_3=\cL_3\cup\cM_2$ in
\eqref{4.25}, \eqref{4.25a} (cf.\ also Remark 2.11\,$(iv)$).

\medskip

Next, we turn to the case of simply periodic meromorphic KdV
potentials of period $\omega\in\bbC\backslash\{0\}$ bounded near the ends
of the period strip $\cS_\omega$. In this case
\begin{equation}
X=\bbC/\Lambda_\omega \, \text{ and } \, \cP(z)=\f{\pi^2}{\omega^{2}}
   \Big([\sin(\pi z/\omega)]^{-2}-\f{1}{3}\Big).
\end{equation}

We denote by $\bbC^*$ the set of nonzero complex numbers
$\bbC\backslash\{0\}$ equipped with the relative topology in $\bbC$.
We start with the following result.

\begin{lemma} \label{l3.9}
Fix $N\in\bbN$, assume $r_0=1$, $(r_1,\dots,r_{N-1})\in\bbC^{N-1}$,
$r_N\in\bbC\backslash\{0\}$ and let
\begin{equation}
\tau_N (u)=\sum_{k=0}^N r_{k}u^k=r_N \prod_{j=1}^N
\big(u-e^{2\pi iz_j/\omega}\big), \quad u\in\bbC
\end{equation}
be a polynomial of degree $N$ with divisor of zeros
$\big[e^{2\pi iz_1/\omega},\dots,e^{2\pi iz_N/\omega}\big]\in X^N/S_N$.
Then all zeros of
$\tau$ are nonzero and  each has a logarithm\footnote{We note that the
logarithm of $e^{2\pi iz/\omega}$ is well-defined for $z\in X$.}. In
particular, $z_j\in X$, $1\leq j\leq N$. Introduce the map
\begin{equation}
\Phi_{\tau_N}\colon\begin{cases} \bbC^{N-1}\times\bbC^*\to X^N/S_N \\
(r_1,\dots,r_N)\mapsto [z_1(r_1,\dots,r_N),\dots,z_N(r_1,\dots,r_N)].
\end{cases}
\end{equation}
Then $\Phi_{\tau_N}$ is a homeomorphism.

Let $\Omega\subseteq\bbC^{N-1}\times\bbC^*$ and
$\Phi_{\tau_N,\Omega}\colon\Omega\to \Phi_{\tau_N}(\Omega)$ be the
restriction of $\Phi_{\tau_N}$ to $\Omega$. If $\Omega$ and
$\Phi_{\tau_N}(\Omega)$ are both equipped with their relative
topologies, then $\Phi_{\tau_N,\Omega}$ is a homeomorphism.
\end{lemma}
\begin{proof}
The proof is similar to that of Lemma \ref{l5.2}.
\end{proof}

In the following let $g\in\bbN$. Introducing the $g\times g$
Vandermonde matrix
\begin{equation}
\cV(a_1,\dots,a_g)=\Big(a_p^{p'-1}\Big)_{1\leq p,p'\leq g}, \quad
a_p\in\bbC, \; 1\leq p\leq g, \lb{5.42}
\end{equation}
and denoting its determinant by
$\vartheta(a_1,\dots,a_g)$, that is,
\begin{equation}
\vartheta(a_1,\dots,a_g)=\det (\cV(a_1,\dots,a_g))
=\prod_{\substack{p,p'=1\\p<p'}}^g (a_{p'}-a_p),
\end{equation}
it is clear that $\vartheta(a_1,\dots,a_g)\neq 0$ if and only if the
$a_p$ are pairwise distinct.

Next, define the sets\footnote{Here $\gcd(n_1,\dots,n_g)$
abbreviates the greatest common divisor of $(n_1,\dots,n_g)\in\bbN^g$.}
\begin{equation}
\cN_g=\{(n_1,\dots,n_g)\in\bbN^g\,|\, n_1<n_2<\cdots<n_g, \;
   \gcd(n_1,\dots,n_g)=1\}
\end{equation}
and
\begin{equation}
\cN=\bigcup_{g=1}^\infty \cN_g.
\end{equation}
For $\ul n=(n_1,\dots,n_g)\in\cN$ we denote the number of its
components by
\begin{equation}
\#(\ul n)=g.
\end{equation}
For $\ul n=(n_1,\dots,n_g)\in\cN_g$ and $\ul
v=(v_1,\dots,v_g)\in\bbC^{*g}$ we also introduce the $g\times g$ matrix
$T(\ul n,\ul v,u)$ by
\begin{equation}
T(\ul n,\ul v,u)=\Big(n_{p'}^{p-1}\Big[v_{p'}
u^{n_{p'}}-(-1)^{p}\Big]\Big)_{1\leq p,p'\leq g}.
\end{equation}
Moreover, we define\footnote{Here $\floor{x}$ denotes the greatest
integer less or equal to $x\in\bbR$.}
\begin{equation}
\tau_N(\ul n,\ul v,u)=(-1)^{\floor{g/2}}\frac{\det (T(\ul n,\ul
v,u))}{\vartheta(n_1,\dots,n_g)}.
\end{equation}

\begin{lemma} \label{l3.10}
Suppose $\ul n\in\cN_g$ and $\ul v\in\bbC^{*g}$. Then
\begin{equation}
\tau_N(\ul n,\ul v,u)=(-1)^{\floor{g/2}}\frac{\det (T(\ul n,\ul
v,u))}{\vartheta(n_1,\dots,n_g)}
=\sum_{k=0}^N r_k(\ul v) u^k, \quad N=\sum_{p=1}^g n_p,
\end{equation}
where
\begin{equation}
r_k(\ul v)=\sum_{\substack{\sigma_1=0,\dots,\sigma_g=0\\
\ul n\cdot\ul\sigma=k}}^{1}
\f{\vartheta\big((-1)^{\sigma_1}n_1,\dots,(-1)^{\sigma_g}
n_g\big)}{\vartheta(n_1,\dots,n_g)} v_1^{\sigma_1}\cdots v_g^{\sigma_g},
\quad 0\leq k\leq N.
\end{equation}
In particular, assigning the weight $n_p$ to $v_p$, one infers the
following properties of the coefficients $r_k(\ul v)$, $0\leq k\leq
N$:\\
$(i)$ $r_k(\ul v)$ is a polynomial of the variables $v_1$,\dots,$v_g$
isobaric of weight $k$.\\
$(ii)$ $r_k(\ul v)$ has degree at most one if it is considered as a
polynomial of $v_p$ only.\\
$(iii)$ $r_0(\ul v)=1$ and $r_N(\ul v)= (-1)^{\floor{g/2}}v_1\cdots
v_g$.\\
$(iv)$ The coefficient of $v_1^{\sigma_1}\cdots v_g^{\sigma_g}$ in
$r_k$ is different from zero for  any $\ul\sigma\in\{0,1\}^g$ such
that $\ul n\cdot\ul\sigma=k$.
\end{lemma}
\begin{proof}
The $p$th column of $T(\ul n,\ul v,u))$ can be written as
\begin{equation}
v_p u^{n_p}
\ul \alpha_p+\ul\beta_p,
\end{equation}
where
\begin{align}
\ul\alpha_p&=\big(1,n_p,n_p^2,\dots,n_p^{g-1}\big)^\top,  \\
\ul\beta_p&=\big(1,-n_p,(-n_p)^2,\dots,
(-n_p)^{g-1}\big)^\top, \quad 1\leq p\leq g.
\end{align}
Hence, for $\sigma_p\in\{0,1\}$, one infers that
\begin{equation}
\ul\gamma_p(\sigma_p)=\sigma_p\ul\alpha_p+(1-\sigma_p)\ul\beta_p
\end{equation}
either equals $\ul\alpha_p$ or $\ul\beta_p$, $1\leq p\leq g$.
The multilinearity of the determinant then implies
\begin{align}
\begin{split}
\det (T(\ul n,\ul v,u)))&=\sum_{\sigma_1=0}^1 \cdots
   \sum_{\sigma_g=0}^1 (v_1 u^{n_1})^{\sigma_1}\cdots
(v_g u^{n_g})^{\sigma_g}
   \det\big(\ul\gamma_1(\sigma_1),\dots,\ul\gamma_g(\sigma_g)\big) \\
&=\sum_{\sigma_1=0}^1 \cdots
   \sum_{\sigma_g=0}^1 v_1^{\sigma_1}\cdots v_g^{\sigma_g}
u^{n_1\sigma_1+\cdots+n_g\sigma_g}
   \det\big(\ul\gamma_1(\sigma_1),\dots,\ul\gamma_g(\sigma_g)\big).
\end{split}
\end{align}
Next, noticing
\begin{align}
\det\big(\ul\gamma_1(\sigma_1),\dots,\ul\gamma_g(\sigma_g)\big)&=
\vartheta\big((-1)^{1+\sigma_1}
n_1,\dots,(-1)^{1+\sigma_g}n_g\big) \no \\
&=(-1)^{\floor{g/2}}\vartheta\big((-1)^{\sigma_1}
n_1,\dots,(-1)^{\sigma_g}n_g\big),
\end{align}
one computes
\begin{align}
\tau_N(\ul n,\ul v,u)&=(-1)^{\floor{g/2}}\frac{\det (T(\ul n,\ul
v,u))}{\vartheta(n_1,\dots,n_g)} \no \\
&=\Big[(-1)^{\floor{g/2}} \vartheta\big(n_1,\dots,n_g\big)\Big]^{-1}
\no \\
& \quad \times \sum_{\sigma_1=0}^1 \cdots
   \sum_{\sigma_g=0}^1 v_1^{\sigma_1}\cdots v_g^{\sigma_g}
u^{\ul n\cdot\ul\sigma}
\det\big(\ul\gamma_1(\sigma_1),\dots,\ul\gamma_g(\sigma_g)\big) \no \\
&=\sum_{k=0}^N\Bigg(
\sum_{\substack{\sigma_1=0,\dots,\sigma_g=0\\
\ul n\cdot\ul\sigma=k}}^{1}
\f{\vartheta\big((-1)^{\sigma_1} n_1,\dots,(-1)^{\sigma_g}
n_g\big)}{\vartheta(n_1,\dots,n_g)} v_1^{\sigma_1}\cdots
v_g^{\sigma_g}\Bigg)u^k.
\end{align}
This implies all statements made in the lemma.
\end{proof}

Next we introduce the map
\begin{equation}
\Theta_{\ul n}\colon \begin{cases} \bbC^{*g}\to \bbC^{N-1}\times \bbC^*
\\ (v_1,\dots,v_g) \mapsto (r_1,\dots,r_N). \end{cases}
\end{equation}

\begin{lemma} \label{l3.11}
$\Theta_{\ul n}$ is a homeomorphism from $\bbC^{*g}$ to
$\Theta_{\ul n}(\bbC^{*g})$.
\end{lemma}
\begin{proof}
Since the $r_k$ are polynomials in terms of the $v_p$, continuity of
$\Theta_{\ul n}$ is obvious. Next we prove by induction that
$v_p\in\bbC[r_1,\dots,r_{n_p}]$. By Lemma \ref{l3.10} one infers
$r_{n_1}=\alpha_1 v_1$ with $\alpha_1\neq0$. Hence
$v_1=r_{n_1}/\alpha_1$ and the claim holds for $p=1$. Assume it holds
for $p=1,\dots,m-1$. Again by Lemma \ref{l3.10} one infers that
$v_m=(r_{n_m}-\tilde r_{n_m})/\alpha_m$, where $\tilde r_{n_m}$ is a
suitable polynomial in $\bbC[v_1,\dots,v_{m-1}]$. By the induction
hypothesis $v_1$,\dots, $v_{m-1}$ are, in turn, polynomials in terms
of $r_1$,\dots,$r_{n_{m-1}}$. This completes the induction proof. This
proves both that $\Theta_{\ul n}$ is injective and that
$\Theta_{\ul n}^{-1}$ is continuous.
\end{proof}

\begin{lemma} \label{l3.12}
Fix $\ul n\in\cN$ with $\#(\ul n)=g$. Then the discriminant of
$\tau_N(\ul n,\ul v,\cdot)$ is a nonzero polynomial in
$v_1$,\dots,$v_g$.
\end{lemma}
\begin{proof}
It is clear that the discriminant of $\tau_N(\ul n,\ul v,\cdot)$ is
a polynomial in $v_1$,\dots, $v_g$. Next we will prove that it is not
identically zero.

Let $\ul v_k=(v_1,\dots,v_k,0,\dots,0)$ and
$f_k(u)=\tau_N(\ul n,\ul v_k,u)$. We will prove by induction that
there is a choice of $v_1$,\dots, $v_k$ such that $f_k$ has
$n_1+\cdots+n_k$ simple zeros. In particular, there  is a choice of
$\ul v$ such that $\tau_N(\ul n,\ul v,\cdot)=f_g$ has
$N$ simple zeros and hence its discriminant is not identically equal to
zero.

First one notes that $f_1(u)=1+c_1 v_1 u^{n_1}$ for some nonzero
constant $c_1$. Thus, $f_1$ has $n_1$ simple zeros for any
$v_1\in\bbC^*$. Next, assume that $f_k$ has
$n_1+\cdots+n_k$ simple zeros. Choose $v_{k+1}=\varepsilon$ and
define
\begin{equation}
\tilde f_{k+1}(\varepsilon,t)=t^{n_1+\cdots+n_{k+1}} f_{k+1}(1/t).
\end{equation}
Then there are polynomials $g_{k+1}$ and $h_{k+1}$ of degree
$n_1+\cdots+n_k$ such that
\begin{equation}
\tilde f_{k+1}(\varepsilon,t)
   =t^{n_{k+1}} g_{k+1}(t)+\varepsilon h_{k+1}(t).
\end{equation}
One notes that $g_{k+1}(0)$ is the coefficient of $u^{n_1+\cdots+n_k}$
in $f_k(u)$ and that $h_{k+1}(0)$ is the coefficient of
$u^{n_1+\cdots+n_{k+1}}$ in $f_{k+1}(u)$. By statement (iv) of
Lemma \ref{l3.10} both $g_{k+1}(0)$ and $h_{k+1}(0)$ are different
from zero.

$\tilde f_{k+1}(0,\cdot)$ has $n_1+\cdots+n_k$ simple zeros away from
zero and a zero of multiplicity $n_{k+1}$ at zero. As the zeros of a
polynomial are continuous functions of the coefficients, one infers for
$\varepsilon$ sufficiently small, that
$\tilde f_{k+1}(\varepsilon,\cdot)$ has $n_{k+1}$ zeros in some small
disk $D_0$ centered at zero and $n_1+\cdots+n_k$ simple zeros outside
$D_0$. The zeros in $D_0$ have Puiseux expansions whose leading term is
given by $\gamma \varepsilon^{1/n_{k+1}}$, where $\gamma$ is any of the
$n_{k+1}$st roots of $-h_{k+1}(0)/g_{k+1}(0)$. This implies that there
are
$n_{k+1}$ simple roots in $D_0$. Thus all roots  of $\tilde f_{k+1}$
and hence all roots of $f_{k+1}$ are simple.
\end{proof}

We briefly illustrate $\tau_N(\ul n, \ul v, u)$ with a few explicit
examples.

\begin{example} \lb{e3.13}  ${}$ \\
$g=1$: Then necessarily $n_1=N=1$ and
\begin{align}
\quad \tau_1(1,v_1,u)&=1+v_1 u. \\
\intertext{$g=2$, $N=n_1+n_2$:}
\tau_{n_1+n_2}(\ul n,\ul v,u)&=1-\frac{n_1+n_2}{n_2-n_1}v_1 u^{n_1}
   +\frac{n_1+n_2}{n_2-n_1}v_2 u^{n_2} - v_1 v_2u^{n_1+n_2}. \\
\intertext{$g=3$, $n_1=1$, $n_2=2$, and $n_3=3$, $N=6$:}
\tau_6(\ul n,\ul v,u)&=1 + 6 v_1 u - 15 v_2 u^2
   + (10v_3-10v_1v_2)u^3 + 15 v_1 v_3 u^4 \no \\
&\quad - 6 v_2 v_3 u^5 - v_1 v_2 v_3 u^6. \\
\intertext{$g=3$, $n_1=1$, $n_2=3$, and $n_3=4$, $N=8$:}
\tau_8(\ul n,\ul v,u)&=1 + 10/3 v_1 u - 14 v_2 u^3 +
      35/3(v_3-v_1 v_2) u^4 + 14 v_1 v_3 u^5 \no \\
& \quad - 10/3 v_2 v_3 u^7 -
      v_1 v_2 v_3 u^8. \\
\intertext{$g=4$, $n_1=1$, $n_2=3$, $n_3=4$, and $n_4=6$, $N=14$:}
\tau_{14}(\ul n,\ul v,u)&=1-14/3v_1u+42v_2u^3-(175/3v_3+49v_1 v_2)u^4
       + 98 v_1 v_3 u^5 \no \\
   &\quad+ 21 v_4 u^6 - 50 (v_2 v_3+v_1 v_4) u^7
      +21 v_1 v_2 v_3 u^8+ 98 v_2 v_4 u^9 \no \\
   &\quad - (175/3 v_1 v_2 v_4 + 49 v_3 v_4) u^{10}
      + 42 v_1 v_3 v_4 u^{11} \no \\
   &\quad - 14/3 v_2 v_3 v_4 u^{13}+ v_1 v_2 v_3 v_4 u^{14}.
\end{align}
\end{example}

Given these preparations we can now characterize the class of
simply periodic meromorphic KdV potentials of period $\omega\in\bbC^*$,
bounded near the ends of the period strip $\cS_\omega$, as follows.

\begin{theorem} \label{t3.8a}
Let $g\in\bbN$ and assume $q$ is a simply periodic, meromorphic
KdV potential of period $\omega\in\bbC^*$, bounded near the ends of the
period strip $\cS_\omega$, corresponding to the singular hyperelliptic
curve $\cK_g$ in \eqref{2.8B}. Then $q$ is of the form
\begin{align}
\begin{split}
&q(z)=e_0+2\big[\ln\big(\tau_N\big(\ul n,\ul v,e^{2\pi
iz/\omega}\big)\big)\big]''
\lb{5.67} \\
&\text{for some $\ul n=(n_1,\dots,n_g)\in\cN_g$,
$\ul v=(v_1,\dots,v_g)\in\bbC^{*g}$, $N=\sum_{p=1}^g n_p$.}
\end{split}
\end{align}
Conversely, every $q$ of the form \eqref{5.67} is a simply
periodic meromorphic KdV potential of period $\omega$, bounded near
the ends of the period strip $\cS_\omega$, corresponding to a
singular hyperelliptic curve $\cK_g$ of the form \eqref{2.8B}.
\end{theorem}
\begin{proof}
Suppose $q$ satisfies the hypotheses of the theorem. By
\cite[Theorem\ 2.3]{GH00} (see also \cite[App.\ G]{GH03}), $g$ Darboux
transformations at the mutually distinct energy parameters $e_p\in\bbC$,
$1\leq p\leq g$ (all different from $e_0\in\bbC$), reduce $q$ to the
constant potential
$q_0=e_0$. Reversing the $g$ Darboux transformations, using the
Crum--Darboux approach discussed, for instance, in \cite[Appendix\
A]{GS95}, then shows that $q$ is of the form
\begin{equation}
q(z)=e_0+2[\ln(W(\psi_1(e_1,z),\dots,\psi_g(e_g,z)))]'', \lb{5.68}
\end{equation}
where
\begin{equation}
\psi_p(e_p,z)=A_pe^{(e_p-e_0)^{1/2}z}+B_pe^{-(e_p-e_0)^{1/2}z}, \quad
e_{p}\neq e_{p'} \text{ for $p\neq p'$}, \, 0\leq p,p'\leq g \lb{5.69}
\end{equation}
for some choice of $A_p, B_p \in \bbC^*$, $1\leq p\leq g$ and
$W(\psi_1,\dots,\psi_g)$ denotes the Wronskian of
$\psi_1,\dots,\psi_g$. Introducing
\begin{align}
v_p&=A_p/B_p, \quad 1\leq p \leq g, \\
\wti T(\ul v,z)&=\Big(\big[(e_{p'}-e_0)^{1/2}\big]^{(p-1)}\Big[v_{p'}
e^{2(e_{p'}-e_0)^{1/2}z}-(-1)^{p}\Big]\Big)_{1\leq p,p'\leq g},
\lb{5.71}
\end{align}
a direct computation confirms that \eqref{5.68} can be rewritten as
\begin{equation}
q(z)=e_0+2\big[\ln\big(\det\big(\wti T(\ul v,z)\big)\big)\big]''.
\lb{5.72}
\end{equation}
In general, expressions such as \eqref{5.72} exhibit no periodicity
properties with respect to $z$. Periodicity of $q$ is obtained as
follows. By
\eqref{5.71},
$\det(\wti T(\ul v,z))$ is of the form
\begin{equation}
F\big(e^{2\pi iz_1/\omega_1},\dots,e^{2\pi iz_g/\omega_g}
\big)\big|_{z_1=\cdots =z_g=z}, \quad \omega_p=\pi i/(e_p-e_0)^{1/2},
\; 1\leq p\leq g
\end{equation}
for some continuous function $F\colon\bbC^g\to\bbC$. Thus, $\det(\wti
T(\ul v,\cdot))$ (and hence $q$) becomes periodic with respect to $z$
of period $\omega\in\bbC^*$ if and only if
\begin{equation}
\omega_p=\omega/n_p, \text{ that is, } (e_p-e_0)^{1/2}=\pi
in_p/\omega, \quad 1\leq p\leq g
\end{equation}
for some integers $n_p\in\bbN$, $1\leq p\leq g$. By \eqref{5.69}, the
integers $n_p$ are necessarily mutually distinct. In addition, $\omega$
is a fundamental period\footnote{$\omega$ is a fundamental period of
$q$ if and only if every period $\Omega$ of $q$ is of the form
$\Omega=m\omega$ for some $m\in\bbZ\backslash\{0\}$} of $q$ if and only
if $\gcd(n_1,\dots,n_g)=1$. Thus, observing
\begin{equation}
\big[\ln\big(\det\big(\wti T(\ul v,z)\big)\big)\big]''
=\big[\ln\big(\tau_N\big(\ul n,\ul v,
e^{2\pi iz/\omega}\big)\big)\big]''
\end{equation}
yields
\begin{equation}
q(z)=e_0+2\big[\ln\big(\tau_N\big(\ul n,\ul v,e^{2\pi
iz/\omega}\big)\big)\big]'' \lb{5.76}
\end{equation}
and hence \eqref{5.67}.

Conversely, suppose $q$ is of the form \eqref{5.67}. Then,
\begin{equation}
q(z)=q^*\big(e^{2\pi iz/\omega}\big),
\end{equation}
where
\begin{equation}
q^*(u)=e_0-8\pi^2\omega^{-2}\frac{u^2\ti\tau_N''(u)\ti\tau_N(u)
+u\ti\tau_N'(u)\ti\tau_N(u) -u^2\ti\tau_N'(u)^2}{\ti\tau_N(u)^2},
\quad u=e^{2\pi iz/\omega}, \lb{5.77}
\end{equation}
and we denoted $\ti\tau_N(u)=\tau_N(\ul n,\ul v, e^{2\pi iz/\omega})$.
We recall that $\ti\tau_N(\cdot)$ is a polynomial of degree $N$ and
hence it has precisely $N$ zeros counting multiplicities. As discussed
above  these correspond to precisely $N$ poles of the meromorphic
function $q$ in each period strip. By inspection of \eqref{5.77} one
confirms that the degree of the numerator of $q^*$ is less  than $2N$
and that $q^*(0)=e_0$. Thus, $q^*-e_0$ tends to zero at the ends of
the period strip $\cS_\omega$. Reversing the argument leading from
\eqref{5.68} to \eqref{5.76} then yields
\begin{equation}
q^*(u)=q(z)=e_0+2[\ln(W(\psi_1(e_1,z),\dots,\psi_g(e_g,z)))]'', \quad
u=e^{2\pi iz/\omega}, \lb{5.78}
\end{equation}
where
\begin{equation}
\psi_p(e_p,z)=\wti
A_pe^{(e_p-e_0)^{1/2}z}+\wti B_pe^{-(e_p-e_0)^{1/2}z} \lb{5.78a}
\end{equation}
for some choice of $\wti A_p, \wti B_p \in \bbC^*$ satisfying
\begin{equation}
v_p=\wti A_p/\wti B_p, \quad 1\leq p \leq g.
\end{equation}
This proves that $q$ is obtained from the constant potential
$q_0=e_0$ by precisely $g$ Darboux transformations. Again applying
\cite[Theorem\ 2.3]{GH00} then shows that $q$ is a simply periodic,
meromorphic KdV potential of period $\omega$, bounded near the ends of the
period strip, and associated with the singular hyperelliptic curve
\eqref{2.8B}.
\end{proof}

Summarizing, one obtains the following result.

\begin{corollary} \label{c3.14}
The set
\begin{equation}
\cS=\{q(z)=C+2[\ln (\tau_N(\ul n,\ul v,\exp(2\pi iz/\omega)))]''\,|\,
   C\in\bbC, \, \ul n\in\cN_g, \, \ul v\in\bbC^{*g}, \, g\in\bbN_0\}
\end{equation}
is precisely the set of simply periodic meromorphic KdV potentials of
period $\omega$, bounded near the ends of the period strip $\cS_\omega$.
\end{corollary}

\begin{theorem} \label{t3.15}
For $N\in\bbN-\{2\}$ there are finitely many $\ul n\in\cN$ such that
$\sum_{p=1}^{\#(\ul n)} n_p=N$. For each of these $\ul n$, the map
$\Phi\circ\Theta_{\ul n}$ is a homeomorphism from $\bbC^{*g}$ to its
image, a closed subset of
$\cA_N$. Furthermore $\cA_N$ is the union of these images over all
possible choices of $\ul n$. In particular, $\cA_N$ is a finite union of
closed connected sets.
\end{theorem}
\begin{proof}
The first statement is obvious. Next, we denote by $g=\#(\ul n)$
the number of components in $\ul n$. By  Lemma \ref{l3.11},
$\Theta_{\ul n}$ is a homeomorphism from $\bbC^{*g}$ to
$\Theta_{\ul n}(\bbC^{*g})$ and by Lemma \ref{l3.9}, $\Phi$ restricted
to this set is also a homeomorphism. Clearly, the image of
$\Phi\circ\Theta_{\ul n}$ is a closed set in $\cA_N$. This proves the
second statement. Finally, pick any element $\Xi$ in
$\cA_N$ and let $q$ be the associated potential. By Corollary
\ref{c3.14}, $q\in\cS$, that is, there exist $C\in\bbC$, $g\in
\bbN$, $\ul n\in\cN_g$ and $\ul v\in\bbC^{*g}$ such that
$q(z)=C+2[\ln(\tau_N(\ul n,\ul v,\exp(2\pi iz/\omega)))]''$. Since the
number of  poles of $q$ per period strip is $N$, we must have
$n_1+\cdots+n_g=N$. Thus, $\Xi=\Phi(\Theta_{\ul n}(\ul v))$.
\end{proof}

\begin{theorem} \label{t3.16}
The DG locus $\hatt\cL_N$ is the closure of the CM locus $\cL_N$ in the
quotient topology $\tau_{S_N}$ of $X^N/S_N$,
\begin{equation}
\cA_N=\hatt\cL_N=\overline{\cL_N}.  \lb{5.79}
\end{equation}
\end{theorem}
\begin{proof}
The statement is trivial if $N=2$ (all sets are empty). Hence we
suppose for the rest of this proof that $N\neq2$. The first equality
in \eqref{5.69} is then the content of Theorem \ref{t3.5}.

Since by Theorem \ref{t3.15}, $\hat\cL_N$ is closed, and since
$\cL_N\subseteq\hat\cL_N$ it follows that $\overline{\cL_N}\subseteq
\hat\cL_N$.

Next we prove that $\hat\cL_N\subseteq\overline{\cL_N}$. Let
\begin{equation}
\Xi=\Bigg[\underbrace{\zeta_1,\dots,\zeta_1}_{s_1(s_1+1)/2},\dots,
\underbrace{\zeta_M,\dots,\zeta_M}_{s_M(s_M+1)/2}\Bigg], \qquad
\sum_{\ell=1}^M s_\ell(s_\ell+1) = 2N  \lb{5.80}
\end{equation}
be an arbitrary point in $\hat\cL_N$. By Theorem \ref{t3.15}, there is
an $\ul n\in\cN$ and a $\ul {\ti v}=(\ti v_1,\dots,\ti
v_g)\in\bbC^{*g}$ such that
$\Xi$ represents $\omega/(2\pi i)$ times the logarithm of the roots
of $\tau_N(\ul n,\ul {\ti v},\cdot)$. The discriminant
$\Delta_{\tau_N}$ of $\tau_N(\ul n,\ul v,\cdot)$ is a polynomial in
$\bbC[v_1,\dots,v_g]$ which is not identically equal to zero according
to Lemma \ref{l3.12}. Let $m$ denote an index for which
$\Delta_{\tau_N}$ actually depends on $v_m$ and define
$\delta\in\bbC[w]$ by
\begin{equation}
\delta(w)=\Delta_{\tau_N}(\ti
v_1,\dots,\ti v_{m-1},w,\ti v_{m+1},\dots,\ti v_g).
\end{equation}
Then there is a neighborhood of $\ti v_m$ which contains only one zero
of $\delta$ (namely, $\ti v_m$). Let
$v_{n,m}\in\bbC^*\backslash\{\ti v_m\}$, $n\in\bbN$, be a sequence
of points in this neighborhood which converges to $\ti v_m$ as
$n\to\infty$. Then
\begin{equation}
\Xi_n=(\Phi\circ\Theta_{\ul n})
(\ti v_1,\dots,\ti v_{m-1},v_{n,m},\ti v_{m+1},\dots,\ti v_g)
\end{equation}
is in $\cL_N$ and converges to $\Xi$ as $n\to\infty$ by the continuity
of $\Phi\circ\Theta_{\ul n}$. This proves the second equality in
\eqref{5.79}.
\end{proof}

To the best of our knowledge, the precise structure of the isospectral
set of simply periodic meromorphic KdV potentials bounded near the ends of
the period strip as described in Theorem \ref{t3.15} and the explicit
characterization of the closure of the simply periodic CM locus are
new.

\begin{remark} \lb{r3.17}
The corresponding results in the elliptic case require different
techniques since elliptic KdV potentials cannot be constructed by
finitely many Darboux transformations starting from constant
potentials.
\end{remark}

\section{Some applications to the time-dependent KdV hierarchy}
\lb{s4}

Rational, simply periodic, and elliptic KdV solutions are frequently
discussed in a time-dependent setting and the dynamics of their poles
is well-known to be in an intimate relationship with completely
integrable systems of the Calogero--Moser-type. In our discussion
below, the time-dependence (and the ensuing isospectral deformations
of the KdV hierarchy) will be approached from the point of view of
tracing trajectories in the DG locus \eqref{4.25} (the appropriate
extension of the CM locus \eqref{4.24}), which permits an efficient
description of the behavior of solutions in a neighborhood of
collisions of their poles.

We start with a brief summary of the time-dependent setup and freely
employ the notation used in Appendix \ref{sA}. Fix $r\in\bbN_0$ and
suppose $q=q(x,t_r)$ satisfies the $r$th time-dependent KdV equation with
initial condition $q^{(0)}=q^{(0)}\big(x,t_r^{(0)}\big)$ a solution of
the $n$th stationary KdV equation for some $n\in\bbN$,
\begin{align}
\wti\KdV_r(q)&=q_{t_r}-2\tilde f_{r+1,z}=0, \lb{4.1} \\
q\big|_{t_r=t_r^{(0)}}&=q^{(0)}, \no \\
\sKdV_n(q^{(0)})&=-2f_{n+1,z}=0, \lb{4.2}
\end{align}
where
\begin{subequations} \lb{4.3}
\begin{align}
q\big(z,t_r^{(0)}\big)=q^{(0)}(z)&=q_0 -\sum_{\ell=1}^{M^{(0)}}
s_\ell^{(0)}\big(s_\ell^{(0)} +1\big) \cP\big(z-\zeta_\ell^{(0)}\big)
\lb{4.3a} \\
&=q_0 -2\sum_{j=1}^{N} \cP\big(z-z_j^{(0)}\big) \lb{4.3b}
\end{align}
\end{subequations}
and
\begin{equation}
s_\ell^{(0)}\in\bbN, \; 1\leq s_\ell^{(0)} \leq M^{(0)}, \quad
\sum_{\ell=1}^{M^{(0)}} s_\ell^{(0)}\big(s_\ell^{(0)}+1\big)=2N.
\lb{4.4}
\end{equation}
Here we assume in accordance with the paragraph following \eqref{A.19}
that the set of integration constants $\tilde c_s$, $1\leq s=1\leq r$ in
$\tilde f_{r+1,z}$ and $c_j=c_j(\ul E)$, $1\leq j\leq n$ (cf.\
\eqref{A.17}) in $f_{n+1,z}$ are independent of each other as
discussed in the paragraph following \eqref{A.19}.

Next, taking advantage of the isospectral property of KdV flows, one can
replace \eqref{4.1}--\eqref{4.4} by
\begin{align}
\wti\KdV_r(q)&=q_{t_r}-2\tilde f_{r+1,z}=0, \lb{4.6} \\
\sKdV_n(q)&=-2f_{n+1,z}=0, \lb{4.7}
\end{align}
or equivalently, by the pair of equations
\begin{align}
&q_{t_r}=\f12 \wti F_{r,zzz}+2(q-E)\wti F_{r,z}+q_z\wti F_r, \lb{4.8} \\
&-\f12 F_{n,zz}F_n+\f14 F_{n,z}^2 +(E-q)F_n^2=\prod_{m=0}^{2n} (E-E_m)
\text{ for some $\{E_m\}_{m=0}^{2n}\subset\bbC$}, \lb{4.9}
\end{align}
where
\begin{subequations} \lb{4.10}
\begin{align}
q(z,t_r)&=q_0 -2\sum_{j=1}^{N} \cP(z-z_j(t_r)) \lb{4.10a} \\
&=q_0 -\sum_{\ell=1}^{M(t_r)} s_\ell(t_r)(s_\ell(t_r)
+1)\cP(z-\zeta_\ell(t_r)) \lb{4.10b}
\end{align}
\end{subequations}
and for each $t_r\in\bbC$,
\begin{equation}
s_\ell(t_r)\in\bbN, \; 1\leq \ell \leq M(t_r), \quad
\sum_{\ell=1}^{M(t_r)} s_\ell(t_r)(s_\ell(t_r)+1)=2N. \lb{4.11}
\end{equation}

Below we will show that $z_j(t_r)$ locally have an algebraic behavior
so that they can be labelled in such a manner that they remain
continuous functions of
$t_r$ even through the process of collisions. On the other hand,
$s_\ell(t_r)\in\bbN$ are integer-valued and discontinuous with respect
to $t_r$ at instances of collisions.

First we turn to a determination of the time-dependence of $z_j(t_r)$
in the absence of collisions.

\begin{lemma} \lb{l4.1}
Let $\Omega\subset\bbC^2$ be open and assume $q$ in \eqref{4.10b}
satisfies \eqref{4.8}, \eqref{4.9} on $\Omega$ for some set of constants
$\tilde c_s$, $1\leq s\leq r$. In addition suppose that
$z_j(t_r)$ are pairwise disjoint for $q|_\Omega$. Then $z_j$ is
analytic with respect to $t_r$ for $(z_j(t_r),t_r)$ in a sufficiently
small neighborhood of any point $\big(z_0,t_r^{(0)}\big)\in\Omega$.
Moreover, introducing the recursion relation
\begin{align}
a_{0,j}(t_r)&=0, \; 1\leq j\leq N, \quad \tilde c_0=1, \no \\
a_{s+1,j}(t_r)&=a_{s,j}(t_r)q_0-\tilde c_s-\sum_{p=1}^s
\tilde c_{s-p}\alpha_pq_0^p \no \\
& \quad -\sum_{\substack{k=1\\k\neq j}}^{N}
\big(a_{s,k}(t_r)+2a_{s,j}(t_r)\big)\cP(z_j(t_r)-z_k(t_r)), \lb{4.11a} \\
& \hspace*{3.95cm} 0\leq s\leq r, \; 1\leq j\leq N \no
\end{align}
with $\alpha_p=2^{-p}((2p-1)!!)/p!$, one obtains,
\begin{equation}
\f{dz_j}{d t_r}(t_r)=a_{r+1,j}(t_r), \quad 1\leq j\leq N. \lb{4.12}
\end{equation}
\end{lemma}
\begin{proof}
By results of \cite{SW85}, the $\tau$-function for
algebro-geometric (and hence for rational, simply periodic, and elliptic)
solutions of the KdV hierarchy is entire with respect to $(z,\ul t)$,
where $\ul t=(t_0,t_1,t_2,\dots)$ comprises all time variables in
Hirota's notation. Thus, choosing $t_s=\tilde c_{r-s}t_r$, $0\leq s\leq
r$ with $\tilde c_0=1$, and $t_s=0$, $s\geq r+1$, the resulting
$\tau$-function is entire in $(z,t_r)$ and hence $q$  is analytic in
$(z,t_r)$ as long as $(z,t_r)\in\Omega$, that is, as long as collisions
of the $z_j$ are avoided. Hence the implicit function theorem applied to
the $\tau$-function \eqref{nu}, \eqref{tau} yields analyticity of $z_j$
with respect to $t_r$ for $(z_j(t_r),t_r)$ in a sufficiently small
neighborhood of any point $\big(z_0,t_r^{(0)}\big)\in\Omega$. Using
\eqref{2.41} and \eqref{4.10a} one then computes
\begin{align}
q_{t_r}(z,t_r)&=2\sum_{j=1}^N \f{d z_j}{d t_r}(t_r) \cP'(z-z_j(t_r))
=2\tilde f_{r+1,z}(z,t_r) \no \\
&=2\sum_{j=1}^N a_{r+1,j}(t_r)\cP'(z-z_j(t_r)), \quad
1\leq j\leq N, \lb{4.13}
\end{align}
implying \eqref{4.12}.
\end{proof}

Next we illustrate Theorem \ref{t2.12} and Lemma \ref{l4.1} with the
simplest nontrivial rational example.

\begin{example} \lb{e4.2}
The genus $g=2$ ($N=3$) example with $r=1$ $($see, e.g., \cite{Ai78},
\cite{DG86}$)$. In this case one verifies
\begin{subequations} \lb{4.13a}
\begin{align}
q(z,t_1)&=q_0+2\partial_z^2 [\ln(z^3-3t_1)] \lb{4.13AA} \\
&=q_0-\frac{6z(z^3+6t_1)}{(z^3-3t_1)^{2}}  \lb{4.13aa}\\
&=q_0-2 \sum_{j=1}^{3} \frac{ 1}{[z- (3t_1)^{1/3}\varepsilon_j]^{2}},
\quad t_1\in\bbC, \lb{4.13ab}
\end{align}
\end{subequations}
and hence
\begin{equation}
z_j(t_1)=(3t_1)^{1/3}\varepsilon_j,\quad \varepsilon_j=\exp(2\pi ij/3),
\;\; j=1,2,3, \lb{4.13c}
\end{equation}
and
\begin{equation}
\tau(z;z_1(t_1),z_2(t_1),z_3(t_1))=\prod_{j=1}^3[z-z_j(t_1)]=z^3-3t_1,
\lb{4.13ca}
\end{equation}
explicitly illustrates the CM (respectively, DG) locus of poles in
\eqref{2.38A} (respectively, \eqref{3.36A}). Moreover, one computes for
the symmetric functions
\begin{equation}
\sigma_k= \sigma_k (z_1(t_1),z_2(t_1),z_3(t_1)), \quad 1\leq k \leq 3
\lb{4.13D}
\end{equation}
and
\begin{equation}
s_\ell=s_\ell (z_1(t_1),z_2(t_1),z_3(t_1)), \quad \ell\in\bbN, \lb{4.13E}
\end{equation}
of $(z_1(t_1),z_2(t_1),z_3(t_1)$ that
\begin{align}
&\sigma_1 =\sigma_2 =0, \; \sigma_3 =3t_1, \lb{3.53} \\
&s_{3k+1} =s_{3k+2} =0, \; s_{3k} =3(3t_1)^k, \; k\in\bbN_0. \lb{3.54}
\end{align}
In addition, one verifies the following facts,
\begin{align}
c_0&=1, \quad c_1= -\frac{5}{2}  q_0, \quad
c_2=  \frac{15}{8}  q_0^2, \lb{4.13d} \\
f_0(z,t_1) &= 1  ,\no\\
f_1(z,t_1)&=  \frac{1}{2} q(z,t_1) + c_1 =
-2 q_0 - \frac{ 3 z\, (z^3+6t_1)}{ (z^3-3t_1)^{2}} ,\no \\
&=-2 q_0 - \sum_{j=1}^{3} \frac{ 1}{[z- (3t_1)^{1/3}\varepsilon_j
]^{2}}, \lb{4.13e} \\
f_2(z,t_1)&=  \frac{1}{8} q_{zz}(z,t_1) +\frac{3}{8}
q^2(z,t_1) - \frac{5}{4} q(z,t_1)   q_0 + \frac{15}{8}  q_0^2 \no \\
&= q_0^2 + \frac{3\,q_0 z\,\left(z^3 + 6\,t_1 \right) }
     {{\left(z^3-3\,t_1 \right) }^2} + \frac{9\,z^2}{{\left(
z^3-3\,t_1 \right) }^2}, \no \\
&=q_0^2+ \sum_{j=1}^3
\frac{q_0+(3t_1)^{-2/3}\varepsilon_j}{[z-(3t_1)^{1/3}\varepsilon_j]^2},
\lb{4.13f}
\\ F_2(E,z,t_1)&= E^2 + f_1(z,t_1) E +  f_2(z,t_1) \no \\
&=E^2 - \bigg(2 q_0 +
\frac{ 3 z\, (z^3+6t_1)}{ (z^3-3t_1)^{2}} \bigg) E \no \\
& \quad +   q_0^2  +
\frac{3\,q_0\,\left( 6\,t_1\,z + z^4 \right) }
     {{\left(z^3-3\,t_1 \right) }^2} + \frac{9\,z^2}{{\left(z^3-3\,t_1
\right) }^2}, \lb{4.13g} \\
&=(E-\mu_1(z,t_1))(E-\mu_2(z,t_1)), \no \\
\mu_{1,2}(z,t_1) &= q_0 + \frac{3 z\,( 6\,t_1+z^3)  \pm
3\, \sqrt{3 z^5 (12\,t_1 - z^3)}}{2\, \left(z^3-3\,t_1 \right)^2}.
\lb{4.13h}
\end{align}
Finally, $q$ satisfies the following $2$nd stationary nonhomogeneous
KdV equation,
\begin{equation}
\sKdV_2(q)=\shKdV_2(q)-\frac{5}{2}q_0\shKdV_1(q)+
\frac{15}{8}q_0^2 \shKdV_0(q)= 0, \lb{4.13i}
\end{equation}
as well as the $1$st nonhomogeneous time-dependent KdV equation
\begin{equation}
\KdV_1(q)=q_{t_1}-\frac14 q_{zzz}-\frac32 q q_{z}+\frac32 q_0q_z=0.
\lb{4.13j}
\end{equation}
\end{example}

The following result explicitly connects the DG locus with rational,
simply periodic, and elliptic solutions of the KdV hierarchy and also
describes the local behavior of $z_j(t_r)$ as a function of $t_r$,
including the case of collisions.

\begin{theorem} \lb{t4.4}
Fix $N\in\bbN$ and suppose $\hatt \cL_N$ to be nonempty. \\
$(i)$ Consider KdV solutions $q=q(z,t_r)$ of
\eqref{4.6}, \eqref{4.7} $($for some set of constants $\tilde c_s$,
$1\leq s\leq r$$)$ of the type \eqref{4.10a}--\eqref{4.11}. Then
\begin{equation}
[z_1(t_r),\dots,z_N(t_r)]\subset\hatt\cL_N, \quad t_r\in\bbC. \lb{4.26}
\end{equation}
$(ii)$ Fix $N\in\bbN$ and $t_r^{(0)}\in\bbC$ and consider KdV solutions
$q=q(z,t_r)$ of \eqref{4.6}, \eqref{4.7} $($for some set of constants
$\tilde c_s$, $1\leq s\leq r$$)$ of the type \eqref{4.10a}--\eqref{4.11},
such that for $t_r=t_r^{(0)}$,
\begin{subequations} \lb{4.27}
\begin{align}
q(z,t_r^{(0)})
&=q_0 -\sum_{\ell=1}^{M^{(0)}} s_\ell^{(0)}\big(s_\ell^{(0)}+1\big)
\cP(z-\zeta_\ell^{(0)}). \lb{4.27b} \\
& \hspace*{-1.25cm}
s_\ell^{(0)}\in\bbN, \; 1\leq \ell \leq M^{(0)},
\quad \sum_{\ell=1}^{M^{(0)}} s_\ell^{(0)}\big(s_\ell^{(0)}+1\big)=2N.
\lb{4.27c}
\end{align}
\end{subequations}
Then, in a sufficiently small neighborhood of
$\big(\zeta_\ell^{(0)},t_r^{(0)}\big)\in\bbC^2$, $1\leq \ell \leq
M^{(0)}$, there exist precisely $s_\ell^{(0)}\big(s_\ell^{(0)}+1\big)/2$
points
$z_{j_k}(t_r)$ $($not necessarily distinct\,$)$ such that
\begin{equation}
q(z,t_{r})\underset{\substack{z\to\zeta_\ell^{(0)}\\ t_r\to
t_r^{(0)}}}{=} -2\sum_{k=1}^{s_\ell^{(0)}(s_\ell^{(0)}+1)/2}
\cP(z-z_{j_k}(t_{r})) +\Oh(1) \lb{4.28}
\end{equation}
and each $z_{j_k}(t_{r})$ has a Puiseux expansion of the type
\begin{equation}
z_{j_k}(t_{r})\underset{t_r\to t_r^{(0)}}{=}\zeta_\ell^{(0)}+
\sum_{p=1}^\infty C_{j_k,\ell,p} \big(t_r-t_r^{(0)}\big)^{p/q_k}
\lb{4.29}
\end{equation}
for some constants $C_{j_k,\ell,p}\in\bbC$, $p\in\bbN$, and appropriate
$q_k\in\big\{1,\dots,s_\ell^{(0)}\big(s_\ell^{(0)}+1\big)/2\big\}$,
$1\leq k\leq s_\ell^{(0)}\big(s_\ell^{(0)}+1\big)/2$,
$1\leq \ell\leq M^{(0)}$. In particular, all elementary symmetric
functions of the $z_{j_k}$,
$1\leq k\leq s_\ell^{(0)}\big(s_\ell^{(0)}+1\big)/2$ $($cf.\
\eqref{4.20},
\eqref{4.21}$)$ are analytic in a neighborhood of $t_r^{(0)}$. Finally, in
the special rational case, the $z_{j_k}$,
$1\leq k\leq s_\ell^{(0)}\big(s_\ell^{(0)}+1\big)/2$, are algebraic
functions $($on an appropriate compact Riemann surface$)$. \\
$(iii)$ $q$ defined by
\begin{equation}
q(z)=q_0 -2\sum_{j=1}^{N} \cP(z-z_j) \lb{4.30}
\end{equation}
satisfies a particular stationary KdV equation $($and hence is an
algebro-geometric KdV potential\,$)$ if and only if
\begin{equation}
[z_1,\dots,z_N]\subset\hatt\cL_N. \lb{4.31}
\end{equation}
\end{theorem}
\begin{proof}
$(i)$ The DG locus $\hatt \cL_N$ as defined in \eqref{4.25} is a
consequence of Theorem \ref{t3.5}, the isospectral property of KdV
flows (cf.\ \eqref{4.7}, \eqref{4.9}), and the fact that any potential
$q=q(z)$ in \eqref{3.35}, \eqref{3.36} can be chosen as the initial value
$q^{(0)}$ in \eqref{4.1}, \eqref{4.2}. Put differently, the poles
$\{z_j(t_r)\}_{1\leq j\leq N}$ of any rational, simply periodic, and
elliptic solution of \eqref{4.6}, \eqref{4.7}, of the form \eqref{4.10b}
for fixed $N\in\bbN$, trace out  curves on the DG locus \eqref{4.25} as
$t_r$ varies in $\bbC$ as described in \eqref{4.26}. \\
$(ii)$ As mentioned in the proof of Lemma \ref{l4.1}, the
$\tau$-function associated with $q(z,t_r)$,
\begin{equation}
\tilde\tau(z,t_r)=
\tau(z;z_1(t_r),\dots,z_N(t_r))=\prod_{j=1}^N \nu(z-z_j(t_r)), \lb{4.32}
\end{equation}
where
\begin{equation}
\nu(z)=\begin{cases} z & \text{in the rational case,} \\
(\omega/\pi) \sin(\pi z/\omega)\exp[\pi^2z^2/(6\omega)^2]
& \text{in the simply periodic case,} \\
\sigma(z) & \text{in the elliptic case,} \end{cases} \lb{4.33}
\end{equation}
is entire in $(z,t_r)$. By \eqref{4.27},
\begin{equation}
\tilde\tau\big(z,t_r^{(0)}\big)=
\big(z-\zeta_\ell^{(0)}\big)^{s_\ell^{(0)}(s_\ell^{(0)}+1)/2}
\tilde{\tilde \tau}\big(z,t_r^{(0)}\big) \lb{4.34}
\end{equation}
and
\begin{equation}
\tilde {\tilde \tau}\big(\cdot,t_r^{(0)}\big) \text{ is analytic and
nonvanishing in a neighborhood of
$\zeta_\ell^{(0)}$.} \lb{4.35}
\end{equation}
Applying the Weierstrass preparation theorem (cf., e.g.,
\cite[Sect.\ III.14]{SZ71}), one obtains
\begin{subequations} \lb{4.36}
\begin{align}
\tilde\tau(z,t_r)&= \sum_{k=0}^{s_\ell^{(0)}(s_\ell^{(0)}+1)/2}
A_k(t_r)\big(z-\zeta_\ell^{(0)}\big)^{s_\ell^{(0)}(s_\ell^{(0)}+1)/2-k}
\, \tilde{\tilde \tau}(z,t_r)  \lb{4.36a} \\
&=\Bigg(\prod_{k=1}^{s_\ell^{(0)}(s_\ell^{(0)}+1)/2} [z-z_{j_k}(t_r)]
\Bigg)\tilde{\tilde \tau}(z,t_r), \lb{4.36b}
\end{align}
\end{subequations}
where
\begin{equation}
A_0(t_r)=1, \quad A_k\big(t_r^{(0)}\big)=0, \;\,
1\leq k\leq s_\ell^{(0)}\big(s_\ell^{(0)}+1\big)/2 \lb{4.37}
\end{equation}
and
\begin{equation}
\tilde {\tilde \tau} \text{ is analytic and nonvanishing in a
neighborhood of $\big(\zeta_\ell^{(0)},t_r^{(0)}\big)$.} \lb{4.37a}
\end{equation}
In particular, the elementary symmetric functions
\begin{equation}
A_k(t_r)=(-1)^k\sigma_k\Big(z_{j_1}(t_r),\dots,
z_{j_{s_\ell^{(0)}(s_\ell^{(0)}+1)/2}}(t_r)\Big) \lb{4.38}
\end{equation}
of the roots $z_{j_k}$ are all analytic in a neighborhood of $t_r^{(0)}$.
By Lemma \ref{l4.3}, also the corresponding symmetric functions $s_j$,
$j\in\bbN$, of the $z_{j_k}$ are analytic at $t_r^{(0)}$. The roots
$z_{j_k}$ of
$\prod_{k=1}^{s_\ell^{(0)}(s_\ell^{(0)}+1)/2} (z-z_{j_k}(t_r))$ then
permit a Puiseux expansion of the type \eqref{4.29} (see, e.g.,
\cite[p.\ 303--304]{Ma85}). In the special rational case the
corresponding $\tau$-function is of the type (cf.\ \cite{AM78})
\begin{subequations}
\begin{align}
\tilde\tau(z,t_r)&= \sum_{k=0}^{s_\ell^{(0)}(s_\ell^{(0)}+1)/2}
A_k(t_r)\big(z-\zeta_\ell^{(0)}\big)^{s_\ell^{(0)}(s_\ell^{(0)}+1)/2-k}
\lb{4.39a} \\
&= \prod_{k=1}^{s_\ell^{(0)}(s_\ell^{(0)}+1)/2} (z-z_{j_k}(t_r))
\lb{4.39b}
\end{align}
\end{subequations}
with $A_k(t_r)$ polynomials in $t_r$. Hence $z_{j_k}$ are
(globally) algebraic functions of $t_r$. \\
Part (iii) is just a reformulation of (parts of) Theorem \ref{t3.5}.
\end{proof}

\begin{remark} \lb{r4.5}
$(i)$ In the elementary case of Example \ref{e4.2}, where $g=2$
($N=3$) and $r=1$, Theorem \ref{t4.4} is explicitly illustrated by the
results \eqref{3.53} and \eqref{3.54}. \\
$(ii)$ In the case of the classical elliptic $N$-particle
Calogero--Moser system on the circle (cf.\ \cite[Ch.\ 2]{Ca01},
\cite{Ta99}), a model that differs from the one describing the motion of
poles of KdV solutions considered in this paper, it was shown in
\cite{GP99} that every symmetric elliptic function of the $N$ coordinates
is meromorphic with respect to time.
\end{remark}

\noindent {\bf ACKNOWLEDGEMENTS.} We are indebted to Gilbert Weinstein
and Finn Faye Knudsen for helpful discussions.


\appendix

\section{The KdV hierarchy} \lb{sA}

\renewcommand{\theequation}{A.\arabic{equation}}
\setcounter{theorem}{0}
\setcounter{equation}{0}

In this section we review basic facts on the KdV hierarchy. Since this
material is well-known, we confine ourselves to a brief account (for a
detailed treatment, see, e.g., \cite[Ch.\ 1]{GH03}). Assuming for
simplicity $q(\cdot,t)$ to be meromorphic in $\bbC$ for all $t\in\bbC$
and $q(x,\cdot)$ to be $C^1$ with respect to $t\in\bbC$ (except
possibly for a discrete set) for all $x\in\bbC$, consider the
recursion relation
\begin{align}
{f}_0(z,t)&=1, \no \\
{f}_{j+1,z}(z,t)&=\tfrac14
{f}_{j,zzz}(z,t) + q(z,t){f}_{j,z}(z,t)+\tfrac12 q_z(z,t){f}_{j}(z,t),
\quad j\in\bbN_0 \lb{A.1}
\end{align}
(with $\bbN_0=\bbN\cup\{0\}$) and the
associated differential expressions (Lax pair)
\begin{align}
L_2(t) &=\dfrac{d^2}{dz^2} + q(z,t), \lb{A.2} \\
P_{2n+1}(t)&=\sum^{n}_{j=0} \left[-\dfrac{1}{2}
{f}_{j,z}(z,t)+{f}_j(z,t) \dfrac{d}{dz}\right] L_2^{n-j}(t),\quad
g\in\bbN_0.
\lb{A.3}
\end{align}
One can show that
\begin{equation}\lb{A.4}
\left[ P_{2n+1}(t), L_2(t) \right]=2{f}_{n+1,z}(\cdot,t)
\end{equation}
([$\cdot,\cdot$] the commutator symbol) and explicitly computes
{}from \eqref{A.1},
\begin{equation}\lb{A.5}
{f}_0=1, \; {f}_1=\tfrac12 q+c_1,\;
{f}_2=\tfrac18q_{zz} + \tfrac38 q^2 +c_1 \tfrac12 q+c_2, \text{ etc.},
\end{equation}
where $c_j\in\bbC$, $j\in\bbN$, are integration constants. For subsequent
purposes we also introduce the corresponding homogeneous coefficients
$\hat f_j$ defined by the vanishing of all integration constants
$c_\ell=0$, $1\leq \ell\leq j$,
\begin{equation}
\hat f_0=f_0=1, \quad
\hat f_j = f_j\big|_{c_\ell=0, \, \ell=1,\dots,j}, \quad j\in\bbN.
\lb{A.5A}
\end{equation}
If one assigns to $q^{(\ell)}=d^{\ell}q/dz^{\ell}$ the degree
$\deg(q^{(\ell)}) = \ell+2, \; \ell\in\bbN_0$, then the homogeneous
differential polynomial $\hat f_j$ with respect to $q$ turns
out to have degree $2j$, that is,
\begin{equation}\lb{A.5B}
\deg (\hat f_j)=2j, \quad j\in\bbN_0.
\end{equation}
Introducing
\begin{equation}
c_0=1, \lb{A.5C}
\end{equation}
one verifies,
\begin{equation}
f_0=\hat f_0=1, \quad
f_j=\sum_{\ell=0}^j c_{j-\ell} \hat f_\ell, \quad j\in\bbN. \lb{A.5D}
\end{equation}
The KdV hierarchy is then defined as the sequence of evolution equations
\begin{equation}
\KdV_n(q)=L_{2,t} -[P_{2n+1}, L_2] =q_t -2 f_{n+1, z}=0, \quad
n\in\bbN_0. \lb{A.6}
\end{equation}
Explicitly one obtains,
\begin{align}
\KdV_0 (q)&=q_t-q_z=0, \no \\
\KdV_1(q)&=q_t -\tfrac14
q_{zzz}-\tfrac32 qq_z+c_1(-q_z)=0, \text{ etc.} \lb{A.7}
\end{align}
with $\KdV_1(.)|_{c_1=0}$ the usual KdV functional. Moreover, one
verifies,
\begin{equation}
\KdV_n(q)=q_t-2\sum_{\ell=0}^{n+1} c_{n-\ell} \hat f_{\ell,z}=0, \quad
n\in\bbN. \lb{A.8A}
\end{equation}
Next, introducing the polynomial ${F}_n(\cdot,z,t)$ of degree $n$,
\begin{equation}
{F}_n(E,z,t) = \sum^{n}_{\ell=0} {f}_{n-\ell}(z,t) E^\ell
=\prod_{p=1}^n [E-\mu_p(z,t)], \lb{A.9}
\end{equation}
\eqref{A.6} becomes
\begin{equation}
q_t =\dfrac12 F_{n, zzz} +2(q-E) F_{n,z}+q_z F_n. \lb{A.10}
\end{equation}
In the following we turn to the stationary case characterized by
$q_t=0$, or equivalently, by
\begin{equation}
[P_{2n+1},L_2]=0. \lb{A.10a}
\end{equation}
The corresponding stationary KdV hierarchy
is then defined as the sequence of equations
\begin{equation}
\sKdV_n(q)=-[P_{2n+1},L_2]=-2f_{n+1,z} =0, \quad n\in\bbN_0. \lb{A.10b}
\end{equation}
Explicitly, this yields
\begin{align}
\sKdV_0(q)&=-q_z =0, \no \\
\sKdV_1(q)&=-\tfrac14 q_{zzz}-\tfrac32 qq_z + c_1 (-q_z)=0, \text{ etc. }
\lb{A.10c}
\end{align}
Similarly, the corresponding homogeneous stationary KdV equations are
then defined by
\begin{equation}
\shKdV_n(q)=-2\hat f_{n+1,z}=0, \quad n\in\bbN_0 \lb{A.10d}
\end{equation}
and one thus obtains from \eqref{A.8A},
\begin{equation}
\sKdV_n(q)=\sum_{\ell=0}^n c_{n-\ell} \shKdV_\ell(q). \lb{A.10D}
\end{equation}
Since $f_0(z)=1$,
\begin{equation}
-\f12 F_{n,zz}(E,z) F_n(E,z)+\f14
F_{n,z}(E,z)^2+(E-q(z)) F_n(E,z)^2 ={R}_{2n+1}(E,z) \lb{A.12}
\end{equation}
is a monic polynomial in $E$ of degree $2n+1$.
However, equations \eqref{A.1} and \eqref{A.10b} imply that
\begin{equation}\lb{A.11}
\f12 {F}_{n,zzz}-2(E-q){F}_{n,z}+q_z {F}_n=0
\end{equation}
and this shows that ${R}_{2n+1}(E,z)$ is in fact independent of $z$.
Hence it can be written as
\begin{equation}\lb{A.13}
{R}_{2n+1}(E)=\prod^{2n}_{m=0} (E-E_m) \text{ for some }
\{E_m\}_{0\leq m\leq 2n} \subset \bbC
\end{equation}
and \eqref{A.12} becomes
\begin{align}
&-\f12 F_{n,zz}(E,z) F_n(E,z)+\f14
F_{n,z}(E,z)^2+(E-q(z)) F_n(E,z)^2 \no \\
&\quad ={R}_{2n+1}(E)=\prod^{2n}_{m=0} (E-E_m). \lb{A.13a}
\end{align}

By \eqref{A.10a} the stationary KdV equation \eqref{A.10b} is
equivalent to the commutativity of $L_2$ and ${P}_{2n+1}$ and therefore,
if $L_2 \psi=E\psi$ one infers $P^2_{2n+1}\psi =R_{2n+1}(E)\psi$. Thus
$[{P}_{2n+1}, L_2]=0$ implies $P^2_{2n+1} = R_{2n+1}(L_2)$
by the Burchnall--Chaundy theorem. This illustrates the intimate
connection between the stationary KdV equation ${f}_{n+1,z}=0$ in
\eqref{A.10b} and the compact (possibly singular) hyperelliptic curve
$\ol{\cK_n}$ of (arithmetic) genus $n$ obtained upon one-point
compactification of the curve
\begin{equation}\lb{A.16}
   \cK_n\colon y^2={R}_{2n+1}(E)=\prod^{2n}_{m=0}(E-{E}_m)
\end{equation}
by joining the point at infinity, denoted by $P_\infty$. Points
$P\in\cK_n$ are denoted by $P=(E,y)$.

The above formalism leads to the following standard definition.

\begin{definition} \label{dA.1}
Any solution $q$ of one of the stationary KdV
equations \eqref{A.10b} is called an {\it algebro-geometric KdV
potential}.
\end{definition}

\noindent For brevity of notation we will occasionally call such $q$
simply {\it KdV potentials}.

Next, denoting ${\ul E}=(E_0,\dots,E_{2n})$, we consider
\begin{equation}
\bigg(\prod_{m=0}^{2n} \bigg(1-\frac{E_m}{z}\bigg)
\bigg)^{1/2}=\sum_{k=0}^{\infty}c_k(\ul E)z^{-k},  \lb{A.16a}
\end{equation}
where
\begin{align}
c_0(\ul E)&=1,\no \\
c_k(\ul E)&=\!\!\!\!\!\sum_{\substack{j_0,\dots,j_{2n}=0\\
   j_0+\cdots+j_{2n}=k}}^{k}\!\!\!\!\!
\f{(2j_0-3)!!\cdots(2j_{2n}-3)!!}
{2^k j_0!\cdots j_{2n}!}E_0^{j_0}\cdots E_{2n}^{j_{2n}},
\,\, k\in\bbN, \label{A.16b}
\end{align}
and hence the first few coefficients explicitly read
\begin{align}
   c_0(\ul E)&=1, \quad
c_1(\ul E)=-\f12\sum_{m=0}^{2n} E_m, \no \\
c_2(\ul E)&=\f14\!\sum_{\substack{m_1,m_2=0\\ m_1< m_2}}^{2n}\!\!\!\!
E_{m_1} E_{m_2}-\f18 \sum_{m=0}^{2n} E_m^2,  \quad \text{etc.} \lb{A.16c}
\end{align}
Assuming that $q$ satisfies the $g$th stationary (nonhomogeneous) KdV
equation \eqref{A.6}, the corresponding integration constants
$c_\ell$ in \eqref{A.5} become symmetric functions of
the branch points $E_0,\dots,E_{2n}$ of the underlying curve \eqref{A.16}
and one verifies (cf., e.g., \cite[Sect.\ 1.2]{GH03})
\begin{equation}
c_\ell=c_\ell(\ul E), \quad 1\leq \ell\leq n. \lb{A.17}
\end{equation}

Finally, we return to the general time-dependent setup and briefly recall
the  algebro-geometric KdV initial value problem, where by definition $q$
satisfies the $r$th time-dependent KdV equation
\begin{subequations} \lb{A.18}
\begin{align}
\wti\KdV_r(q)&=q_{t_r}-2\tilde f_{r+1,z}=0, \quad (z,t_r)\in\bbC^2,
\lb{A.18a} \\
q\big|_{t_r=t_r^{(0)}}&=q^{(0)}  \lb{A.18b}
\end{align}
\end{subequations}
with initial value $q^{(0)}$ satisfying the $n$th stationary KdV equation
\begin{equation}
\sKdV_n(q^{(0)})=-2f_{n+1,z}=0 \lb{A.19}
\end{equation}
for fixed $n,r\in\bbN_0$ and some $t_{r}^{(0)}\in\bbC$. Here we replaced
$t$ by $t_r$ to emphasize the $r$th KdV flow. Moreover, since the
integration constants in \eqref{A.18a} and \eqref{A.19} are independent
of each other, we denote the ones in $f_k$ by $c_\ell$, $1\leq \ell\leq
k$ as before and the ones in the right-hand side of \eqref{A.18a} by
$\tilde c_s$, $1\leq s
\leq r$. Similarly, $\tilde f_j$, $\wti F_r$, $\wti P_{2r+1}$,
$\wti\KdV_r$ are constructed as $f_j$, $F_r$, $P_{2r+1}$, $\KdV_r$ in
\eqref{A.1}, \eqref{A.3}, \eqref{A.6}, \eqref{A.9}, replacing $c_\ell$
by $\tilde c_s$, etc. The isospectral property of KdV flows then permits
one to replace \eqref{A.18} and
\eqref{A.19} by the following pair of equations
\begin{align}
&q_{t_r}=\f12\wti F_{r,zzz}+2(q-E)\wti F_{r,z}+ q_z\wti F_r, \lb{A.20a}
\\
&-\f12 F_{n,zz}F_n+\f14 F_{n,z}^2+(E-q)F_n^2 =R_{2n+1}, \lb{A.20b}
\end{align}
or in terms of Lax differential expressions, by
\begin{subequations} \lb{A.21}
\begin{align}
L_{2,t_r}(t_r)-[\wti P_{2r+1}(t_r),L_2(t_r)]&=0, \lb{A21a} \\
[P_{2n+1}(t_r),L_2(t_r)]&=0. \lb{A.21b}
\end{align}
\end{subequations}
Because of \eqref{A.21}, the common eigenfunction $\psi(P)$ of $L_2$ and
$P_{2n+1}$, the Baker--Akhiezer function, will satisfy
\begin{align}
L_2(t_r)\psi(P,z,t_r)&=E\psi(P,z,t_r), \lb{A.22} \\
P_{2n+1}(t_r)\psi(P,z,zt_r)&=y\psi(P,z,t_r), \lb{A.23} \\
\psi_{t_r}(P,z,t_r)&=\wti P_{2r+1}(t_r)\psi(P,z,t_r) \lb{A.24} \\
&=\wti
F_r(E,z,t_r)\psi_z(P,z,t_r)-\f12 \wti F_{r,z}(E,z,t_r)\psi(P,z,t_r),
   \lb{A.25} \\
& \hspace*{6cm} \quad P=(E,y). \no
\end{align}

\section{A Few Basic Results on Elliptic Functions} \label{sB}

\renewcommand{\theequation}{B.\arabic{equation}}
\setcounter{theorem}{0}
\setcounter{equation}{0}

For convenience of the reader we recall some theorems representing an
arbitrary elliptic function in terms of $\sigma$- and $\zeta$-functions
which are used in  this text. For general references see,
for instance, Akhiezer \cite{Ak90}, Chandrasekharan \cite{Ch85},
Markushevich \cite{Ma85}, and Whittaker and Watson \cite{WW86} (for
connoisseurs we recommend, in particular, Krause's two volume treatise
\cite{Kr95}, \cite{Kr97}).

A function $f:\bbC\to\bbC\cup\{\infty\}$ with two periods
$a,b\in\bbC\backslash\{0\}$, the ratio of which is not real, is called
doubly  periodic. If all its periods are of the form $m_1 a+m_2 b$,
where $m_1,m_2\in\bbZ$, then $a$ and $b$ are called fundamental periods
of $f$. A doubly periodic meromorphic function is called {\it elliptic}.
It is customary to denote the fundamental periods  of an elliptic
function by $2\omega_1$ and $2\omega_3$ with $\Im(\omega_3/\omega_1)>0$.
We also introduce $\omega_2=\omega_1+\omega_3$ and $\omega_4=0$. The
numbers $\omega_1,\dots,\omega_4$ are called half-periods.  The
fundamental period parallelogram $\Delta$ is the half-open
region consisting of the line segments $[0,2\omega_1)$, $[0,2\omega_3)$
and the interior of the parallelogram with vertices $0$, $2\omega_1$,
$2\omega_2$, and $2\omega_3$.

The function
\begin{equation}
\wp(z;\omega_1,\omega_3)=\frac1{z^2}+
\sum_{\substack{(m,n)\in\bbZ^2 \\ (m,n)\neq(0,0)}}
\left(\frac1{(z-2m\omega_1-2n\omega_3)^2} -
   \frac1{(2m\omega_1+2n\omega_3)^2}\right), \lb{B.1}
\end{equation}
or $\wp(z)$ for short, was introduced by Weierstrass.
It is an even
elliptic function of order 2 with fundamental periods
$2\omega_1$ and
$2\omega_3$. Its derivative $\wp'$ is an odd elliptic
function of order
3 with fundamental periods $2\omega_1$ and $2\omega_3$.
Every elliptic
function may be written as $R_1(\wp(z))+R_2(\wp(z))\wp'(z)$
where $R_1$
and $R_2$ are rational functions of $\wp$.

The numbers
\begin{equation}
   g_2 = 60\sum_{\substack{(m,n)\in\bbZ^2\\(m,n)\neq(0,0)}}
   \frac1{(2m\omega_1+2n\omega_3)^4}, \quad
   g_3 = 140\sum_{\substack{(m,n)\in\bbZ^2\\(m,n)\neq(0,0)}}
   \frac1{(2m\omega_1+2n\omega_3)^6}    \lb{B.2}
\end{equation}
are called the invariants of $\wp$. Since the coefficients of the
Laurent expansions of $\wp(z)$ and $\wp'(z)$ at $z=0$ are
polynomials of $g_2$ and $g_3$ with rational coefficients, the function
$\wp(z;\omega_1,\omega_3)$ is also uniquely characterized by its
invariants $g_2$ and $g_3$. One also frequently uses the notation
$\wp(z|g_2,g_3)$.

The function $\wp(z)$ satisfies the first-order
differential equation
\begin{equation} \label{B.3}
\wp'(z)^2 = 4\wp(z)^3-g_2\wp(z)-g_3
\end{equation}
and hence the equations
\begin{equation}
\wp''(z) = 6\wp(z)^2-g_2/2 \text { and } \wp'''(z)=
12\wp'(z)\wp(z). \lb{B.4}
\end{equation}
Thus, $-2\wp$ is a stationary solution of the first
KdV equation, $\sKdV_1(q)=0$ in \eqref{A.7} with $c_1=0$.

The function $\wp'$, being of order $3$, has three zeros
in $\Delta$.
Since $\wp'$ is odd and elliptic it is obvious that these
zeros are the
half-periods $\omega_1,\omega_2=\omega_1+\omega_3$ and
$\omega_3$. Let
$e_j=\wp(\omega_j)$, $j=1,2,3$. Then \eqref{B.3} implies that
$4e^3_j-g_2e_j-g_3=0$ for $j=1,2,3$. Therefore
\begin{align}
   0 &= e_1+e_2+e_3, \lb{B.5} \\
   g_2 &=-4(e_1e_2+e_1e_3+e_2e_3)=2(e^2_1+e^2_2+e^2_3), \lb{B.6} \\
   g_3 &= 4e_1e_2e_3 =\frac43 (e^3_1+e^3_2+e^3_3).  \lb{B.7}
\end{align}

Weierstrass also introduced two other functions denoted by
$\zeta$ and
$\sigma$. The Weierstrass $\zeta$-function is defined by
\begin{equation}
\frac{d}{dz}\zeta(z) = -\wp(z), \quad
   \lim_{z\to0} \bigg(\zeta(z)-\frac1z\bigg) = 0. \lb{B.8}
\end{equation}
It is a meromorphic function with simple poles at
$2m\omega_1+2n\omega_3, m,n\in\bbZ$ having residues $1$.
It is not
periodic but quasi-periodic in the sense that
\begin{equation}
\zeta(z+2\omega_j)=\zeta(z)+2\eta_j, \quad 1\leq j\leq 4, \lb{B.9}
\end{equation}
where $\eta_j=\zeta(\omega_j)$ for $j=1,2,3$ and $\eta_4=0$.

The Weierstrass $\sigma$-function is defined by
\begin{equation}
\frac{\sigma'(z)}{\sigma(z)}=\zeta(z), \quad
   \lim_{z\to0} \frac{\sigma(z)}{z}=1. \lb{B.10}
\end{equation}
$\sigma$ is an entire function with simple zeros at the
points
$2m\omega_1+2n\omega_3, m,n\in\bbZ$. Under translation
by a period
$\sigma$ behaves according to
\begin{equation}
\sigma(z+2\omega_j)=-\sigma(z)e^{2\eta_j(z+\omega_j)},
\quad j=1,2,3. \lb{B.11}
\end{equation}

\begin{theorem} \label{tb0} $($\cite{GW98a}$)$
Given numbers $\alpha_1,\dots,\alpha_m$ and $\beta_1,\dots,\beta_m$
such that $\beta_k\neq\beta_\ell\,({\rm mod}\,\Delta)$ for
$k\neq\ell$, the
following identity holds
\begin{equation}
\prod_{j=1}^m \f{\sigma(z-\alpha_j)}{\sigma(z-\beta_j)}
   =\sum_{j=1}^m \f{\prod_{k=1}^m \sigma(\beta_j-\alpha_k)}
   {\prod_{\ell=1,\ell\neq j}^m \sigma(\beta_j-\beta_\ell)}
   \f{\sigma(z-\beta_j+\beta-\alpha)}
   {\sigma(z-\beta_j)\sigma(\beta-\alpha)},  \lb{B.12}
\end{equation}
where
\begin{equation}
\alpha=\sum_{j=1}^m \alpha_j \text{ and }\beta=
\sum_{j=1}^m \beta_j \lb{B.13}
\end{equation}
and $\sigma$ is constructed {}from the fundamental periods
$2\omega_1$ and $2\omega_3$.
\end{theorem}

\begin{theorem} \label{tb1} $($\cite[p.\ 182, Theorem 5.12]{Ma85}$)$
Given an elliptic function $f$ of order $n$ with fundamental periods
$2\,\omega_1$ and $2\,\omega_3$, let $a_1,\dots,a_n$ and
$b_1,\dots,b_n$ be the
zeros and poles of $f$ in the fundamental period parallelogram $\Delta$
repeated according to their
multiplicities. Then
\begin{equation}
f(z) = C \frac{\sigma(z-a_1)\cdots\sigma(z-a_n)}
      {\sigma(z-b_1)\cdots\sigma(z-b_{n-1}) \sigma(z-b_n')}, \lb{B.14}
\end{equation}
where $C\in\bbC$ is a suitable constant, $\sigma$ is constructed {}from
the fundamental periods $2\,\omega_1$ and $2\,\omega_3$, and where
\begin{equation}
b_n'-b_n=(a_1+\cdots+a_n)-(b_1+\cdots+b_n) \lb{B.15}
\end{equation}
is a period of $f$.
Conversely, every such function is an elliptic function.
\end{theorem}

\begin{theorem} \label{tb2} $($\,\cite[p.\ 182, Theorem 5.13]{Ma85}$)$
Given an elliptic function $f$ with fundamental periods $2\, \omega_1$
and $2\, \omega_3$, let $b_1,\dots,b_r$ be the distinct poles of  $f$
in $\Delta$. Suppose the principal part of the Laurent expansion near
$b_k$ is given by
\begin{equation}
\sum^{\beta_k}_{j=1} \frac{A_{j,k}}{(z-b_k)^j}, \quad 1\leq k\leq r.
\lb{B.16}
\end{equation}
Then
\begin{align}
f(z) = A_0+\sum^r_{k=1} \sum^{\beta_k}_{j=1}(-1)^{j-1}
\frac{A_{j,k}}{(j-1)!} \zeta^{(j-1)}(z-b_k), \lb{B.17}
\end{align}
where $A_0\in\bbC$ is a suitable constant and $\zeta$ is constructed
{}from the fundamental periods $2\,\omega_1$ and $2\,\omega_3$.
Conversely, every such function is an elliptic function if
$\sum^{r}_{k=1} A_{1,k}=0$.
\end{theorem}

One notes that this theorem resembles the partial fraction expansions
for rational functions.

Finally, we turn to elliptic functions of the second kind, the central
object in our analysis. A meromorphic function
$\psi:\mathbb{C}\to\mathbb{C}\cup\{\infty\}$ for which there exist
two complex constants $\omega_1$ and $\omega_3$ with non-real ratio and
two complex constants $\rho_1$ and $\rho_3$ such that
\begin{equation}
\psi(z+2\, \omega_j)=\rho_j \psi(z), \quad j=1,3,  \lb{B.18}
\end{equation}
is called elliptic of the second kind. We call $2\,\omega_1$ and
$2\,\omega_3$ the quasi-periods of $\psi$. Together with $2\,\omega_1$ and
$2\,\omega_3$, $2\,m_1\omega_1+2\,m_3\omega_3$ are also quasi-periods of
$\psi$ if $m_1,m_3\in\bbZ$. If every quasi-period of $\psi$
can be written as an integer linear combination of $2\,\omega_1$ and
$2\,\omega_3$, then these are called fundamental quasi-periods.

\begin{theorem} \label{tb3}
A function $\psi$ which is elliptic of the second kind and
has fundamental quasi-periods $2\,\omega_1$ and $2\,\omega_3$ can
always be put into the form
\begin{equation}
\psi(z) = C \exp(\lambda z) \frac{\sigma(z-a_1)
\cdots\sigma(z-a_n)}
      {\sigma(z-b_1)\cdots\sigma(z-b_{n})} \lb{B.19}
\end{equation}
for suitable constants $C$, $\lambda$, $a_1,\dots,a_n$ and
$b_1,\dots,b_n$. Here $\sigma$ is constructed {}from the fundamental
periods $2\,\omega_1$ and $2\,\omega_3$. Conversely, every such
function is elliptic of the second kind.
\end{theorem}

\section{Symmetric products} \label{sC}

\renewcommand{\theequation}{C.\arabic{equation}}
\setcounter{theorem}{0}
\setcounter{equation}{0}

Let $X$ be a Riemann surface. In addition to the cartesian product
$X^N=X\times\cdots\times X$ ($N$ factors), $N\in\bbN$, we also
introduce the $N$th symmetric product of $X$ defined as the quotient
space
\begin{equation}
X^N/S_N. \lb{C.1}
\end{equation}
Here $S_N$ denotes the symmetric group on $N$ letters acting as the group
of permutations of the factors in the cartesian product $X^N$, that is,
\begin{equation}
\pi(x_1,\dots,x_n)=(x_{\pi(1)},\dots, x_{\pi(N)}), \quad \pi\in S_N.
\lb{C1.a}
\end{equation}
Thus, the points in $X^N/S_N$ can be considered as $N$-tuples of points
of $X$ without regard to their order. $X^N/S_N$ inherits the topology
from $X^N$ (the quotient topology) and the canonical projection (quotient
map)
\begin{align}
\nu\colon \begin{cases} X^N \to X^N/S_N \\
(x_1,\dots,x_N)\mapsto [x_1,\dots,x_N]=
\{\pi(x_1,\dots,x_N)\in X^N \,|\,\pi\in S_N\} \end{cases} \lb{C.2}
\end{align}
defines a complex structure on $X^N/S_N$ as follows. Consider a point
$[p_1,\dots,p_N]\in X^N/S_N$, let $x_j$ be a local coordinate in an open
neighborhood $U_j$ of $p_j\in X$, assuming $U_j\cap U_k=\emptyset$ if
$p_j\neq p_k$ and $x_j=x_k$ in $U_j=U_k$ for $p_j=p_k$. Denote by
$\sigma_1,\dots,\sigma_N$ the elementary symmetric functions of
$x_1,\dots,x_N$, then the map
\begin{align}
\begin{split}
&\nu(U_1\times \cdots \times U_N) \to \bbC^N  \\
&[q_1,\dots,q_N] \mapsto (\sigma_1(x_1(q_1),\dots,x_N(q_N)),\dots,
\sigma_N(x_1(q_1),\dots,x_N(q_N))) \lb{C.3}
\end{split}
\end{align}
provides a coordinate chart on $\nu(U_1\times \cdots \times U_N)$.
In this manner, $X^N/S_N$ (like $X^N$) becomes an $N$-dimensional
complex manifold with $X^N$ an $N!$-sheeted branched analytic covering of
$X^N/S_N$.

Away from the branch locus the map $\nu$ is a covering map and one can
take
\begin{equation}
(x_1(q_1),\dots,x_N(q_N)) \lb{C.4}
\end{equation}
as coordinates on $X^N/S_N$ (here the points $p_j$, corresponding to
the charts $(U_j,x_j)$, are mutually distinct). At the other extreme,
where $p_1=p_2=\dots=p_N$, local coordinates are given by
$(\sigma_1(x_1(q_1),\dots,x_N(q_N)),\dots, \sigma_N(x_1(q_1),\dots,
x_N(q_N)))$, that is, by
\begin{equation}
\Bigg(\sum_{j=1}^N x_j(q_j),\dots,\prod_{j=1}^N x_j(q_j)\Bigg). \lb{C.5}
\end{equation}

Next, assume the topological space $(X^N,\tau)$ is generated by the
metric $d$ on $X^N$. We then write $\tau=(d)$ and hence
$(X^N,\tau)=(X^N,(d))$. In addition, let $(X^N/S_N,\tau_{S_N})$ denote
the topological space equipped with the quotient topology of
$X^N/S_N$ relative to $(X,\tau)$,
\begin{equation}
\tau_{S_N}=\{U\subseteq X^N/S_N \,|\, \nu^{-1}(U)\in\tau\}, \lb{C.6}
\end{equation}
We now investigate a case in which $(X^N/S_N,\tau_{S_N})$ is also
generated by a metric $D$ on $X^N/S_N$. For this purpose we now
assume that the metric $d$ is such that each permutation in $S_N$ is an
isometry\footnote{This holds for $X=\bbC$, $X=\bbC/\Lambda_\omega$, and
$X=\bbC/\Lambda_{2\omega_1,2\omega_3}$ and the usual metrics on them
(cf.\ Remark \ref{rC.2}).}, that is,
\begin{equation}
\text{for all $\pi\in S_N$: } \, d(\pi(x),\pi(y))=d(x,y),
\quad x,y \in X^N \lb{C.7}
\end{equation}
(here $x=(x_1,\dots,x_N)\in X^N$, etc.). A standard situation in which
\eqref{C.7} can be verified is as follows: Suppose $\delta$ is a
metric on $X$. Then for any fixed $r\in [1,\infty)$, $d_r \colon
X^N\times X^N\to [0,\infty)$, defined as
\begin{equation}
d_r (x,y)=\bigg(\sum_{j=1}^N \delta (x_j,y_j)^r\bigg)^{1/r},
\quad x=(x_1,\dots,x_N), \, y=(y_1,\dots,y_N)\in X^N,
\end{equation}
defines a metric on $X^N$ satisfying \eqref{C.7} (and similarly in the
case $r=\infty$ using the supremum over $j\in\{1,\dots,N\}$).

Since $S_N$ is transitive, the expression ${\min}_{\sigma,\rho\in
S_N}\{d(\sigma(x),\rho(y))\}$ does not change  when $x$ and $y$ are
replaced by other representatives in their respective equivalence
classes, that is, it depends only on $[x]$ and
$[y]$. Hence, we may define
\begin{equation}
D([x],[y])={\min}_{\sigma,\rho\in S_N}\{d(\sigma(x),\rho(y))\}, \quad
[x], [y] \in X^N/S_N. \lb{C.8}
\end{equation}
The assumption that the permutations are isometries then yields
\begin{equation}
D([x],[y])={\min}_{\rho\in S_N}\{d(x,\rho(y))\}, \quad
[x], [y] \in X^N/S_N. \lb{C.9}
\end{equation}

\begin{theorem} \lb{tC.1}
Let $(X^N,d)$ be a metric space and suppose that every permutation in
$S_N$ is an isometry on $X^N$. Define $D$ as in \eqref{C.9}. Then
$(X^N/S_N,D)$ is a metric space and the topology $(D)$ induced by the
metric $D$ on $X^N/S_N$ is the quotient topology $\tau_{S_N}$,
$(X^N/S_N,(D))=(X^N/S_N,\tau_{S_N})$.
\end{theorem}
\begin{proof}
Clearly $D$ assumes non-negative real values only and symmetry of $D$
follows immediately from \eqref{C.8}. If $[x]=[y]$ then there is
a $\rho\in S_N$ such that $x=\rho(y)$. Hence $d(x,\rho(y))=0$ and thus
$D([x],[y])=0$. Next, suppose that $D([x],[y])=0$. Then there exists a
$\rho\in S_N$ such that $x=\rho(y)$, that is, $y$ is equivalent to $x$
and hence $[x]=[y]$.  For the triangle inequality one notes that, given
$z\in X^N$,
\begin{align}
D([x],[y])&={\min}_{\rho\in S_N}\{d(x,\rho(y))\} \no \\
   &\leq {\min}_{\rho\in S_N}\{d(x,\sigma(z))+d(\sigma(z),\rho(y))\} \no \\
   &=d(x,\sigma(z))+{\min}_{\rho\in S_N}\{d(\sigma(z),\rho(y))\} \no \\
   &=d(x,\sigma(z))+D([z],[y]), \quad \sigma\in S_N. \lb{C.10}
\end{align}
In particular \eqref{C.10} holds for that $\sigma$ which yields the
minimum of the right-hand side of \eqref{C.10} and hence $D$ is a metric
on $X^N/S_N$.

The metric $D$ induces a topology $\tilde\tau$ on $X^N/S_N$ and we
denote  the resulting topological space by $(X^N/S_N,\tilde\tau)$.
Let $\nu:X^N\to X^N/S_N,\, x\mapsto [x]$ denote the canonical
projection. We will next show that the map
$\nu:(X^N,d)\to (X^N/S_N,\tilde\tau)$ is  open and continuous. It is
obviously surjective. By \cite[Theorem 6.5.1]{Wi83} we then conclude that
$\tau_{S_N}=\tilde\tau$.

To prove that $\nu$ is continuous, let $U$ be an open set in
$(X^N/S_N,\tilde\tau)$. We want to show that
$\nu^{-1}(U)$ is open. Let $x$ be a point in $\nu^{-1}(U)$. Then $[x]$
is in $U$ and there is an $\varepsilon>0$ such that
$B([x],\varepsilon)$, the ball of radius $\varepsilon$ centered at
$[x]$, is a subset of $U$. Pick $y\in B(x,\varepsilon)\subset X^N$. We
note that
\begin{equation}
D([x],[y])\leq d(x,y)<\varepsilon, \lb{C.11}
\end{equation}
that is, $[y]\in B([x],\varepsilon)\subset U$ and thus $y\in\nu^{-1}(U)$.
Since $y$ is arbitrary, one infers $B(x,\varepsilon)\subset\nu^{-1}(U)$.

To prove that $\nu$ is open, let $V$ be an open set in $X^N$. We want to
show that $\nu(V)$ is open. Let $[x]$ be a point in $\nu(V)$. Then there
is a point in the equivalence class of $x$ which is in $V$. Without loss
of generality  we may assume that $x$ is that point. In addition, there
is an $\varepsilon>0$ such that $B(x,\varepsilon)$ is a subset of $V$.
Pick $[y]\in B([x],\varepsilon)\subset (X^N/S_N,\tilde\tau)$. Note that
this is equivalent to $D([x],[y])<\varepsilon$, which in turn means that
there is a $\rho$ in $S_N$ such that $d(x,\rho(y))<\varepsilon$. Hence
$\rho(y)\in B(x,\varepsilon)\subset V$ and thus
$[y]=\nu(y)=\nu(\rho(y))\in\nu(V)$. Since $[y]$ is arbitrary, one
concludes $B([x],\varepsilon)\subset\nu(V)$.
\end{proof}

\begin{remark} \lb{rC.2}
The results of this appendix apply in the three cases $X=\bbC$,
$X=\bbC/\Lambda_\omega$, $X=\bbC/\Lambda_{2\omega_1,2\omega_3}$
considered in Section \ref{s3}. For brevity we just take a quick look at
the simply periodic case $X=\bbC/\Lambda_\omega$:  Consider the
equivalence classes
$[x]=\{x+m\omega \,|\, x\in\bbC, \, m\in\bbZ\}\in \bbC/\Lambda_\omega$,
then the quotient topology on $\bbC/\Lambda_\omega$ is seen to be
generated by the following metric
$\delta\colon \bbC/\Lambda_\omega\times \bbC/\Lambda_\omega\to
[0,\infty)$ on $\bbC/\Lambda_\omega$,
\begin{equation}
\delta([x],[y])=\inf_{m,n\in\bbZ} |x+m\omega - (y+n\omega)|, \quad
[x], [y]\in \bbC/\Lambda_\omega.
\end{equation}
Analogous considerations apply to the elliptic case
$X=\bbC/\Lambda_{2\omega_1,2\omega_3}$.
\end{remark}

\section{The proof of Theorem \ref{t2.13}} \label{sD}

\renewcommand{\theequation}{D.\arabic{equation}}
\setcounter{theorem}{0}
\setcounter{equation}{0}

In this appendix we provide the proof of Theorem \ref{t2.13}.

\begin{theorem} \lb{tD.1}
Assume $M\in\bbN$, $s_\ell\in\bbN$, $1\leq\ell\leq M$, $q_0\in\bbC$,
and suppose $\zeta_\ell\in\bbC$, $\ell=1,\dots,M$, are pairwise
distinct. Consider
\begin{equation}
q(z)=q_0-\sum_{\ell=1}^M s_\ell(s_\ell+1)\cP(z-\zeta_\ell), \lb{2.66a}
\end{equation}
and suppose the DG locus conditions
\begin{equation}
\sum_{\substack{\ell^\prime=1\\ \ell^\prime\neq \ell}}^M
    s_{\ell^\prime}(s_{\ell^\prime}+1) \cP^{(2k-1)}(\zeta_{\ell}
-\zeta_{\ell^\prime})=0
    \text{ for $1\leq k\leq s_{\ell}$ and $1\leq \ell\leq M$} \lb{2.67a}
\end{equation}
are satisfied. Then
\begin{equation}
f_0=1, \quad
f_j(z)=d_j+\sum_{\ell=1}^M \sum_{k=1}^{\min(j,s_\ell)} a_{j,\ell,k}
\cP(z-\zeta_\ell)^k, \quad j\in\bbN \lb{2.68a}
\end{equation}
for some $\{a_{j,\ell,k}\}_{1\leq k \leq \min(j,s_\ell), 1\leq \ell\leq
M} \subset\bbC$ and $d_j\in\bbC$, $j\in\bbN$.
\end{theorem}
\begin{proof}
By equation \eqref{WP} we can treat the rational, simply periodic, and
elliptic cases simultaneously. \\
\noindent $(1)$ $j=1$: Then
\begin{align}
f_1(z) = c_1 +\frac{1}{2}\, q(z)
=c_1 + \frac{1}{2}\, q_0 - \sum_{\ell=1}^M  \frac{1}{2}
s_\ell(s_\ell+1)\, \cP(z-\zeta_\ell)
\label{}
\end{align}
is of the form \eqref{2.68a} with $d_1=c_1 + \frac{1}{2}q_0$ and
$a_{1,\ell,1}=  -\frac{1}{2} s_\ell(s_\ell+1) $. \\
\noindent $(2)$ We assume \eqref{2.68a} holds for some $j\in\bbN$, that
is,
\begin{equation}
f_j(z) =d_j+\sum_{\ell=1}^M \sum_{k=1}^{\min(j,s_\ell)}  a_{j,\ell,k}
\cP(z-\zeta_\ell)^k,
\end{equation}
or equivalently,
\begin{align}
f_j'(z)= \sum_{\ell=1}^M \sum_{k=1}^{\min(j,s_\ell)} a_{j,\ell,k}\,
k \, \cP(z-\zeta_\ell)^{k-1}\,  \cP'(z-\zeta_\ell). \label{6aa}
\end{align}
We now start the proof of \eqref{2.68a} for $j+1$: First, we recall the
recurrence relation \eqref{A.1},
\begin{align}
f_{j+1}'(z) &=\frac{1}{4} f_j(z)'''+ q(z) f_j'(z)+\frac{1}{2} q'(z)
f_j(z) \label{np1} \\
& =\frac{1}{4} f_j(z)'''+  (q(z) f_j (z))'- \frac{1}{2} q'(z) f_j(z)
\lb{10A} \\
&=\frac{1}{4} f_j(z)'''+ \frac{1}{2} q(z) f_j'(z)+\frac{1}{2} (q(z)
f_j(z))'. \label{10a}
\end{align}
Since $q$ is elliptic, so are $f_k$ for all $k\in\bbN$ by the recursion
relation \eqref{A.1} as the latter implies that each
$f_k$ is a differential polynomial in $q$. Equations \eqref{10A} and
\eqref{10a} then imply that as $z\to \zeta_\ell$, none of the terms in
\eqref{np1} can have a constant term or a term of the form
$(z-\zeta_{\ell})^{-1}$ in the Laurent expansion around $\zeta_\ell$.
This fact will be used repeatedly in the remainder of this proof.

Next we separately investigate each of the three terms on the right-hand
side of \eqref{np1}. For brevity we denote $\min(j,s_\ell)$ by $m$ in
the following. \\
$(i)$ Considering $f'''_j$ one computes
\begin{equation}
f'''_j(z) = \sum_{\ell=1}^M  \sum_{k=1}^{m} a_{j,\ell,k}
\frac{d ^3}{d z^3} \cP(z-\zeta_\ell)^k
\end{equation}
and
\begin{align}
\frac{d ^3}{d z^3} \cP(z-\zeta_\ell)^k
= & \big[k (2 k+1) (2k+2) \cP(z-\zeta_\ell)^{k} \cP' (z-\zeta_\ell) \no \\
& -g_2 k (k- \frac{1}{2}) (k-1)\cP(z-\zeta_\ell)^{k-2}
\cP' (z-\zeta_\ell) \no   \\
& -g_3 k (k-1) (k-2)\cP(z-\zeta_\ell)^{k-3} \cP'(z-\zeta_\ell)\big],
\lb{2.76}
\end{align}
using \eqref{B.3} and \eqref{B.4}. (For $k=2$ the term
$\cP(z-\zeta_\ell)^{k-3}$ does not occur in \eqref{2.76}, for $k=1$ the
terms $\cP(z-\zeta_\ell)^{k-3}$ and $\cP(z-\zeta_\ell)^{k-2}$ do not
occur in \eqref{2.76}.) Thus, $\tfrac{1}{4}f'''_j$ is of the expected
form \eqref{2.68a},
\begin{align}
\frac{1}{4}f'''_j(z)=\frac{1}{4}\sum_{\ell=1}^M \sum_{k=1}^{m+1} \tilde
a_{j+1,\ell,k} \cP(z-\zeta_\ell)^{k-1} \cP'(z-\zeta_\ell). \label{6a}
\end{align}
Moreover, the highest-order pole of $\tfrac{1}{4} f_j'''$ at
$\zeta_\ell$ reads
\begin{align}
\frac{1}{4} \, m ( 4 m +2) (m+1)\,  \frac{ (-2)
a_{j,\ell,m}}{(z-\zeta_\ell)^{2 m +3}} . \label{11a}
\end{align}
$(ii)$ Considering $qf_j'$ one obtains
\begin{align}
&q(z) f_j'(z) \no \\
& = \bigg( q_0-\sum_{\ell=1}^M
s_\ell(s_\ell+1)\cP(z-\zeta_\ell) \bigg)
\bigg(\sum_{\ell=1}^M  \sum_{k=1}^{m} a_{j,\ell,k}
k \cP(z-\zeta_\ell)^{k-1} \cP'(z-\zeta_\ell) \bigg) \no \\
& = q_0  \sum_{\ell=1}^M  \sum_{k=1}^{m} a_{j,\ell,k}
k \cP(z-\zeta_\ell)^{k-1} \cP'(z-\zeta_\ell)  \no \\
& \quad -  \sum_{\ell=1}^M   \sum_{k=1}^{m} s_\ell(s_\ell+1) a_{j,\ell,k}
k \cP(z-\zeta_\ell)^{k} \cP'(z-\zeta_\ell)  \label{2.79} \\
&\quad -\sum_{\ell=1}^M \bigg[\bigg( \sum_{k=1}^{m} a_{j,\ell,k}
k \cP(z-\zeta_\ell)^{k-1} \cP'(z-\zeta_\ell) \bigg)
\bigg( \sum_{\substack{\ell'=1\\ \ell'\neq \ell}}^M
s_{\ell^\prime}(s_{\ell^\prime}+1)\cP(z-\zeta_{\ell'}) \bigg) \bigg].
\no
\end{align}
The first two terms on the right-hand side of \eqref{2.79} are already of
the expected form \eqref{2.68a}. Next, we investigate the third term in
\eqref{2.79}. Let
\begin{equation}
g_{1, \ell}(z) =  \sum_{k=1}^{m} a_{j,\ell,k}\,
k \, \cP(z-\zeta_\ell)^{k-1}\,  \cP'(z-\zeta_\ell), \;\;
h_{1, \ell}(z) =   \sum_{\substack{\ell'=1\\ \ell'\neq \ell}}^M
s_{\ell^\prime}(s_{\ell^\prime}+1)
\cP(z-\zeta_{\ell'}).
\end{equation}
Then the third term in \eqref{2.79} equals
$-\sum_{\ell=1}^M g_{1, \ell} h_{1, \ell}$. Next we recall (cf.\
\eqref{B.17}) that any elliptic function $f$ can be written in the form
\begin{align}
f(z) & = A_0 +\sum^M_{\ell=1} \sum^{ s}_{k=1}(-1)^{k-1}
\frac{A_{\ell,k}}{(k-1)!} \zeta^{(k-1)}(z-\zeta_\ell), \ \ s \in
\bbN
\end{align}
for appropriate $M,s\in\bbN$, $A_0, A_{\ell,k}\in\bbC$, $1\leq \ell\leq
M$, $1\leq k\leq s$.  Here $\zeta(\cdot)=\zeta(\cdot\,|g_2,g_3)$
abbreviates the Weierstrass $\zeta$-function in the elliptic case
associated with the invariants $g_2$ and $g_3$ (see \cite[Sect.\
18.1]{AS72}) and
\begin{equation}
\zeta(z)=\begin{cases} \zeta(z|0,0)=1/z & \text{in the rational
case,} \\
\zeta\big(z|[2\pi^2/\omega^2]^2/3,[2\pi^2/\omega^2]^3/27\big) \\
=[\pi^2z/(3\omega^2)]+ (\pi/\omega)\cot(\pi z/\omega) & \text{in the
simply periodic case} \end{cases}
\end{equation}
(cf.\ \cite[p.\ 652]{AS72}). Since $g_{1,\ell}$ and $ h_{1,\ell}$ are
elliptic, we thus have
\begin{align}
    \sum_{\ell=1}^M g_{1, \ell}(z)   & = G_{1,0}+\sum^M_{\ell=1} \sum^{2
m +1}_{k=1}(-1)^{k-1}
\frac{G_{1,\ell,k}}{(k-1)!} \zeta^{(k-1)}(z-\zeta_\ell),  \\
    \sum_{\ell=1}^M g_{1, \ell}(z) h_{1, \ell}(z) & = B_0 +
\sum^M_{\ell=1} \sum^{2 m+1}_{k=1}(-1)^{k-1}
\frac{ B_{\ell,k}}{(k-1)!} \zeta^{(k-1)}(z-\zeta_\ell) .
\end{align}
To calculate $B_{\ell,k}$ we expand $g_{1, \ell}$ and $h_{1,\ell}$ at
$z=\zeta_{\ell}$ using \eqref{2.67a}. First we recall (cf.\ \eqref{P} and
\cite[Sect.\ 18.5]{AS72})
\begin{align}
\cP (z) =  \frac{1}{z^2} + \sum_{r=2}^{\infty} c_{r} z^{ 2 r-2}.
\label{}
\end{align}
Thus, $\cP^k$ admits the Laurent expansion
\begin{align}
(\cP (z))^k =  \frac{1}{z^{2k}} +
\frac{1}{z^{2k-4}}\sum_{s=0}^{\infty} d_{s} z^{ 2 s} \label{pk}
\end{align}
with only even orders of $z$ occurring in the expansion of $\cP^k$
since $\cP$ is an even function. For the derivative of $\cP^k$ one
computes
\begin{align}
\frac{d}{dz}(\cP (z))^k = (-2k) \frac{1}{z^{2k+1}} +
(-2k+4)\frac{1}{z^{2k-3}}\sum_{s=0}^{\infty} d_{s} z^{ 2 s}
+\frac{1}{z^{2k-4}}\sum_{s=1}^{\infty} d_{s} 2 s  z^{ 2 s-1}
\label{}  \no
\end{align}
and hence only odd orders of $z$ occur in the expansion of
$\frac{d}{dz} (\cP (z) )^k$. Thus, one concludes that only odd orders
of $z$ occur in the expansion of $g_{1,\ell}$ at $z=\zeta_\ell$.

On the other hand any elliptic function $f$, whose residue at
$\zeta_\ell$ vanishes and whose principal part of its Laurent expansion
at $z=\zeta_\ell$ contains only odd terms, can be written in the form
\begin{equation}
f(z)=\sum_{k=1}^{n_\ell} \tilde d_k \frac{d}{dz} (\cP (z-\zeta_\ell) )^k
+ O(1) \lb{2.87}
\end{equation}
for $z$ in a neighborhood of $\zeta_\ell$. Here $n_\ell\in\bbN$
depends on the order of the pole of $f$ at $\zeta_\ell$.

By \eqref{2.67a}  the odd powers of $(z-\zeta_\ell)^j$ in
the  expansion of $h_{1, \ell}(z)$
at $z=\zeta_\ell$ up to order   $(2 s_\ell -1)$  are zero and hence
\begin{align}
&h_{1, \ell}(z) = h_{1,\ell,0} + \sum_{k=1}^{\infty}
\frac{h_{1,\ell}^{(k)}(\zeta_{\ell}) }{k!} (z-\zeta_{\ell})^k
= \sum_{\substack{\ell'=1\\ \ell'\neq \ell}}^M
s_{\ell^\prime}(s_{\ell^\prime}+1)
\cP(\zeta_{\ell}-\zeta_{\ell'})   \no\\
& \quad +  \sum_{\substack{\ell'=1\\ \ell'\neq \ell}}^M
s_{\ell^\prime}(s_{\ell^\prime}+1)
\cP'(\zeta_{\ell}-\zeta_{\ell'}) (z-\zeta_{\ell})
   +  \frac{1}{2} \sum_{\substack{\ell'=1\\ \ell'\neq \ell}}^M
s_{\ell^\prime}(s_{\ell^\prime}+1)
\cP''(\zeta_{\ell}-\zeta_{\ell'}) (z-\zeta_{\ell})^2 \no \\
& \quad + \frac{1}{6} \sum_{\substack{\ell'=1\\ \ell'\neq \ell}}^M
s_{\ell^\prime}(s_{\ell^\prime}+1)
\cP^{(3)}(\zeta_{\ell}-\zeta_{\ell'}) (z-\zeta_{\ell})^3 + \ldots\no\\
&= h_{1,\ell,0} +  h_{1,\ell,2} (z-\zeta_{\ell})^{   2  } +
h_{1,\ell,4} (z-\zeta_{\ell})^{   4  } +
\ldots + h_{1,\ell,2 s_\ell}
(z-\zeta_{\ell})^{   2 s_\ell }  \no \\
& \quad + O\big((z-\zeta_{\ell})^{2 s_\ell+1 }\big).
\end{align}

Expanding $ g_{1,\ell} h_{1,\ell}$ at  $z= \zeta_\ell$ then yields
\begin{align}
g_{1,\ell}(z) h_{1,\ell}(z)  &= b_{ -2 m-1}
\frac{1}{(z-\zeta_{\ell})^{ 2 m+1}}+
b_{ -2 m+1} \frac{1}{(z-\zeta_{\ell})^{ 2 m-1}}+ \ldots \no\\
&\quad +  b_{2 s_\ell- 2m -1}  (z-\zeta_{\ell})^{2 s_\ell -2 m-1}
+O \big((z-\zeta_{\ell})^{2 s_\ell -2 m }\big) .
\label{14a}
\end{align}
By \eqref{2.87} we can write (\ref{14a}) as
\begin{align}
    g_{1, \ell}(z) h_{1, \ell}(z)&=
    \sum_{k=1}^{m} e_{j,\ell,k}
k \cP(z-\zeta_\ell)^{k-1}\,  \cP'(z-\zeta_\ell)
+ \frac{c_{1,\ell}}{z-\zeta_\ell} + c_{0,\ell} \no \\
& \quad + \Oh\big((z-\zeta_{\ell})^{ 1 }\big). \label{2.90}
\end{align}
Since no terms of the form $(z-\zeta_\ell)^{-1}$ and no constant term
can occur in $\sum_{\ell=1}^M g_{1, \ell} h_{1, \ell}$ by the comment
following \eqref{10a}, the coefficients $c_{1,\ell}$ of
$(z-\zeta_{\ell})^{-1}$, $\ell=1,\dots,M$, in \eqref{2.90}, as well as
the constant term $\sum_{\ell=1}^{M} c_{0,\ell}$, must be zero and we
arrive at the expected form \eqref{2.68a},
\begin{align}
    \sum_{\ell=1}^M g_{1, \ell}(z) h_{1, \ell}(z)=
    \sum_{\ell=1}^M \sum_{k=1}^{m} e_{j,\ell,k}\,
k \, \cP(z-\zeta_\ell)^{k-1}\,  \cP'(z-\zeta_\ell) \label{}
\end{align}
of the third term in \eqref{2.79}. The highest-order pole of $q f_j'$ at
$\zeta_\ell$ reads
\begin{align}
- s_\ell(s_\ell+1)m\frac{(-2)\, a_{j,\ell,m}}{(z-\zeta_\ell)^{2m+3}}.
\label{12a}
\end{align}
$(iii)$ Considering $\frac{1}{2}q'f_j$ one obtains
\begin{align}
\frac{1}{2} q'(z) f_j(z)  &=-\frac{1}{2} \,
    \bigg(   \sum_{\ell=1}^M s_\ell(s_\ell+1)\cP '(z-\zeta_\ell) \bigg)
\bigg(d_j + \sum_{\ell=1}^M  \sum_{k=1}^{m} a_{j,\ell,k}\,
    \cP(z-\zeta_\ell)^{k}    \bigg) \no \\
& = -\frac{1}{2} \, d_j \sum_{\ell=1}^M s_\ell(s_\ell+1)
\cP '(z-\zeta_\ell) \no \\
& \quad - \frac{1}{2} \, \sum_{\ell=1}^M  \sum_{k=1}^{m}
s_\ell(s_\ell+1)  a_{j,\ell,k}\,
     \cP(z-\zeta_\ell)^{k}\,  \cP'(z-\zeta_\ell) \lb{2.93} \\
& \quad - \frac{1}{2} \, \sum_{\ell=1}^M \bigg[\bigg( \sum_{k=1}^{m}
a_{j,\ell,k}\,
    \cP(z-\zeta_\ell)^{k} \bigg) \bigg(
\sum_{\substack{\ell'=1\\ \ell'\neq
\ell}}^M s_{\ell^\prime}(s_{\ell^\prime}+1)
\cP'(z-\zeta_{\ell'}) \bigg)\bigg]. \no
\end{align}
The first two terms in \eqref{2.93} are already of the expected form
\eqref{2.68a}. Next we investigate the third term in \eqref{2.93}. Let
\begin{align}
g_{2,\ell}(z) =   \sum_{k=1}^{m} a_{j,\ell,k}\,
    \cP(z-\zeta_\ell)^{k}, \quad
h_{2,\ell}(z) = \bigg( \sum_{\substack{\ell'=1\\ \ell'\neq \ell}}^M
s_{\ell^\prime}(s_{\ell^\prime}+1)
\cP'(z-\zeta_{\ell'}) \bigg).
\label{}
\end{align}
Then the third term in \eqref{2.93} equals $-\f{1}{2}\sum_{\ell=1}^M
g_{2,\ell} h_{2,\ell}$. Since $g_{2,\ell}$ and $ h_{2,\ell}$  are
elliptic, one has
\begin{align}
\sum^M_{\ell=1} g_{2,\ell}(z) & = G_{2,0}+\sum^M_{\ell=1} \sum^{2 m
}_{k=1}(-1)^{k-1}
\frac{G_{2,\ell,k}}{(k-1)!} \zeta^{(k-1)}(z-\zeta_\ell), \\
\sum^M_{\ell=1} g_{2,\ell} (z) h_{2,\ell} (z) & = D_0
+\sum^M_{\ell=1} \sum^{ 2 m-1}_{k=1}(-1)^{k-1}
\frac{ D_{\ell,k}}{(k-1)!} \zeta^{(k-1)}(z-\zeta_\ell)   .
\end{align}
  From \eqref{pk} one concludes that only even orders in $z$ can occur in
the expansion of $g_{2,\ell}$ at $z=\zeta_\ell$. Next we expand
$h_{2,\ell}$ at $z=\zeta_{\ell}$. By \eqref{2.67a}, the even powers of
$(z-\zeta_\ell)^k$ in the  expansion of $h_{2, \ell}$ at $z=\zeta_\ell$
up to order $(2 s_\ell -2)$  are zero and hence,
\begin{align}
h_{2,\ell}(z)& =  \sum_{k=0}^{\infty}
\frac{h_{2,\ell}^{(k) }(\zeta_{\ell})}{k!} (z-\zeta_{\ell})^k \no\\
&=  h_{1,\ell,1} (z-\zeta_{\ell}) + h_{1,\ell,3} (z-\zeta_{\ell})^{3} +
\ldots + h_{1,\ell,2 s_\ell-1}(z-\zeta_{\ell})^{   2 s_\ell -1} \no \\
& \quad + \Oh \big((z-\zeta_{\ell})^{2 s_\ell}\big).
\end{align}
Expanding $ g_{2,\ell} \, h_{2,\ell}$ at  $z= \zeta_\ell$ then yields
\begin{align}
g_{2,\ell}(z) h_{2,\ell}(z)  &= \tilde b_{ -2 m+1} \frac{1}{
(z-\zeta_{\ell})^{ 2 m-1}}+
\tilde b_{ -2 m+3} \frac{1}{(z-\zeta_{\ell})^{ 2 m-3}}+ \ldots \no\\
& \quad + \tilde b_{2 s_\ell- 2m -1}  (z-\zeta_{\ell})^{2 s_\ell -2 m-1}
+O \big((z-\zeta_{\ell})^{2 s_\ell -2 m }\big). \label{20a}
\end{align}
By \eqref{2.87} we can write \eqref{20a} as
\begin{align}
    g_{2, \ell}(z) h_{2, \ell}(z)=
    \sum_{k=1}^{m-1} \tilde e_{j,\ell,k}\,
k \, \cP(z-\zeta_\ell)^{k-1}\,  \cP'(z-\zeta_\ell)
      + \frac{\tilde c_{1,\ell}}{z-\zeta_\ell} +\tilde c_{0,\ell} + O
\big((z-\zeta_{\ell})^{ 1 }\big).
\label{2.97}
\end{align}
Since no terms of the form $(z-\zeta_\ell)^{-1}$ and no constant term
can occur in $\sum_{\ell=1}^M g_{2, \ell} h_{2, \ell}$ by the comment
following \eqref{10a}, the coefficients $\ti c_{1,\ell}$ of
$(z-\zeta_{\ell})^{-1}$, $1\leq \ell\leq M$, in \eqref{2.97}, as well as
the constant term $\sum_{\ell=1}^{M} \ti c_{0,\ell}$, must vanish and we
arrive at the expected form \eqref{2.68a},
\begin{align}
    \sum_{\ell=1}^M g_{2, \ell}(z) h_{2, \ell}(z)=
    \sum_{\ell=1}^M \sum_{k=1}^{m-1} \tilde e_{j,\ell,k}
k \cP(z-\zeta_\ell)^{k-1} \cP'(z-\zeta_\ell)
       \no . \label{}
\end{align}
The highest-order pole of $\frac{1}{2}q'f_n$ at $\zeta_\ell$ reads
\begin{align}
- \frac{1}{2}\, s_\ell(s_\ell+1)\frac{ (-2)\, a_{j,\ell,m}
}{(z-\zeta_\ell)^{2 m +3}}. \label{15a}
\end{align}
Summing up (\ref{11a}), (\ref{12a}), and (\ref{15a}) yields
\begin{align}
\bigg[ \frac{1}{4} \, m ( 4 m +2) (m+1)   - s_\ell(s_\ell+1) m -
\frac{1}{2} s_\ell(s_\ell+1)
\bigg]  \frac{-2  a_{j,\ell,m}}{(z-\zeta_\ell)^{2 m +3}}. \label{18a}
\end{align}
This term becomes zero as soon as $m= s_\ell$. Summing up our analysis of
the three terms in \eqref{np1}, each term has the form \eqref{2.68a} and
the index $k$ does not exceed $\min(j+1,s_\ell)$, because of \eqref{18a}.
\end{proof}


\end{document}